\documentclass[prd,aps,showpacs,floats,letterpaper,floatfix,groupedaddress,eqsecnum,nofootinbib]{revtex4}

\usepackage{amssymb,amsmath}
\usepackage[dvips]{graphicx}

\usepackage{longtable}
\usepackage{dcolumn,epsfig}

\def\np{({\bf n}\cdot{\bf p})}
\def\pp{{\bf p}^2}
\def\ppp{({\bf p}^2)}

\def\laq{\raise 0.4ex\hbox{$<$}\kern -0.8em\lower 0.62ex\hbox{$\sim$}}
\def\gaq{\raise 0.4ex\hbox{$>$}\kern -0.7em\lower 0.62ex\hbox{$\sim$}}

\newlength{\sizeonefig}
\newlength{\sizetwofig}
\newcommand{\hL}{\mbox{\boldmath${\hat{L}}$}}

\newcommand{\vx}{\mbox{\boldmath${x}$}}
\newcommand{\vp}{\mbox{\boldmath${p}$}}
\newcommand{\vq}{\mbox{\boldmath${q}$}}
\newcommand{\vv}{\mbox{\boldmath${v}$}}
\newcommand{\vV}{\mbox{\boldmath${V}$}}

\newcommand{\vJ}{\mbox{\boldmath${J}$}}
\newcommand{\vF}{\mbox{\boldmath${F}$}}

\newcommand{\vO}{\mbox{\boldmath${\Omega}$}}
\newcommand{\vP}{\mbox{\boldmath${P}$}}

\newcommand{\vS}{\mbox{\boldmath${S}$}}
\newcommand{\vN}{\mbox{\boldmath${N}$}}
\newcommand{\vX}{\mbox{\boldmath${X}$}}
\newcommand{\vL}{\mbox{\boldmath${L}$}}
\newcommand{\vlb}{\mbox{\boldmath${\lambda}$}}
\newcommand{\vs}{\mbox{\boldmath${s}$}}
\newcommand{\beq}{\begin{equation}}
\newcommand{\eeq}{\end{equation}}
\newcommand{\bea}{\begin{eqnarray}} 
\newcommand{\eea}{\end{eqnarray}}
\newcommand{\ba}{\begin{array}}
\newcommand{\ea}{\end{array}}

\newcommand{\comment}[1]{}

\begin{document}

\title{Transition from inspiral to plunge in precessing binaries of spinning black holes}

\author{Alessandra Buonanno}

\affiliation{AstroParticule et Cosmologie (APC),
11, place Marcelin Berthelot, 75005 Paris, France}

\altaffiliation{UMR 7164 (CNRS, Universit\'e Paris7, CEA, Observatoire
de Paris). Also: Institut d'Astrophysique de Paris, 98$^{\rm bis}$ 
Boulv. Arago, 75013 Paris, France.}

\author{Yanbei Chen}

\affiliation{Max-Planck-Institut f\"ur Gravitationsphysik
(Albert-Einstein-Institut), Am M\"uhlenberg 1, D-14476 Golm bei
Potsdam, Germany}

\author{Thibault Damour}

\affiliation{Institut des Hautes Etudes Scientifiques, 91440 Bures-sur-Yvette, France}

\begin{abstract}
We investigate the non-adiabatic dynamics of spinning black hole
binaries by using an analytical Hamiltonian completed with a radiation-reaction force,
containing spin couplings, which matches the known rates of energy and
angular momentum losses on quasi-circular orbits. We consider
both a straightforward post-Newtonian-expanded Hamiltonian
(including spin-dependent terms), and a version of
the resummed post-Newtonian
Hamiltonian defined by the Effective One-Body approach.
We focus on the influence of spin terms onto the dynamics and waveforms.
We evaluate the energy and angular momentum released during the final
stage of inspiral and plunge. For an equal-mass
binary the energy released between 40\,Hz and the frequency beyond
which our analytical treatment becomes unreliable is found to be,
when using the more reliable Effective One-Body dynamics:
 $0.6\% M$ for  anti-aligned maximally spinning black holes,
 $5\% M$ for aligned maximally spinning black hole, and
$1.8\% M$ for non-spinning configurations. 
In confirmation of previous results, we find that, for all binaries considered,
the dimensionless rotation
parameter $J/E^2$ is always smaller than unity at the end of the inspiral, 
so that a Kerr black hole can form right after the inspiral phase.
By matching a quasi-normal mode ringdown to the last reliable stages
of the plunge, we construct complete waveforms approximately describing the
gravitational wave signal emitted by the entire process of coalescence
of precessing binaries of spinning black holes.
\end{abstract}

\maketitle

\section{Introduction}
\label{sec1}

An international network of kilometer-scale laser-interferometric
gravitational-wave detectors, consisting of the Laser-Interferometer
Gravitational-wave Observatory (LIGO) \cite{LIGO}, of VIRGO
\cite{VIRGO}, of GEO\,600 \cite{GEO} and of TAMA\,300 \cite{TAMA}, has
by now begun the science operations. TAMA\,300 reached its full
design sensitivity in 2001, VIRGO is in its commissioning phase
and plans to start the first scientific runs by the end of 2005, 
while LIGO has already completed three science runs (two of them 
in coincidence with GEO\,600) with increasing sensitivity and duty 
cycle. LIGO and GEO\,600 are expected to reach their full design 
sensitivity in 2005.

Binary black holes are among the most promising sources for these
detectors. Among black hole binaries, it was emphasized in \cite{TD} that
there is a bias towards first detecting mostly aligned spinning
binaries with high masses, whose last stable spherical orbits are drawn, 
by spin effects, to larger binding energies, yet due to their high masses 
these energies are still emitted through gravitational waves in the sensitive 
band of the detectors. Studying in detail the waveforms emitted during the 
last stages of dynamical evolution of such heavy spinning black hole binaries,
with explicit consideration of the crucial transition between
adiabatic inspiral and plunge, is a demanding theoretical challenge.
The aim of the present paper is to provide a first attack on this problem
by using some of the best analytical tools currently available to describe 
the transition from adiabatic inspiral to plunge, and notably
the Effective One Body (EOB) approach \cite{BD1,BD2}.

So far, most theoretical and data-analysis studies on precessing binaries of spinning black holes
assumed {\it adiabatic evolution}. Thus they were restricted to considering {\it only} 
the inspiral phase~\cite{ACST94,K,apostolatos0,apostolatos1,apostolatos2,
GKV,GK,bcv2,pbcv1,bcpv1,Gpc}. Actually, even for non-spinning
binary configurations, most theoretical studies confined themselves 
to considering the adiabatic inspiral phase, though a lot 
of effort was spent to improve the accuracy of the phasing during the last stages 
of the inspiral, see e.g. \cite{DIS98, DIS00}.

For heavy black hole binaries, most of the signal-to-noise ratio in the 
ground detectors will come from the very last stages of the
inspiral, and from the non-adiabatic transition between
inspiral and plunge. It is therefore essential to be able
to describe, with acceptable accuracy, this non-adiabatic evolution.
In Refs.~\cite{BD1,BD2} a new way of describing the dynamics of
binary systems was introduced: the Effective One-Body (EOB)
approach. The EOB approach uses both a specific resummation of the
post-Newtonian Hamiltonian, and a resummed version of
radiation reaction. This was shown to lead to a rather
robust formalism, which is likely to provide a reliable description
of non-adiabatic effects, of the transition between inspiral and
plunge, and of the beginning of the plunge. It was also used in
\cite{BD2} to model the full merger phase of
non-spinning binaries, by matching the natural end
of the EOB plunge with the ring-down phase. The EOB approach was used
in Refs.~\cite{BD2,DJS} to derive non-adiabatic template waveforms emitted
by {\it non-spinning} black hole binaries. It was shown in \cite{DIS01}
that these new EOB templates led to significantly enhanced signal to noise
ratios in current detectors (mainly because of the inclusion of
the plunge signal). The EOB Hamiltonian was extended to the
case of spinning black holes in \cite{TD}.
The analytical predictions made by the EOB method (including spin)
were found to agree remarkably well \cite{DGG} with the
 numerical results obtained by means of the helical Killing
vector approach \cite{GGB} for circular orbits of corotating
black holes. Several other studies showed that the EOB method
provides phasing models which are more reliable and robust
than other (adiabatic or non-adiabatic) models \cite{bcv1,DIS03}.

The main purpose of this paper is to extend the use of the EOB approach 
to the case of precessing binaries of spinning black holes,
both by including spin-dependent terms in radiation-reaction effects,
and by studying the waveforms generated beyond the adiabatic
approximation, i.e. taking into account the transition between
inspiral and plunge, and the plunge itself. Let us emphasize again
that the EOB approach has the advantage of providing an {\it analytical}
description of the transition from inspiral to plunge. Recently,
some attempts have been made to tackle, by means of 3D {\it numerical}
simulations (combined with a perturbative approach), the
gravitational radiation emitted by a very tight black hole binary both
in non-spinning~\cite{BBCLT}, and moderately spinning,
but non-precessing~\cite{BCLT} configurations. These three-dimensional (3D)
simulations concluded to the emission of a significantly larger amount of
energy in the form of gravitational radiation than what we shall
find from our analytical, EOB approach. It should be mentioned in
this respect that the energy released in the form of gravitational
waves depends very much on the choice of initial data, and that the
amount by which the initial data chosen in \cite{BBCLT,BCLT}
differ from  the physically correct ``no-incoming-radiation'' data
is unclear. This crucial issue will be further discussed below.

Recent simulations~\cite{Gpc} based on population synthesis codes predict that
$\sim 50 \% -80\%$ of neutron star-black hole (NS-BH) binaries in the Galactic 
field may have tilt angles (i.e., the angle between the black hole (BH) spin and the orbital
angular momentum) between $0$ and $40^o$ and $10-20 \%$ of
NS-BH binaries between $40^o$ and $50^o$. By studying the formation of close compact
binaries and the misalignment angle that can occur after the second core-collapse
event, Kalogera~\cite{K00} predicted that the majority of BH-BH binaries
in Galactic binaries may have a tilt angle
smaller than $30^{o}$. All these results assume that the misalignement is entirely due to
the recoil velocity (``kick'') imparted to the NS (or the smaller BH in the binary)
at birth by the core-collapse.
However, the spin properties of NS-BH and BH-BH binaries in dense environment
and centers of globular clusters could be very different than in
the Galactic field.
Considering the low event rates, $\sim 1$ per 2 years, of binary coalescences in
first generation of ground-based detectors, it is worthwhile to adopt a conservative
point of view and investigate waveforms for generic spin configurations.
Little is known about the magnitude of the spin of NSs and BHs.
{}From the observed pulsars the dimensionless rotation parameter $a_{\rm NS}$
takes values in the range $0.005 - 0.02$. Our analysis will focus on BH-BH binaries and
we shall consider arbitrary spins: $0 < a_{\rm BH} < 1$.

For completeness, we investigate the two-body dynamics by adopting two approaches:
the straightforward post-Newtonian (PN)-expanded Hamiltonian~\cite{DS88,JS98}
and the PN-resummed Hamiltonian \`a la EOB~\cite{BD1}, \cite{DJS}, \cite{TD} [henceforth
referred to simply as the EOB-Hamiltonian]. For simplicity, instead of using
the Kerr-deformed EOB-Hamiltonian derived by Damour in Ref.~\cite{TD}, we shall use as Hamiltonian
in this paper the {\it sum} of the purely orbital (Schwarzschild-deformed) 
EOB-Hamiltonian~\cite{BD1,DJS} and of the {\it separate} contributions
due to spin-orbit and spin-spin effects. [Note that, among the
spin-spin terms, there are the terms due to monopole-quadrupole interactions~\cite{EP}, \cite{TD}.]

The paper is organized as follows. Section~\ref{sec2} summarizes the main 
results of the conservative part of the two-body dynamics in the Hamiltonian
formalism, and contains formula for the PN-expanded and EOB Hamiltonians
up to 3PN order. In Sec.~\ref{sec3} we augment the dynamics with radiation-reaction 
effects. We derive the radiation-reaction force which includes spin effects and 
matches known rates of energy and angular momentum losses on quasi-circular orbits. 
[Our result agrees with the recent results of Will~\cite{Will} that appeared after we had completed our work.]
In Sec.~\ref{sec4.1} we define the two-body approximants and discuss
the initial conditions used to evolve 
the precessing two-body dynamics; we compare the
(lowest-order) waveforms
obtained using PN-expanded and (PN-resummed) EOB
dynamics by computing the overlaps between these two types of
waveforms. In view of the greater a priori reliability of the
EOB approach, we use them (and only them) to estimate the energy and angular momentum released
during the last stages of evolution. Section~\ref{sec5} contains our 
main conclusions.

We leave to future work a thorough application of
our results to data analysis purposes.

\section{Conservative Hamiltonian including spin-orbit and spin-spin effects}
\label{sec2}

\subsection{Orbital third-post-Newtonian expanded Hamiltonian}
\label{sec2.1}

The  purely {\it orbital} (non-spinning)third-post-Newtonian Hamiltonian
$H^{\rm 0}=H^{\rm 0}(\vX,\vP)$ (in the center of mass frame,
and after subtraction of the total rest-mass term)
was derived in Ref.~\cite{JS98} (completed by Refs.~\cite{DJSPoincare,DJSd}). In scaled variables, and
written as a straigthforward PN-expansion, it reads
(see Ref.~\cite{DJS}):
\beq
\label{3.1}
H^{\rm 0}(\vX,\vP)_{\rm EXP} = \mu\,\widehat{H}^{\rm 0}(\vq,\vp) = \mu\,\left [\widehat{H}_{\rm Newt}(\vq,\vp) +
\widehat{H}_{\rm 1PN}(\vq,\vp) +\widehat{H}_{\rm 2PN}(\vq,\vp) +
\widehat{H}_{\rm 3PN}(\vq,\vp) \right ]
\eeq 
where $\mu = m_1\,m_2/M$, $M=m_1+m_2$ and $(\vq,\vp)$ denote the
canonical variables ${\bf p}\equiv {\bf P_1}/\mu=-{\bf P_2}/\mu$, and 
${\bf q} \equiv {\bf X}/M =({\bf X}_1 - {\bf X}_2)/M$, where ${\bf X_1}$ and ${\bf X_2}$ 
are the positions of the black hole centers of mass in quasi--Cartesian Arnowitt-Deser-Misner 
(ADM) coordinates. The Newtonian term and the PN contributions read
(denoting $\eta \equiv \mu/M = m_1m_2/M^2$):
\bea
\widehat{H}_{\rm Newt}\left({\bf q},{\bf p}\right) &=& \frac{\pp}{2} -
\frac{1}{q}\,, \label{eq:hfirst}\\
\widehat{H}_{\rm 1PN}\left({\bf q},{\bf p}\right) &=& \frac{1}{8}(3\eta-1)\ppp^2
- \frac{1}{2}\left[(3+\eta)\pp+\eta\np^2\right]\frac{1}{q} + \frac{1}{2q^2}\,,\\
\widehat{H}_{\rm 2PN}\left({\bf q},{\bf p}\right)
&=& \frac{1}{16}\left(1-5\eta+5\eta^2\right)\ppp^3
+ \frac{1}{8} \left[
\left(5-20\eta-3\eta^2\right)\ppp^2-2\eta^2\np^2\pp-3\eta^2\np^4 \right]\frac{1}{q}
\nonumber \\
&& + \frac{1}{2} \left[(5+8\eta)\pp+3\eta\np^2\right]\frac{1}{q^2}
- \frac{1}{4}(1+3\eta)\frac{1}{q^3}\,,
\eea
\bea
\widehat{H}_{\rm 3PN}\left({\bf q},{\bf p}\right)
&=& \frac{1}{128}\left(-5+35\eta-70\eta^2+35\eta^3\right)\ppp^4
\nonumber \\
&& + \frac{1}{16}\left[
\left(-7+42\eta-53\eta^2-5\eta^3\right)\ppp^3
+ (2-3\eta)\eta^2\np^2\ppp^2
+ 3(1-\eta)\eta^2\np^4\pp - 5\eta^3\np^6
\right]\frac{1}{q}
\nonumber \\
&& +\left[ \frac{1}{16}\left(-27+136\eta+109\eta^2\right)\ppp^2
+ \frac{1}{16}(17+30\eta)\eta\np^2\pp + \frac{1}{12}(5+43\eta)\eta\np^4
\right]\frac{1}{q^2} \\
&& +\left\{ \left[ -\frac{25}{8} + \left(\frac{1}{64}\pi^2-\frac{335}{48}\right)\eta 
- \frac{23}{8}\eta^2 \right]\pp
+ \left(-\frac{85}{16}-\frac{3}{64}\pi^2-\frac{7}{4}\eta\right)\eta\np^2 
\right\}\frac{1}{q^3}
\nonumber \\
&& + \left[ \frac{1}{8} + \left(\frac{109}{12}-\frac{21}{32}\pi^2\right)\eta 
\right]\frac{1}{q^4}\,, \label{eq:hlast}
\eea
where the scalars $q$ and $p$ are the
(coordinate) lengths of the two vectors $\vq$ and $\vp$; and the vector ${\bf n}$ is
just ${\bf q}/q$.

\subsection{Orbital third-post-Newtonian effective-one-body Hamiltonian}
\label{sec2.2}

As was emphasized in previous work (see e.g. \cite{KWW,WS}), and as
we shall confirm below, the non-resummed PN-expanded Hamiltonian
(or the non-resummed PN-expanded equations of motion) do not lead to
a reliable description of the evolution near the last stable
circular orbit, nor, {\it a fortiori} during the transition
between inspiral and plunge. On the other hand, it was argued
in \cite{BD1,BD2,DJS} that the EOB approach
defines a specific resummation of the PN-expanded Hamiltonian
which leads to a much more reliable description of the dynamical
evolution near the last stable circular orbit, and of the
transition between inspiral and plunge.

The explicit expression of the purely
orbital, EOB-Hamiltonian is~\cite{BD1}:

\beq
\label{himpr}
H^{\rm 0}_{\rm EOB}(\vX^\prime,\vP^\prime) = M\,\sqrt{1 + 2\eta\,\left ( \frac{H_{\rm eff}(\vX^\prime,\vP^\prime) 
-\mu}{\mu}\right )} -M\,.  
\eeq
where $H_{\rm eff}$ is given by~\cite{BD1,DJS}:
\bea 
\label{eq:genexp}
&& H_{\rm eff}(\vX^\prime,\vP^\prime) = \mu\, 
\widehat{H}_{\rm eff}({\mathbf q}^\prime,{\mathbf p}^\prime) \nonumber \\
&=& \mu\,\sqrt{A (q^\prime) \left[ 1 + 
{\mathbf p}^{\prime\,2} +
\left( \frac{A(q^\prime)}{D(q^\prime)} - 1 \right) ({\mathbf n}^\prime \cdot {\mathbf p}^\prime)^2
+ \frac{1}{q^{\prime\,2}} \left( z_1 ({\mathbf p}^{\prime\,2})^2 + z_2 \, {\mathbf p}^{\prime\,2}
({\mathbf n}^\prime \cdot {\mathbf p}^{\prime\,})^2 + z_3 ({\mathbf n}^\prime \cdot {\mathbf
p}^{\prime})^4 \right) \right]} \,,
\eea
with ${\bf q}^\prime$ and ${\bf p}^\prime$ being the reduced canonical variables obtained by rescaling 
$\vX^\prime$ and $\vP^\prime$ by $M$ and $\mu$, respectively, ${\bf n}^\prime = \vq^\prime/q^\prime$ 
where we set $q^\prime = |{\bf q}^\prime|$. 
The coefficients $z_1,z_2$ and $z_3$ are arbitrary, subject to the constraint
\beq 8z_1 + 4z_2 +3z_3 = 6(4-3\eta)\,\eta\,.  
\eeq
Setting (as in Ref.~\cite{bcv1})  $z_1 = \eta \tilde{z}_1,
z_2 =\eta \tilde{z}_2$ and $z_3 =\eta \tilde{z}_3$, the coefficients $A(q^\prime)$ and $D(q^\prime)$ in Eq.~(\ref{eq:genexp}) read:
\bea
\label{coeffA}
A(q^\prime) &=& 1 - \frac{2}{q^\prime}+\frac{2\eta}{q^{\prime\,3}}+ \left [ \left (
\frac{94}{3}-\frac{41}{32}\pi^2\right ) -\tilde{z}_1\right ]\,\frac{\eta}{q^{\prime\,4}}\,,\\
\label{coeffD}
D(q^\prime) &=& 1 -\frac{6\eta}{q^{\prime\,2}}+\left [7\tilde{z}_1 +\tilde{z}_2+ (3\eta-26)\right ]\,
\frac{\eta}{q^{\prime\,3}}\,. 
\eea
As done in Ref.~\cite{DJS}, 
we restrict ourselves to the case $\tilde{z}_1=\tilde{z}_2=0$ and improve the behavior
\footnote{As shown in \cite{DJS}, the use of the
straightforward PN-expanded, 3PN-{\it accurate} ``effective potential''
$A(q^\prime)$ does not lead to a well-defined last stable circular orbit 
(contrary to what happens in the 2PN-accurate case \cite{BD1}). This
is due to the rather large value of the 3PN coefficient 
$\frac{94}{3}-\frac{41}{32}\pi^2 \simeq 18.688$ entering the PN expansion
of $A(q^\prime)$. Replacing the PN-expanded form of $A(q^\prime)$
by a Pad\'e approximant cures this problem.[See also \cite{DGG} for
a comparison of the physical consequences of various possible
 resummations of $A(q^\prime)$.]} by replacing
 the ``effective potential'' $A(q^\prime)$ with the Pad\'e approximants
\beq
A_{P_2}(q^\prime) = \frac{q^\prime\,(-4+2q^\prime+\eta)}{2q^{\prime\,2}+2\eta+q^\prime\,\eta}\,,
\label{coeffPA2}
\eeq
at 2PN order and
\beq
\label{coeffPA}
A_{P_3}(q^\prime) = \frac{q^{\prime\,2}\,[(a_4(\eta,0)+8\eta-16) + q^\prime\,(8-2\eta)]}{
q^{\prime\,3}\,(8-2\eta)+q^{\prime\,2}\,(a_4(\eta,0)+4\eta)+q^\prime\,(2a_4(\eta,0)+8\eta)+4(\eta^2+a_4(\eta,0))}\,,
\eeq
at 3PN order where
\beq \label{a4}
a_4(\eta,\tilde{z}_1) = \left [\frac{94}{3}-\frac{41}{32}\pi^2
-\tilde{z}_1\right ]\,\eta\,.
\eeq

\subsection{Adding spin couplings}
\label{sec2.3}

There are several ways of including spin effects in the Hamiltonian
dynamics. When considering the PN-expanded form of the
orbital Hamiltonian $H^{\rm 0}_{\rm EXP}$,
it is natural to add the spin-dependent terms as further additional
contributions: $ H_{\rm TOT} = H^{\rm 0}_{\rm EXP} +H^{\rm SPIN}$ .
On the other hand, when considering the EOB-resummed
form of the Hamiltonian $H^{\rm 0}_{\rm EOB}$,
it has been argued in Ref.~\cite{TD} that
it is probably better to incorporate most of the spin effects
within a suitably generalized (\`a la  Kerr) EOB-Hamiltonian,
whose explicit form will be found in \cite{TD}.
In the present work, we shall, for technical simplicity,
adopt a uniform way of including spin effects. Namely,
we shall simply include them a linearly added contributions
to the basic (PN-expanded or EOB-resummed) orbital Hamiltonian.
We shall verify below that, in the EOB-resummed case, the
two different ways (\`a la \cite{TD}, or as in the following
equation) of incorporating spin effects lead to
very similar physical effects.

Finally, the explicit forms we shall use of the ``spinning''
Hamiltonian read:

\beq
H_{\rm EXP}(\vX,\vP,\vS_1,\vS_2) = H^{\rm 0}_{\rm EXP}(\vX,\vP) +
H_{\rm SO}(\vX,\vP,\vS_1,\vS_2)+ H_{\rm SS}(\vX,\vP,\vS_1,\vS_2)\,,
\label{Hspinexp}
\eeq
or
\beq
H_{\rm EOB}(\vX,\vP,\vS_1,\vS_2) =  H^{\rm 0}_{\rm EOB}(\vX,\vP) +
H_{\rm SO}(\vX,\vP,\vS_1,\vS_2)+ H_{\rm SS}(\vX,\vP,\vS_1,\vS_2)\,,
\label{Hspineob}
\eeq

where~\cite{BGH,BO,DS88}:
\bea
\label{3.2}
&& H_{SO} = 2\frac{\vS_{\rm eff}\cdot \vL}{R^3}\,, \quad 
\vS_{\rm eff} \equiv \left ( 1+ \frac{3}{4}\,\frac{m_2}{m_1} \right )\,\vS_1 + 
\left ( 1 + \frac{3}{4}\,\frac{m_1}{m_2} \right )\,\vS_2\,,\\
\label{3.3}
&& H_{SS} = H_{S_1 S_1} + H_{S_1 S_2} + H_{S_2 S_2} =
\frac{1}{2R^3}\,\frac{\mu}{M}\,\left [ 3 (\vS_0 \cdot \vN) (\vS_0
\cdot \vN) - (\vS_0 \cdot \vS_0) \right ]\,, \\ && \vS_0 = \left ( 1 +
\frac{m_2}{m_1} \right )\,\vS_1 + \left ( 1 + \frac{m_1}{m_2} \right
)\,\vS_2\,,\\ && H_{S_1 S_2} = \frac{1}{R^3}\,\left [ 3 (\vS_1 \cdot
\vN) (\vS_2 \cdot \vN) - (\vS_1 \cdot \vS_2) \right ]\,,\\ && H_{S_1
S_1} + H_{S_2 S_2}= \frac{1}{2R^3}\,\left [ 3 (\vS_1 \cdot \vN) (\vS_1
\cdot \vN) - (\vS_1 \cdot \vS_1) \right ]\,\frac{m_2}{m_1} +
\frac{1}{2R^3}\,\left [ 3 (\vS_2 \cdot \vN) (\vS_2 \cdot \vN) - (\vS_2
\cdot \vS_2) \right ]\,\frac{m_1}{m_2} \,, 
\eea
with $\vN = \vX/R$, $R = |\vX|$ and $\vL = \vX \times \vP$.
The spin-spin term $H_{S_1 S_2}$ was derived in Ref.~\cite{BO}, 
while the spin-spin terms $H_{S_1 S_1}, H_{S_2 S_2}$ which are valid {\it only} 
for a BH binary were derived in Ref.~\cite{TD} 
[see discussion around Eqs.~(2.51)--(2.55) in Ref.~\cite{TD} and also
Ref.~\cite{EP}]. 
They originate from the interaction of the monopole $m_2$ with the spin-induced 
quadrupole moment of the spinning black hole of mass $m_1$ and viceversa. 
The spin-induced quadrupole moment of a NS depends
on the equation of state. So, if we were applying our approch to 
NS binaries we could take into account the 
monopole-quadrupole interaction by multiplying  $H_{S_1 S_1}, H_{S_2 S_2}$
by some equation-of-state-dependent coefficient $\gamma$ [see Ref.~\cite{EP}].

\subsection{Equations of motion and conserved quantities}

The time evolution of any dynamical quantity 
$f(\vX,\vP,\vS_1,\vS_2)$ is given by~\cite{TD}:
\beq
\label{3.4}
\frac{d }{dt} f(\vX,\vP,\vS_1,\vS_2) = \{f,H\}\,,
\eeq
where with $\{...,...\}$ we indicated the Poisson brackets
$\{X^i,P_j\}= \delta^i_j$. The Hamilton equations
of motion are:
\beq
\label{3.5}
\frac{d \vX}{d t} = + \frac{\partial H}{\partial \vP}\,, \quad \quad
\frac{d \vP}{d t} = - \frac{\partial H}{\partial \vX} \,.
\eeq
The equations of motion for the spins are easily derived as~\cite{TD,TH,BO}: 
\bea
\label{3.6}
\frac{d }{dt} \vS_1 = \{\vS_1,H \} = \frac{\partial H}{\partial \vS_1} \times \vS_1 = 
\vO_1 \times \vS_1\,,\\
\label{3.7}
\frac{d }{dt} \vS_2 = \{\vS_2,H \} = \frac{\partial H}{\partial \vS_2} \times \vS_2 = 
\vO_2 \times \vS_2\,,
\eea
with
\bea
\label{3.8}
\vO_1 &=& \left ( 2 + \frac{3}{2}\,\frac{m_2}{m_1} \right
)\,\frac{\vL}{R^3} + \frac{1}{R^3}\,\left [3\vN\,(\vS_2 \cdot \vN) -
\vS_2 \right ] + \frac{3}{R^3}\,\frac{m_2}{m_1}\,\vN\,(\vS_1 \cdot \vN)
\,, \\
\label{3.9}
\vO_2 &=& \left ( 2 + \frac{3}{2}\,\frac{m_1}{m_2} \right
)\,\frac{\vL}{R^3} + \frac{1}{R^3}\,\left [3\vN\,(\vS_1 \cdot \vN) -
\vS_1 \right ] + \frac{3}{R^3}\,\frac{m_1}{m_2}\,\vN\,(\vS_2 \cdot \vN)\,.  
\eea 
Note that a consequence of the above spin-evolution equations
is the constancy of the lengths of the spin vectors:
$\vS_1^2 =$ cst., $\vS_2^2 =$ cst.

When using the EOB Hamiltonian Eq.~(\ref{Hspineob})
 we should in principle
consider the canonical transformation between $(\vX,\vP)$ and 
$(\vX^\prime,\vP^\prime)$ which is explicitly given as a PN expansion in 
Refs.~\cite{BD1,DJS}.
However, since the Hamilton equations are valid in any canonical coordinate 
system, when we evolve the EOB dynamics we write the Hamilton equations 
in terms of $(\vX^\prime,\vP^\prime)$ and for the spinning part we 
neglect the differences between $(\vX^\prime,\vP^\prime)$ and $(\vX,\vP)$ 
which are of higher PN order. 
When in the following sections we compare the results between PN-expanded and PN-resummed 
Hamiltonians, we will always compare quantities
which are gauge invariant to lowest order.

One of the advantages of using an Hamiltonian formalism is that
one can immediately derive from the fundamental symmetries
of the relative dynamics (time translation and spatial
rotations) two {\it exact} conservation laws: that of the total
energy $E = H$, and that of the total angular momentum
$\vJ = \vL + \vS_1 + \vS_2$.
Using Eqs.~(\ref{3.6}), (\ref{3.7}) it is easy to
check the conservation of the total energy,
\beq
\label{3.10}
\frac{d H}{d t} = \frac{\partial H}{\partial \vX}\,\frac{d \vX}{d t} +  
\frac{\partial H}{\partial \vP}\,\frac{d \vP}{d t} +  \frac{\partial H}{\partial \vS_1}\,
\frac{d \vS_1}{d t} + \frac{\partial H}{\partial \vS_2}\,\frac{d \vS_2}{d t}=0\,.
\eeq

Similarly, one easily checks the conservation of the total
angular momentum $\vJ = \vL + \vS_1 + \vS_2$. Note the
remarkable fact that the orbital contribution to
 $\vJ$ is exactly given, at any PN-order, by the ``Newtonian-looking''
 expression $\vL = \vX \times \vP$. This is contrary to what
 happens when working within a Lagrangian formalism, where the
 expression for the conserved orbital angular momentum gets
 modified at each PN order: $\vL =\mu \vX \times \vv + O(c^{-2})$.
 Here, all the needed PN contributions are included in the
 Hamiltonian $H$ (and thereby imply that
$ \vV =  \partial H/\partial \vP$ is a complicated function
of $\vV$).

\subsection{Spin-orbit interaction and ``spherical orbits''}

 For most of the dynamical evolution, the spin-spin terms are
 much smaller than the spin-orbit ones.
If we restrict ourselves to spin-orbit interactions, the equation of motion for
the orbital angular momentum reads
\beq
\label{3.11}
\frac{d \mathbf{L}}{d t} = \{\mathbf{L},H\} = 
\frac{2}{R^3} \mathbf{S}_{\rm eff} \times \mathbf{L}\,.
\eeq
A useful consequence of this approximate evolution law is
 that $\vL^2$ is conserved.
 Under the same approximation, the total spin $\vS = \vS_1 + \vS_2$ satisfies
the equation:
\beq
\label{3.12}
\frac{d \vS}{d t} = - \frac{2}{R^3}\,\vS_{\rm eff} \times \vL\,.
\eeq
The above (approximate) evolution equations exhibit clearly the (exact) conservation
of the total angular momentum $\vJ = \vL + \vS$
($d\vJ/dt =0$).

It is also easily checked that
$\vS_{\rm eff} \cdot \vL$ is a conserved quantity under the
above, approximate evolution equations.
Therefore, as emphasized by Damour~\cite{TD}, when
only spin-orbit terms are included the orbital dynamics can
be reduced to a simple ``radial Hamiltonian''
 $H(R,P_R)=H(R,P_R,\vL^2={\rm const.},\vS_{\rm eff} \cdot
\vL = {\rm const.})$ describing the radial motion.
 Here $P_R \equiv N^i\,P_i$ is the momentum
canonically conjugated to $R$ ($\{R,P_R\}=1$).  In this case there
exists a class of {\it spherical orbits} satisfying
\beq
\label{3.13}
R = {\rm const.}\,, \quad \quad P_R = 0\,, \quad \quad 
\frac{\partial H(R,P_R=0,\vL^2, \vL \cdot \vS_{\rm eff})}{\partial R}=0\,.
\eeq

\subsection{Characteristics of the last stable spherical orbit (LSSO)}

When spin-spin interactions are included, those
spherical orbits no longer exist as exact solutions. However, as
spin-spin effects are always smaller than spin-orbit ones, one expects
that the above {\it spherical} orbits will play the same important role
as the usual {\it circular} orbits in the non-spinning case.
In particular, the {\it Last Stable Spherical Orbit} (LSSO) should play
the important role of delineating the transition between adiabatic
inspiral and plunge.

The LSSO for the spinning conservative dynamics is determined by setting 
\beq
\frac{\partial H_0}{\partial R} = 0=\frac{\partial^2 H_0}{\partial R^2}\,,
\label{eq:isco}
\eeq
where $H_0(R,P_R,P_\phi, \cdots) = H(R,0,P_\phi, \cdots)$.

The physical characteristics of the LSSO for
(aligned or anti-aligned) spinning configurations were studied in detail
in Ref.~\cite{TD} within the more fully resummed Kerr-like EOB-Hamiltonian
introduced in that reference. It was also shown in \cite{DGG} that the
predictions from the latter Kerr-like EOB-Hamiltonian were in good
agreement with the numerical results on corotating
black hole (BH) binaries obtained by means of the
helical Killing vector approach \cite{GGB}. See Table I of \cite{DGG}.
[Note that the agreement with EOB is better when one considers the
3PN accuracy.]
The latter Table also shows that the numerical results on irrotational
BH binaries obtained by means of other approaches based on considering
only the initial value problem (e.g. \cite{PTC}) significantly differ
both from the numerical helical-Killing-vector (HKV) results, and the
analytical EOB-ones. 
For recent work improving  the numerical implementation of the HKV approach
(which is closely related to the ``conformal thin-sandwich'' decomposition)
and confirming that it yields results that agree well with the
EOB approach, see~\cite{Cook,CookPfeiffer}. As all the
currently published numerical
estimates of the physical characteristics of close binaries
made of {\it spinning } BH's (such as \cite{PTC}) use
initial-value-problem approaches rather than the physically
better motivated HKV one, we shall not try to compare here
the analytical EOB predictions for spinning configurations
with numerical results. On the other hand, it is interesting
to compare several different, PN-rooted, analytical approaches
in their predictions for the binding energy of close BH binaries.

The most straightforward PN-based analytical approach to the
physical characteristics of close BH binaries consists of
starting from the fully PN-expanded Hamiltonian (\ref{Hspinexp}),
considered as defining an exact dynamics, and then to
deduce from it the energy and angular frequency of
spherical orbits. [We consider here only configurations
where the spins are parallel (or antiparallel) to the orbital
angular momentum, so that it makes sense, even in presence of
spin-spin interactions, to consider spherical (and actually circular)
 orbits.] The binding energy of such ``PN-expanded'' spherical
 orbits is plotted in the left panel of Fig.~\ref{fig:eomega} as a function
 of the orbital angular frequency, for equal-mass BH binaries.
 As we see from this Figure, the straightforward PN-expanded
 Hamiltonian does not exhibit any minimum of the
 binding energy, i.e. does not lead to any {\it Last} Stable Spherical
 Orbit (LSSO) in the non-spinning, or aligned maximally spinning
 cases. As the best current numerical results on BH binaries
 clearly indicate the existence of such LSSO's, this disqualifies
 the use of the fully PN-expanded Hamiltonian (\ref{Hspinexp})
 for describing close binaries.

It has, however, been pointed out \cite{LB} that more reasonable results, 
close to the numerical HKV results, can be obtained by plotting, instead 
of the prediction coming straightforwardly from (\ref{Hspinexp}), 
the PN-expansion of the analytically computed function ${E}(\Omega)$ 
giving the binding energy $E$ as a function of the orbital frequency $\Omega$.
Indeed, one can derive from (\ref{Hspinexp}) the following
explicit PN-expansion of the (invariant) function
${E}(\Omega)$ \cite{DJSinvar}, \cite{DJS}, \cite{LB}:
\bea
\label{s1}
E_{\rm 2PN}(\Omega) &=& -\frac{\mu}{2}\,(M\Omega)^{2/3}\,\left \{
1 - \frac{(9+\eta)}{12}\,(M\Omega)^{2/3} + \frac{8}{3} \frac{\hL\cdot \vS_{\rm eff}}{M^2}\,
(M\Omega)+
\frac{1}{24}(-81 + 57 \eta - \eta^2)\,(M\Omega)^{4/3} \right . \nonumber \\
&& \left. + \frac{1}{\eta}\,\frac{1}{M^4}\,
\left [\vS_1\cdot \vS_2 - 3 (\hL \cdot \vS_1) (\hL \cdot \vS_2)
 \right ]\,(M\Omega)^{4/3} \right \}\,,\\
\label{s2}
E_{\rm 3PN}(\Omega) &=& E_{\rm 2PN}(\Omega) -\frac{\mu}{2}\,(M\Omega)^{2/3}\,\left \{
\left [-\frac{675}{64} + \left (\frac{34445}{576}-\frac{205}{96}\pi^2 \right )\eta
-\frac{155}{96}\eta^2-\frac{35}{5184}\eta^3 \right ]\,(M\Omega)^{2}\right \}\,.
\eea
These functions are plotted in the right panel of Fig.~\ref{fig:eomega}. 
Visibly, they are much better behaved than the results plotted on the left 
panel, which came directly from the PN-expanded Hamiltonian. 
They are also close to the numerical HKV results~\cite{LB}. The fact 
that two expressions, which can both be called ``PN-expanded'', and which 
are supposedly equivalent ``modulo higher PN terms'', lead to physically 
markedly different results lead to conclude that the PN-expanded Hamiltonian 
cannot be used to describe the transition from adiabatic inspiral to plunge. 
Let us also recall that if we consider, in absence of spins,  
the test-mass limit $\eta \to 0$, instead of the equal-mass one $\eta \to 1/4$,  the
PN-expansions (\ref{s1}),(\ref{s2}) have been shown (see
\cite{DJS}) to be quite inaccurate representations of the known exact expression
of the function ${E}(\Omega)$. Indeed, the 2PN-accurate function
(\ref{s1}) predicts in this limit an LSO frequency which is
$82\%$ larger than the exact one, while the 3PN-accurate one
(\ref{s2}) predicts an LSO frequency $27\%$ larger than the exact one.

By contrast with these problematic features of PN-expanded
results \footnote{Let us recall here that, in order to be able to describe
the transition between inspiral and plunge, we cannot use just
the function ${E}(\Omega)$, but we need a full description of the
binary dynamics. Therefore, if we wanted to confine ourselves
to a ``PN-expanded'' approach, we would have to use either
the PN-expanded Hamiltonian (\ref{Hspinexp}), or the corresponding
(appropriately truncated) PN-expanded equations of motion.
The left panel of Figure \ref{fig:eomega} shows that this would not be
a reliable thing to do. This is also clear from some of the
results of Ref.\cite{bcv1}.}, the EOB-approach leads to
uniformly better behaved results (even if we use it not
in the Kerr-like form advocated in \cite{TD}, but in the
form (\ref{Hspineob}) used in the present paper). We show
in Fig.~\ref{fig:eomegaEOB} the EOB analog of Fig.~\ref{fig:eomega}, 
i.e. the function ${E}(\Omega)$ deduced from the EOB Hamiltonian 
(\ref{Hspineob}) in the (anti-)aligned case. In this EOB case, we  have none of
the problems entailed by the ``PN-expanded'' approach, and the
uniquely defined curve ${E}(\Omega)$ was shown in \cite{DGG}
to agree well with the HKV  numerically determined curve
(for corotating holes). Note, however, that {\it aligned}
maximally rotating holes lead to a curve which, especially
in the 3PN case, reaches a minimum (not shown on Figure \ref{fig:eomegaEOB})
only for a rather high angular velocity.

This property of the aligned configurations (as well as
the significant difference between the 2PN-EOB result
and the 3PN-EOB one) was already emphasized in \cite{TD}.
As it will be important for the present paper, 
we study it further by plotting in Fig.~\ref{fig:e-a} 
the dependence on the $\vL$- projected effective spin parameter
\beq
\chi_L \equiv \frac{\vS_{\rm eff} \cdot \hL}{M^2}
\eeq
(where $\vS_{\rm eff}$ was defined in Eq.~(\ref{3.2}) above) of the binding
energy $E$ and the angular frequency $\Omega$ {\it at the LSSO},
i.e. at the minimum of the ${E}(\Omega)$ curve.
This Figure shows four results obtained for
equal-mass and equal-spin configurations within
the EOB approach: (i) the result obtained from the
Hamiltonian (\ref{Hspineob}) when using the 2PN-accurate
orbital EOB Hamiltonian, (ii) the result obtained from
(\ref{Hspineob}) when using the 3PN-accurate
orbital EOB Hamiltonian, (iii) the result obtained from
the  2PN-accurate Kerr-like Hamiltonian introduced in \cite{TD},
and (iv) the result obtained from the  3PN-accurate
Kerr-like Hamiltonian introduced in \cite{TD}. [The latter two
Hamiltonians are referred to in the caption as ``SO Resummed'',
because they include a resummation \`a la EOB of the spin-orbit
interactions.] In addition, as we cannot show
on this plot the minimum of the ${E}(\Omega)$ curve deduced
from the PN-expanded Hamiltonian, because the left panel
of Fig.~\ref{fig:e-a} shows that it does not exist, we show instead,
 for comparison purposes, the characteristics
of the  minimum of the PN-expanded functions (\ref{s1}), (\ref{s2})
[i.e. the right panel of Fig.~(\ref{fig:eomega})].

It is interesting to note on Fig.~\ref{fig:e-a} that the effect of
resumming (\`a la Kerr) or not the spin-orbit interaction
seems to be rather small. We see also that, when
considering anti-aligned configurations, all calculations
give very similar results. This is not surprising as
the {\it attractive}  ($H_{SO} <0$) nature of the {\it anti-aligned}
spin-orbit (and spin-spin) interaction has the effect
of pushing the LSSO {\it outwards}, i.e. toward larger-radius,
lower-frequency, less bound and therefore less relativistic
 configurations. On the other hand, working
at the 2PN or the 3PN level induces,as already pointed out in \cite{TD},
 a very significant difference for the LSSO characteristics in the {\it aligned} case (positive
$\chi_L$). In this case, because of the {\it repulsive}($H_{SO} >0$)
nature of the {\it aligned} spin-orbit (and spin-spin) interaction, the LSSO is drawn
towards closer, higher-frequency, more bound and more
relativistic configurations. For such very compact
configurations the {\it repulsive} sign ($a_4 >0$)
of the 3PN contribution to the effective potential $A(q)$
further amplifies, by a ``snow-ball effect'', the tendency
toward closer, and more bound configurations.
We think that this could be a physically real effect due (as confirmed independently by Refs.~\cite{BDE,Itoh}) 
to the large positive value of the crucial 3PN coefficient entering  $a_4$, Eq.~(\ref{a4}).  
This large positive value for $a_4$ is also needed to improve (with respect to the 2PN case)
the  agreement between the HKV corotating results and the 3PN EOB ones \cite{DGG}.
It would be interesting to have numerical HKV studies of 
the LSSO for moderately- and fast-spinning aligned BH's to test the predictions made by the EOB approach.
[The less reliable numerical results of the initial-value-problem of Ref.\cite{PTC},
which extend up to mildly positive values of $\chi_L \sim 0.17$
are in rough qualitative agreement (especially 
for the dependence of $\Omega_{LSSO}$ on $\chi_L$) with the
EOB predictions, but their quantitative agreement is too
poor to reach a firm conclusion).]

Let us note in passing that the significant dependence of the
LSSO frequency on the effective spin parameter $\chi_L$
makes it desirable for data-analysis purposes, when one is content
with using adiabatic templates~\cite{bcv2}, to use at least
templates whose ending frequency is not fixed say to the usually considered 
Schwarzschild LSO, but varies with masses and spins as suggested by the EOB 
approach.

Finally, another consequence of the significant dependence of the
LSSO frequency on the effective spin parameter $\chi_L$, is 
drawn in Fig.~\ref{snr}, where we compare the signal-to-noise ratios (SNRs) as 
function of the binary total mass for an optimally oriented equal-mass binary at 100 Mpc. 
We use LIGO design sensitivity noise curve~\cite{DIS98}.  
The SNRs are obtained  observing the inspiral from 40 Hz until the LSSO predicted by 
the EOB approach at 3PN order. The three curves refer to 
non-spinning binaries and binaries with $\chi_L = -0.875, 0.25$. 
Figure~\ref{snr} reveals a bias towards 
first detecting mostly aligned spinning binaries with high masses, 
as pointed out in \cite{TD}.

\section{Radiation reaction, including spin-effects}
\label{sec3}

The previous section has reviewed various ways of describing the
conservative dynamics of binary systems (including spin effects). In
the present section, we discuss the inclusion of radiation reaction
effects, with emphasis on determining the spin-modifications of
radiation reaction. Within the Hamiltonian approach, that we use here,
radiation reaction can be incorporated by modifying the usual Hamilton
equations in the following way
$$
\frac{dX^i}{dt} = \{ X^i , H \} = \frac{\partial H}{\partial P_i} \, ,
$$
\beq
\label{n1}
\frac{dP_i}{dt} = \{ P_i , H \} + F_i = - \frac{\partial H}{\partial X^i} + F_i \, .
\eeq
Here, $F_i$ denotes a ``non-conservative force'', which is added to
the evolution equation of the (relative) momentum to take into account
radiation-reaction (RR) effects. This Radiation-Reaction (RR)
 force $\vF$ depends, a
priori, both on the (relative) orbital variables $\vX$, $\vP$ and on
the spin variables $\vS_1$, $\vS_2$. In the present paper, our aim
will be limited to determining $F_i$ under the following two
simplifying assumptions: (i) we consider only {\it quasi-circular
orbits}, and (ii) we shall retain only the {\it leading spin-dependent
terms}. After the completion of the work reported in this section,
there appeared a work of Will \cite{Will} dealing with
spin-dependent radiation reaction effects in general orbits.
As the derivations are not the same, and yield results which
we have checked to be equivalent (for circular orbits), but
expressed in different variables
(Hamiltonian $\vX$, $\vP$ here vs. Lagrangian $\vX$, $\vV$
for \cite{Will}), we feel it worth to briefly report our
derivation.

Because of the assumption (ii) we look for terms in $F_i$
which are {\it linear} in the spin-tensors $S_{ij}^a \equiv
\varepsilon_{ijk} \, S_a^k$ ($a=1,2$) of the two considered compact
bodies. [Note that spin effects enter the metric only through the spin
tensors $S_{ij}^a$, rather than through the axial spin vectors
$S_a^k$.] As the time derivative of $S_{ij}^a$ contains a ``small''
post-Newtonian factor $G/c^2$, the leading spin-dependent terms in
$F_i$ will contain only the undifferentiated spin tensors.

Using Euclidean invariance, the spin-dependent terms in the force $F_i
(\vX , \vP , \vS_a)$ must be a combination of three types of
contributions: $c_1 \, S_{ij} \, X^j$, $c_2 \, S_{ij} \, P^j$ and $c_i
\, S_{jk} \, X^j P^k$, where $c_1 (\vX , \vP)$, $c_2 (\vX , \vP)$ are
some scalar functions of $\vX$, $\vP$, while $c_i$ is a vector
function of $\vX$, $\vP$. [Here, $S_{ij}$ denotes one of the two spin
vectors. We shall sum over the two possible spins at the end.]
Imposing that the radiation reaction force $F_i$ be {\it odd under
time reversal}, i.e. odd under the simultaneous changes $X^i \to X^i$,
$P_i \to -P_i$, $S_{ij} \to - S_{ij}$, tells us that: $c_1 (\vX ,
\vP)$ must be an {\it even} function of $\vP$, $c_2 (\vX , \vP)$ must
be an {\it odd} function of $\vP$, and the vector $c_i (\vX , \vP)$
must be an {\it odd} function of $\vP$. [Note also that $c_i$ must be
a true vector, not an axial vector. By parity invariance no
$\varepsilon_{ijk}$ can appear, except in combination with $S_a^k$.]
Therefore, if we further decompose $c_i = c_3 (\vX , \vP) \, P_i + c_4
(\vX , \vP) \, X_i$, the coefficient $c_3 (\vX , \vP)$ must be {\it
even} in $\vP$, while the coefficient $c_4 (\vX , \vP)$ must be {\it
odd} in $\vP$.

At this point, our simplifying assumption (i) above (quasi-circular
motion) will bring a drastic simplification. Indeed, a time-odd scalar
must contain an odd power of the combination $X^k P_k$. However, this
combination vanishes along circular orbits (and is therefore
subleadingly small along adiabatically inspiralling orbits). To
leading order the scalar coefficients $c_2$ and $c_4$ therefore
vanish, and we conclude that $F_i$ contains only two independent spin
contributions: $c_1 (\vX , \vP) \, S_{ij} \, X^j + c_3 (\vX , \vP) \,
P_i \, S_{jk} \, X^j P^k$. It will be convenient in the following to
further decompose the vector $S_{ij} \, X^j$ entering the first
contribution (which is orthogonal to $X^i$) into its component along
the direction of $P_i$, and its component orthogonal to $P_i$, say
\beq
\label{n2}
(S_{ij} \, X^j)^{\perp} \equiv (\delta_{ik} - P_i \, P_k / \vP^2) \,
S_{kj} \, X^j = S_{ij} \, X^j + \frac{P_i}{\vP^2} \, S_{jk} \, X^j P^k
\, .  \eeq
It is easily checked that, along circular orbits $(X^i \, P_i = 0)$,
the vector (\ref{n2}) is orthogonal {\it both} to $\vP$ and to
$\vX$. Therefore, $(S_{ij} \, X^j)^{\perp}$ is parallel to the orbital
angular momentum (axial) vector
\beq
\label{n3}
L_i \equiv \varepsilon_{ijk} \, X^j \, P_k \, .
\eeq
It is easily checked that
\beq
\label{n4}
(S_{ij} \, X^j)^{\perp} = \frac{R^2}{\vL^2} \, (\vP \cdot \vS) \, L_i
= \frac{1}{\vP^2} \, (\vP \cdot \vS) \, L_i \, .  \eeq

Finally, adding the usual spin-independent radiation reaction
(parallel to $P_i$ for circular orbits), and summing over the two
bodies, we conclude that the RR force can be written as
\beq
\label{n5}
F_i (\vX , \vP , \vS_a) = B \, P_i + \sum_{a = 1,2} A_a (S_{ij}^a \,
X^j)^{\perp} = B \, P_i + \sum_{a=1,2} \frac{A_a}{\vP^2} \, (\vP \cdot
\vS_a) \, L_i \, , \eeq
with
\beq
\label{n6}
B \equiv B_0 + \sum_{a=1,2} C_a \, S_{jk}^a \, X^j P^k = B_0 + \sum_{a=1,2} C_a \, \vL \cdot \vS_a \, ,
\eeq
where $B_0$, $C_a$ and $A_a$ are some time-even functions of $\vX$ and $\vP$.

To determine the coefficients $B_0$, $C_a$ and $A_a$, we now impose
that there be a balance between the losses of mechanical energy and
angular momentum of the system due to the additional force $F_i$ in
the Hamilton equations of motion (\ref{n1}) and the losses of energy
and angular momentum at infinity due to the emission of gravitational
radiation. Let us first recall that, in the Hamiltonian formalism, the
quantities
$$
E(t) \equiv H(\vX (t) , \vP (t) , \vS_a (t)) \, ,
$$
\beq
\label{n7}
J_{ij} (t) \equiv X^i \, P_j - X^j \, P_i + S_{ij}^1 + S_{ij}^2 \, ,
\eeq
are exact constants of the motion in absence of RR force in
Eqs. (\ref{n1}) and in the Hamiltonian equations for spin
evolution.
When adding the RR force $\vF$ in Eqs. (\ref{n1}) (and no
corresponding RR torque in the spin evolution equations) we find that
$E$ and $\vJ$ evolve as
\beq
\label{n8a}
\frac{dE}{dt} = \frac{\partial H}{\partial P_i} \, F_i = \dot X^i \, F_i \, ,
\eeq
\beq
\label{n8b}
\frac{d \, J_{ij}}{dt} = X^i \, F_j - X^j \, F_i \, .
\eeq
Inserting in Eqs. (\ref{n8a},\ref{n8b}) the expression (\ref{n5}) for
the RR force $F_i$, we can easily evaluate the {\it averaged} losses
of energy and angular momentum. Along (quasi) circular orbits the
various scalar coefficients $B_0$, $C_a$, $A_a$ are time-independent
(because all basic scalars, $\vX^2 , \vP^2 , \vX \cdot \vP = 0$, are
constant). One then finds that $dE/dt$ is time-independent, while $d
\, J_{ij} / dt$ depends on the orbital phase only in spin-dependent
terms and through the tensor $X^i \, P_j$. Decomposing the latter
tensor into
\beq
\label{n9}
X^i \, P_j = \frac{1}{2} (X^i \, P_j + X^j \, P_i) + \frac{1}{2} (X^i
\, P_j - X^j \, P_i) \simeq \frac{d}{dt} \left( \frac{1}{2} \, \mu \,
X^i \, X^j \right) + \frac{1}{2} \, L_{ij} \, , \eeq
one easily sees that its orbital average is simply $\langle X^i \, P_j
\rangle = \frac{1}{2} \, L_{ij}$. [We consider averages over the
orbital period, considering all more slowly evolving quantities, such
as $\vL$, as fixed during one orbital period.]

When evaluating Eq. (\ref{n8a}) along circular orbits, we cannot use
the Newtonian approximation $\dot X^i \simeq P^i / \mu$ because we
wish to obtain the coefficient $B$ with a high post-Newtonian
accuracy. However we can instead use $\dot X^i \, P_i = \dot\phi \,
P_{\phi} = \omega \, \vert \vL \vert$ where $\omega = \dot\phi =
V/R$ denotes the orbital angular frequency. We finally obtain
\beq
\label{n10a}
\frac{dE}{dt} = B \, \omega \, \vert \vL \vert \, ,
\eeq
\beq
\label{n10b}
\left\langle \frac{d \vJ}{dt} \right\rangle = B \, \vL - \frac{1}{2}
\sum_{a=1,2} A_a \, R^2 \, [\vS_a - \vlb (\vlb \, \vS_a)] \, , \eeq
where $\vlb \equiv \vL / \vert \vL \vert$ denotes the unit vector
along the orbital angular momentum. Note that Eqs. (\ref{n10a}),
(\ref{n10b}) predict a link between energy loss and angular momentum
loss, namely
\beq
\label{n11}
\frac{dE}{dt} = \omega \, \vlb \cdot \left\langle \frac{d \vJ}{dt} \right\rangle \, .
\eeq

To obtain the values of the coefficients $B$ and $A_a$ (and to test
the prediction (\ref{n11})), we need to compare Eqs. (\ref{n10a}),
(\ref{n10b}) with the values of the averaged fluxes of energy and
angular momentum at infinity. The spin contributions to the latter
losses have been computed by Kidder~\cite{K}. However, one must be
careful with the fact that Kidder expressed most of his results in
terms of {\it harmonic coordinates}, with the choice of a covariant
spin supplementary condition: $S_{\mu\nu}^a \, u_a^{\nu} = 0$.

First, using the results of Ref.~\cite{K} as they are, one
straightforwardly checks that the relation (\ref{n11}) is
satisfied. This is a check that it is enough to include RR effects in
the orbital equations of motion (\ref{n1}), without modifying the spin
equations of motion. [For a direct dynamical check, see \cite{Will}.]
 Then, to obtain the value of the coefficient $B$
we can simply use the result (\ref{n10a}), namely
\beq
\label{n12}
B = \frac{1}{\omega \, \vert \vL \vert} \ \frac{dE}{dt} \, ,
\eeq
where it remains to express $dE/dt$ (computed as a flux at infinity,
using Ref.~\cite{K}) in terms of our basic (Hamiltonian) dynamical
variables. One way to proceed would be to transform the
harmonic-coordinates results of \cite{K} into ADM coordinates (with
the corresponding spin condition $S_{i0} + \frac{1}{2} \, S_{ij} \,
v^j = 0$ \cite{DS88}). The transformation linking the two coordinates
has been worked out in \cite{DS88} (for the spin-dependent terms) and
in \cite{DJS01,ABF01} for the spin-independent parts. However, a
simpler way to proceed is to eliminate references to specific
coordinates by expressing $dE/dt$ (for circular orbits) in terms of
the gauge-invariant orbital frequency $\omega$. Adding also, for
better accuracy the recently completed 3PN flux contribution
\cite{BIJ02,B98,BDEI04}, we have
\beq
\label{n13}
\frac{dE}{dt} = - \frac{32}{5} \, \eta^2 \, v_{\omega}^{10} \{ 1 + f_2
(\eta) \, v_{\omega}^2 + [f_3 (\eta) + f_{3{\rm SO}}] \, v_{\omega}^3
+ [f_4 (\eta) + f_{4{\rm SS}}] \, v_{\omega}^4 + f_5 (\eta) \,
v_{\omega}^5 + f_6 (\eta) \, v_{\omega}^6 + f_{\ell 6} \, v_{\omega}^6
\, \ln (4 v_{\omega}) + f_7 (\eta) \, v_{\omega}^7 \} \, , \eeq
where $v_{\omega} \equiv (GM \omega)^{1/3}$, where the
spin-independent flux coefficients $f_2 (\eta) , \ldots , f_7 (\eta)$,
are given by
\beq
\label{n14a}
f_2 (\eta) = - \frac{1247}{336} - \frac{35}{12} \, \eta \, ,
\eeq
\beq
\label{n14b}
f_3 (\eta) = 4\pi \, ,
\eeq
\beq
f_4(\eta) = -\frac{44711}{9072}+\frac{9271}{504}\eta+\frac{65}{18}\eta^2\,, \quad \quad 
f_5(\eta) = -\left ( \frac{8191}{672} + \frac{583}{24}\eta \right )\pi\,, 
\eeq
\beq
f_6(\eta) = \frac{6643739519}{69854400}
+\frac{16}{3}\,\pi^2 -\frac{1712}{105}\,\gamma_E  +
\left (-\frac{134543}{7776}+\frac{41}{48}\,\pi^2\right )\eta 
-\frac{94403}{3024}\,\eta^2 - \frac{775}{324}\,\eta^3 \,,
\eeq
\beq
f_{\ell 6} = -\frac{1712}{105} 
\eeq
\beq
f_7(\eta) = \left (-\frac{16285}{504}+\frac{214745}{1728}\eta+\frac{193385}{3024}\eta^2 \right )
\,\pi\,,
\eeq
with $\gamma_E $ being Euler's gamma, and where the spin-dependent
corrections to the latter flux coefficients are
\beq
\label{n15a}
f_{3{\rm SO}} = - \left( \frac{11}{4} + \frac{5}{4} \, \frac{m_2}{m_1}
\right) \frac{\vlb \cdot \vS_1}{GM^2} - \left( \frac{11}{4} +
\frac{5}{4} \, \frac{m_1}{m_2} \right) \frac{\vlb \cdot \vS_2}{GM^2}
\, , \eeq
\beq
\label{n15b}
f_{4{\rm SS}} = \frac{\eta}{48 G^2 m_1^2\,m_2^2} \,
[ 289 (\vlb \cdot \vS_1) (\vlb \cdot \vS_2) - 103 \, \vS_1 \cdot
\vS_2 ] + \mathcal{O}(\vS_1^2) + \mathcal{O}(\vS_2^2)\,.
\eeq
The present work was aimed at determining the leading spin-dependent
terms, i.e. the ones linear in $\vS_1$ and $\vS_2$, as exemplified in
the correction $f_{3{\rm SO}}$, Eq. (\ref{n15a}), to the coefficient
$f_3 = 4\pi$. For completeness, as the link (\ref{n12}) between the
``longitudinal'' part of RR, $F_i^{\rm long} = B \, P_i$, and the
energy loss, is clearly general, we have also used Kidder's results
\cite{K} to write down the part of $B$ which depends on the product
$\vS_1^i \, \vS_2^j$. The numerically similar
contributions which depend on $\vS_1^i \, \vS_1^j$ and $\vS_2^i \,
\vS_2^j$ have not yet been determined. Only partial results
are known. For instance, Poisson \cite{EP} has derived a
contribution to $f_{4{\rm SS}}$ of the form

\beq
\left[\frac{3(\vlb \cdot \vS_1)^2 -\vS_1^2}{G^2 m_1^2\,M^2}
+\frac{3 (\vlb \cdot \vS_2)^2-\vS_2^2}{G^2 m_2^2 M^2}
\right]
\,.
\eeq
but many other additional contributions $\mathcal{O}(\vS_1^2) + \mathcal{O}(\vS_2^2)$ have not yet been computed.

\medskip

Let us finally turn to the determination of the other spin-related
coefficients, $A_a$, in Eq. (\ref{n10b}). Again, we have the technical
problem that Ref.~\cite{K} expressed its results in terms of
harmonic-coordinate quantities. Namely, Eq. (4.11) of Ref.~\cite{K}
expresses the total angular momentum loss $d \vJ / dt$ in terms of the
harmonic distance $r$ and of the harmonic-coordinate ``Newtonian
orbital momentum'' $\vL_N \equiv \mu \, \vx \times \vv$ (where $\vx$
and $\vv$ denote the relative harmonic position and velocity). A
simple way to convert this result to our ADM distance $R$ and our ADM
total orbital momentum $\vL \equiv \vX \times \vP$ is to relate
$\vL_N$ to $\vL$ by comparing the expression (4.7) of Ref.~\cite{K}
for the (gauge-invariant) conserved total angular momentum $\vJ$ with
the corresponding simple ADM expression (\ref{n7}). This yields a
result of the form
\beq
\label{n16}
\hat\vL_N \equiv \frac{\vL_N}{\vert \vL_N \vert} = c \, \vL + \left(
\frac{GM}{r} \right)^{\frac{3}{2}} \sum_{a=1,2} \chi_a \, \hat\vs_a
\left( \frac{m_a^2}{M^2} + \frac{1}{4} \, \eta \right) \, , \eeq
where the coefficient $c$ is not needed for our present purpose, and
where, following the notation of \cite{K}, $\vS_a \equiv \chi_a \,
m_a^2 \, \hat\vs_a$. Inserting Eq. (\ref{n16}) in Eq. (4.11) of
\cite{K} allows one to compute easily the part of $d \vJ / dt$ which
is proportional to the projection of $\vS_a$ orthogonally to $\vL$.

This yields the following expression for the coefficients $A_a$ in Eq. (\ref{n5})
\beq
\label{n17}
A_a = \frac{8}{15} \, \eta^2 \, \frac{v_{\omega}^8}{R^3} \left( 61 + 48 \, \frac{m_{a'}}{m_a} \right) \, ,
\eeq
where $a' \ne a$ (e.g. $a' = 2$ when $a=1$). Summarizing, the
radiation reaction force to be added to the Hamiltonian equations of
motion (\ref{n1}) reads
\beq
\label{n18}
F_i = \frac{1}{\omega \, \vert \vL \vert} \, \frac{dE}{dt} \, P_i +
\frac{8}{15} \, \eta^2 \, \frac{v_{\omega}^8}{\vL^2 R} \left\{
\left(61 + 48 \, \frac{m_2}{m_1} \right) \vP \cdot \vS_1 + \left( 61 +
48 \, \frac{m_1}{m_2} \right) \vP \cdot \vS_2 \right\} L_i \, , \eeq

where the energy loss (expressed in terms of the orbital frequency
$\omega$, or equivalently of $v_{\omega} \equiv (GM \omega)^{1/3}$,
and of the spin variables) is given by Eqs. (\ref{n13})--(\ref{n15b}).
We have checked that, after taking into account the relation between
the Hamiltonian variables $\vX, \vP$ and the Lagrangian ones $\vX, \vV$
(which involves spin-dependent terms because of the
first Eq.~(\ref{3.5}), Eq.~(\ref{n18}) agrees with the circular
limit of Eq.~(1.6) of \cite{Will} (which assumes the same spin
condition as we do).

Refs.~\cite{DIS98,BD2,DIS01,PS} have shown that (at least 
in the test-mass limit where one can compare analytical and numerical estimates)
it is generally advantageous to replace the Taylor series
in curly brackets on the right hand side (R.H.S.) of Eq. (\ref{n13})
by its (suitably defined) Pad\'e resummation.
In particular, Porter and Sathyaprakash \cite{PS} have compared
``Taylor'' and ``Pad\'e''  approximants for the flux function 
of a test particle around a Kerr black hole with the exact numerical 
estimates~\cite{fluxS} and concluded
that Pad\'e approximants are, when considering
all values of the spin parameter, both more {\it effectual}
(i.e., larger overlaps with the exact signal) and more
{\it faithful} (i.e., smaller biases in parameter estimates)
than Taylor approximants. [We use here the terminology
introduced in \cite{DIS98}.] In view of this,
and as was already advocated in \cite{BD2}, we consider that
that the best way to incorporate a radiation reaction force in
the EOB approach is to
insert Pad\'e approximants of the flux function (R.H.S.
of (\ref{n13})) in (\ref{n18}). However, for added generality,
we shall also consider the case where we leave the flux function
as a plain Taylor series. Note that, when considering arbitrary
values of the dimensionless spin parameters for the two holes
$\chi_1 \equiv S_1 / G m_1^2$, $ \chi_2 \equiv S_2 / G m_2^2$
we used the normal ``direct'' (i.e., lower-diagonal) Pad\'e-approximants, 
instead of the ``inverse''(i.e., upper-diagonal) ones used in \cite{PS}. 
For some values of the spin parameters both the lower and upper diagonal 
Pad\'e-approximants have poles. When this occurs, we apply the Pad\'e-approximant 
only to the non-spinning part of the flux and add the spinning terms separately. 
There exist other Pad\'e-approximants in which poles are absent and it 
would be very desirable to determine them in the entire parameter space. 
This is beyond the scope of this paper. 

In Figs.~\ref{Fig6}, \ref{Fig7} we show T- and (lower-diagonal) P-approximants at 3.5PN 
order for an equal-mass binary and several values of the dimensionless spin 
parameter $\chi=\chi_1=\chi_2$. We notice that the T- and P-approximants are much 
closer in the anti-aligned cases than in the aligned one. 
Since the calculation of the non-spinning flux 
function at 3.5PN order has been completed only recently~\cite{BDE}, 
in Fig.~\ref{Fig5} we contrast the T- and (lower-diagonal) P-approximants at 3PN and 
3.5PN order for an equal-mass non-spinning binary.

\section{Definitions of the initial and ending conditions of two-body models}
\label{sec4.1}

As clear from the comparison of the left and right panel of Fig.~\ref{fig:eomega}, 
because of the bad behaviour of the PN Hamiltonian near LSSO, we propose 
as our best bet, for describing in a physically reliable manner the non-adiabatic
evolution of BH binaries, and their transition between inspiral
and plunge, to use an EOB-resummed Hamiltonian, we shall,
for more generality, consider, and compare, in this Section several types of two-body models.

To define a specific model we must make  various choices:
(i) choice of a PN-expanded (or ``Taylor-expanded'') Hamiltonian
(say ``TH'') versus an EOB-resummed Hamiltonian (say ``EH'');
(ii) choice of a Taylor-expanded flux function (say ``TF'')
versus a Pad\'e-resummed one (say ``PF''), and finally,
(iii) choice of the PN accuracies used both in the Hamiltonian
 (say $ n $ PN) and the flux function (say $m$ PN). This leads
 to models denoted, for instance,${\rm THTF}(n, m)$, ${\rm EHTF}(n ,m)$, ${\rm EHPF}(n, m)$.
 In addition, as we are here mainly considering
 the evolution of {\it spinning} binaries, we shall add
 an initial letter S to recall that fact. This leads to
 models denoted as ${\rm STHTF}(n, m)$,..., ${\rm SEHPF}(n, m)$.
 To simplify, we shall only consider the PN accuracies $(2, 2.5)$ or $(3,3.5)$.
 To further simplify, we shall focus on comparing ``fully Taylor'' models 
(i.e., ${\rm STHTF}$), to ``fully resummed '' ones
(i.e., ${\rm SEHPF}$). Finally, this leads us to considering
only four models: ${\rm STHTF(2, 2.5)}$, ${\rm STHTF}(3, 3.5)$, ${\rm SEHPF}(2, 2.5)$, and 
${\rm SEHPF}(3, 3.5)$.
[Note, as discussed above, that because of the appearance of spurious poles 
in a few tests, we applied in those cases, the Pade resummation only to 
the non-spinning part of the flux.]

An important parameter in our present model building is to
decide when to stop the evolution. This issue was already
tackled in Ref.~\cite{BD2}. There, because we were using
an EOB Hamiltonian, and were considering non-spinning BH's,
we found that we could follow the ``plunge'' (after LSO crossing)
down to a (Schwarzschild-like) radius $\simeq 3 M$, at which
point we could match to a ring-down signal made of least-damped
quasi-normal modes. It was found in \cite{BD2} that,
contrary to what the usually employed word ``plunge'' suggests,
the inspiral motion after crossing the LSO was
staying ``quasi-circular'', with a kinetic energy in the
radial motion staying small
in absolute value, and smaller than $0.3$ times the
kinetic energy in the azimuthal motion down to $ R \simeq 3 M$.
In our ``spinning'' evolutions the situation is more complicated
(notably when considering large aligned spins, and also,
for evident reasons, when considering Taylor-expanded Hamiltonians).
We leave to future work a detailed discussion of the
matching to ring-down. We decided to stop
the evolution  as soon as one of the following inequalities
ceased to be fulfilled:

\begin{subequations}
\begin{eqnarray}
%\label{cond2}
%\omega &>& \omega_{\rm LSSO}^{\rm EOB}\,,\\
\label{cond3}
|\dot{R}| &<& 0.3 |\mathbf{V}_t|\,, \\
\label{cond4}
P_R^2/B(R) &<& 0.3 P^2_\phi/R^2\,, \\
\label{cond5}
|\dot{E}_{RR}|  &>& 0.1 |\dot{E}_{RR}^{\rm Newt}|\,,\\
\label{cond6}
R&>& \alpha M \,,
\end{eqnarray}
\end{subequations}
 where $B(R) = D(R)/A(R)$ [see Eqs.~(\ref{coeffPA}), (\ref{coeffPA2})].
Criteria \eqref{cond3}--\eqref{cond4} ensure that the evolution does
not extend too much beyond circularity, on which our formulation for
radiation-reaction force relies.
The quantity $\mathbf{V}_t$ is the tangential velocity
(i.e., orthogonal to the relative separation vector $\mathbf{X}$).  
Criterion~\eqref{cond5} is used to
avoid going into regimes where the GW energy flux goes to zero (e.g.,
for Taylor-expanded flux at 2.5PN order). Criterion \eqref{cond6}
(in which $\alpha \sim 1$ when using the ADM-coordinate Taylor-expanded
Hamiltonian, and $\alpha \sim 2$ when using the Schwarzschild-like
EOB Hamiltonian)
terminates the evolution at a very small radius, in case all of the
above criteria fail to take effect.

In all  cases, the
instantaneous GW frequency at the time when the integration is stopped
defines the \emph{ending frequency} for these waveforms. We shall
also consider below extended waveforms obtained by matching
a ring-down signal when this ending frequency is reached.

\subsection{Initial conditions: quasi-spherical orbits}
\label{sec4.2}

[In this section we shall use natural units $c=1=G$ and set $M=1$.]

In absence of radiation reaction (RR), spherical orbits with constant
radius and orbital frequency exists under spin-orbit interactions, but
cease to exist when spin-spin interactions are present (except in
special situations when the spins and the orbital angular momentum are
all aligned/anti-aligned). When radiation reaction is treated
adiabatically, an initial spherical orbit will evolve into a sequence
of spherical orbits, due to Eq.~\eqref{n11}. 
In this section, we formulate a prescription to construct
initial conditions for non-adiabatic evolutions, which lead to
quasi-spherical orbits under spin-orbit interaction.

With spin terms kept only up to the spin-orbit order, the Hamiltonian
can be re-written into a simpler form,
\begin{equation}
H(R,P_R,L,\chi_L)=H_{\rm no\,spin}(R,P_R,L)+2\frac{L\chi_L}{R^3}\,.
\end{equation}

Here $H_{\rm no\,spin}$ are terms in the Hamiltonian that do not involve spins, and
\begin{equation}
L\equiv |\mathbf{L}| \,, \quad \chi_L \equiv  \mathbf{S}_{\rm eff} \cdot\hat{\mathbf{L}}.
\end{equation}
In this form, the Hamiltonian depends on four quantites,
$\{R,P_R,L,\chi_L\}$, in which $L$ and $\chi_L$ both depend on
$\{\theta,\phi,P_\theta,P_\phi\}$, while $\chi_L$ also depends on the
spins. In absence of radiation reaction, the conditions 
for spherical orbits written in terms of partial derivatives (indicated by a subscript $i$) with respect 
to the four independent variables $\{R,P_R,L,\chi_L\}$, read
\begin{eqnarray}
\label{eq:circ1}
\big[\dot{R}\big]_0=0 &\Rightarrow&  \left[P_R\right]_0=0\,, \\
\label{eq:circ2}
\big[\dot{P}_R\big]_0= 0 &\Rightarrow &\left[\left(\frac{\partial H}{\partial R}\right)_c\right]_0 =\left[\left(\frac{\partial H}{\partial R}\right)_i\right]_0 =0\,.
\end{eqnarray}
[Here the subscript $c$ indicates canonical partial derivatives. In
the rest of this section, we shall continue to use $i$ and $c$ to
distinguish between these two types of partial derivatives.] With $L$
and $\chi_L$ being conserved quantities, conditions \eqref{eq:circ1}
and \eqref{eq:circ2} will remain satisfied if they are initially
satisfied --- which proves the existence of spherical orbits.

We now construct initial conditions for spherical orbits, {\it in
absence of radiation reaction}, based on Eqs.~\eqref{eq:circ1} and
\eqref{eq:circ2}.  In numerical evolutions, given a source coordinate
frame, $\{\mathbf{e}_x,\mathbf{e}_y,\mathbf{e}_z\}$, we specify spherical 
orbits with the following initial {\it kinetic
parameters}: the orbital frequency $\omega_0$, the orbital orientation (i.e., 
the normal direction to the orbital plane $[\hat{\mathbf{L}}_{\rm N}]_0 = (\vX \times 
\dot{\vX})/|\vX \times \dot{\vX}|$), the spins 
$[\mathbf{S}_{1,2}]_0$, and the direction of initial orbital separation
$\mathbf{N}=\vX/|\vX|$, which can in turn be given by an initial orbital
phase $\phi_{\rm orb}$, calculated with respect to the reference direction of
$[\mathbf{S}_{\rm tot} \times\hat{\mathbf{L}}_{\rm N}]_0$,
\beq
\mathbf{N}_0 = \frac{[\mathbf{S}_{\rm tot} \times\hat{\mathbf{L}}_{\rm N}]_0}{|[\mathbf{S}_{\rm tot} \times\hat{\mathbf{L}}_{\rm N}]_0|}\cos\phi_{\rm orb} +
 \frac{[\hat{\mathbf{L}}_{\rm N}]_0\times[\mathbf{S}_{\rm tot} \times\hat{\mathbf{L}}_{\rm N}]_0}{|[\mathbf{S}_{\rm tot} \times\hat{\mathbf{L}}_{\rm N}]_0|}\sin\phi_{\rm orb}\,.
\eeq  
We calculate initial
values for $\{\mathbf{X},\mathbf{P}\}$ in three steps:
\begin{enumerate} 
\item
We first apply a rotation $\mathcal{R}$ such that 
$[\hat{\mathbf{L}}_{\rm N}]_0 \rightarrow \mathbf{e}_z$ and 
$\mathbf{N}_0 \rightarrow \mathbf{e}_x$.
\item In spherical polar coordinates, the above step implies 
\beq
\label{init:norr1}
\phi_0=0\,,\;\;\;\; \theta_0=\pi/2\,.
\eeq
[The $\phi_0$ here should not to be confused with the orbital phase
$\phi_{\rm orb}$ above.] Then, we specify the initial frequency
$\omega_0$ and impose
\begin{eqnarray}
\label{init:norr2}
\omega_0 = \dot\phi_0=\left[\left(\frac{\partial H}{\partial P_\phi}\right)_c\right]_0\,,&& 
 0=\dot\theta_0 =\left[\left(\frac{\partial H}{\partial P_\theta} \right)_c\right]_0;\\
\label{init:norr3}
\left[P_R\right]_0=0\,, && \left[\left(\frac{\partial H}{\partial R}\right)_c\right]_0=0\,,
\end{eqnarray}
and solve for the four variables $\{R,P_R,P_\theta,P_\phi\}_0$. \item
Finally, we apply the inverse rotation $\mathcal{R}^{-1}$ to the
entire system, obtaining a set of initial spherical-orbit
conditions consistent with the specified initial kinetic parameters.
\end{enumerate}

When radiation reaction is included, we proceed as in Ref.~\cite{BD2} and modify Eq.~\eqref{eq:circ1} 
to include a non-zero $\dot{R}$, according to the prediction from adiabatic evolution, 
\begin{equation}
\label{RdotfromdEdR}
\big[\dot{R}\big]_0= \left[\frac{\dot{E}_{\rm RR} }{(dE/dR)_{\rm sph}}\right]_0\,,
\end{equation}
in order to prevent radial oscillations. [The subscript sph in Eq.~(\ref{RdotfromdEdR}) and 
below denote quantities evaluated along spherical orbits.]
Equations~\eqref{eq:circ2} can be kept unchanged, since $\dot{P}_R$ is second order in radiation
reaction.  We now calculate $(dE/dR)_{\rm sph}$ in terms of the
simplified set of independent variables,
$\{R,P_R,L,\chi_L\}$. Consider neighboring spherical orbits in an
adiabatic sequence, we have
\begin{equation}
dH = \left(\frac{\partial H}{\partial R}\right)_i dR
+
 \left(\frac{\partial H}{\partial P_R}\right)_i dP_R
 +\left(\frac{\partial H}{\partial L}\right)_i dL
 +\left(\frac{\partial H}{\partial \chi_L}\right)_i d\chi_L \,, 
 \end{equation}
 \begin{equation}
\left(\frac{\partial H }{\partial R}\right)_i = 0,\;\;
d\left(\frac{\partial H }{\partial R}\right)_i =0, \;\;
P_R=0,\;\; dP_R=0\,.
 \end{equation}
It is straightforward to deduce that
\begin{equation}
\label{dEdRfull}
\left(\frac{dE}{dR}\right)_{\rm sph} =
-\frac{\displaystyle \bigg(\frac{\partial H}{\partial L}\bigg)_i\bigg(\frac{\partial^2 H}{\partial R^2}\bigg)_i}
{\displaystyle \bigg(\frac{\partial^2 H}{\partial R\partial L}\bigg)_i}
+
\underbrace{
\left[
\left(\frac{\partial H}{\partial \chi_L}\right)_i
-
\frac
{\displaystyle \bigg(\frac{\partial H}{\partial L}\bigg)_i
\bigg(\frac{\partial^2 H}{\partial R \partial \chi_L}\bigg)_i
}
{\displaystyle \bigg(\frac{\partial^2 H}{\partial R\partial L}\bigg)_i}
\right]
\left(
\frac{d\chi_L}{d R}
\right)_{\rm sph}}_{\mbox{will be ignored}}\,.
\end{equation}
The second term on the right-hand side can be ignored, as we argue
later in this section, because $\chi_L$ is still conserved to a high
accuracy even in presence of radiation reaction. In special
configurations with $\hat{\mathbf{L}} =\mathbf{e}_z$ (or equivalently
$\theta=\pi/2$, $P_\theta=0$) we can re-write Eq.~\eqref{dEdRfull} in
terms of canonical variables in spherical-polar coordinates:
\begin{equation}
\label{dEdRsimple}
\left(\frac{dE}{dR}\right)_{\rm sph}
=
-\left[\frac{(\partial H/\partial P_\phi)_c (\partial^2 H/\partial R^2)_c}
{(\partial^2 H / \partial R \partial P_\phi)_c}\right]_{\theta=\pi/2,P_\theta=0}
\,.
\end{equation}
We also note that when $\hat{\mathbf{L}}$ is known to be $\mathbf{e}_z$, we can 
calculate $\dot{E}_{\rm RR}$, from Eq.~\eqref{n13} right away using only $\omega_0$ and $[\mathbf{S}_{1,2}]_0$.

We can now construct quasi-spherical initial conditions when radiation
reaction is present. As done in Ref.~\cite{BD2}, up to leading order in 
radiation reaction, we only need to augment our no-radiation-reaction 
initial conditions with a {\it non-zero} $P_R$, with initial values for all other canonical
variables unchanged. In order to do so, we {\it insert} three more
steps between steps 2 and 3 above:
\begin{itemize}
\item[2a] During step 2, we have obtained a set of spherical-polar-coordinate initial conditions, for a {\it rotated system} with $\{\mathbf{N}_0,[\hat{\mathbf{L}}_{\rm N}]_0\}=\{\mathbf{e}_x,\mathbf{e}_z\}$.  The canonical angular momentum, $[\mathbf{L}]_0$ of this system, though, will not in general be along $\mathbf{e}_z$. However, 
being orthogonal to $\mathbf{N}_0=\mathbf{e}_x$, it must be within the $\mathbf{e}_y-\mathbf{e}_z$ plane.  We now apply a further rotation $\mathcal{R}'$ around $\mathbf{N}_0 = \mathbf{e}_x$ to the entire system, such that afterwards  
we have $\{\mathbf{N}_0,[\hat{\mathbf{L}}]_0\}=\{\mathbf{e}_x,\mathbf{e}_z\}$, i.e., $\theta_0=\pi/2$ and $[P_\theta]_0=0$.
\item[2b]  Now that Eq.~\eqref{dEdRsimple} is applicable and $\dot{E}_{\rm RR}$ is readily obtainable, we insert them into Eq.~\eqref{RdotfromdEdR} and obtain $[\dot{R}]_0$. [Note that in this process we use the set of initial conditions already obtained for a spherical orbit in absence of radiation reaction, with $P_R=0$.] From this $[\dot{R}]_0$ , we obtain the initial value of $[P_R]_0$ to insert into our existing set of initial conditions:
\begin{equation}
\left[P_R\right]_0 = \frac{[\dot{R}]_0}{\displaystyle \left[\frac{1}{P_R}\left(\frac{\partial H}{\partial P_R}\right)_c\right]_{P_R \rightarrow 0}}\,.
\end{equation}
\item[2c] Gathering our new set of $\{R,\theta,\phi,P_R,P_\theta,P_\phi\}_0$, we obtain the Cartesian-coordinate variables, and apply a inverse rotation $(\mathcal{R}')^{-1}$ to the entire system. [Now again we have $\{N_0,[\hat{\mathbf{L}}_{\rm N}]_0\}=\{\mathbf{e}_x,\mathbf{e}_z\}$, and are ready to proceed to step 3.]
\end{itemize}

A straightforward analysis of the various error terms allows us 
to conculde that the fractional error of assuming that  $\chi_L \equiv \mathbf{S}_{\rm
eff}\cdot \hat{\mathbf{L}} $ is constant along the adiabatic evolution is 
of 3PN order. 

We note that our steps 1, 2, (2a, 2b, 2c), and 3 can still be applied
to give initial conditions, even if the Hamiltonian contains spin-spin
terms, although the orbits that follow will in general have
oscillatory radius and orbital frequency, due to the non-existence of
quasi-spherical orbits.  In Fig.~\ref{fig:rdot} we show the evolutions
of $\dot{r}/(r\omega)$ with (dark curves) and without (light curves)
spin-spin terms, for $(10+10)M_{\odot}$ (left panel) and
$(15+5)M_{\odot}$ (right panel) binaries. We start evolution at
40\,Hz, with 
$(\theta_{S_1},\phi_{S_1}\theta_{S_2},\phi_{S_2})=(60^\circ,90^\circ;60^\circ,0^\circ)$,
and show the evolution up to 200\,Hz.

\section{Comparison of waveforms and evaluation of overlaps}
\label{sec4.3}

In harmonic coordinates, the gravitational wave emitted by a binary system at the 
leading quadrupole order, in terms of metric perturbation at a distance $D$, is
\beq
h_{i j} = \frac{H_{ij}}{D} \equiv \frac{2\mu}{D}\,\frac{d^2}{d t^2} (X_i\,X_j)\,.
\eeq
Using the leading-order equation of motion, $\ddot{X}_k = - 
{M\,X_k}/{R^3}$, we re-write the normalized perturbation $H_{ij}$ as: 
\beq
\label{hGW0}
H_{i j} = {4\mu}\,\left (V_i\,V_j -M \frac{X_i\,X_j}{R^3} \right )\,.
\eeq
Here $X_i$ and $V_i \equiv \dot{X}_i$ can be obtained straightforwardly
by solving the Hamilton equations. 
Depending on the wave propagation direction and the orientation of the
detector, the metric perturbation $h_{ij}$ has to be contracted with
an appropriate ``detection tensor" to give the actually detected
waveform. For this we refer, for instance, to Sec.~IIIC of Ref.~\cite{pbcv1} and
Sec.~II of Ref.~\cite{bcv2} (in particular see Eq.(15)). 

Following Ref.~\cite{bcv2}, the parameters in precessing binaries can be 
distinguished in {\it binary local parameters} $\{m_1, m_2, S_1, S_2,
\theta_{\rm S1}, \theta_{\rm S2}, \phi_{\rm S1} - \phi_{\rm S2}\}$, 
{\it binary directional parameters} $\{ \theta_{\rm L}, \phi_{\rm L}, \phi_{\rm S1} + \phi_{\rm S2}\}$ 
(which determine the orientation of the binary as a whole in space) 
and {\it directional parameters} $\{\Theta, \varphi, \theta, \phi, \psi \}$, 
describing the GW-propagation and the detector orientation 
[see Table I in Ref.~\cite{bcv2} and discussion around it].
To these parameters we need to add the initial time and the initial orbital phase. 

In the {\it precessing} convention introduced in Ref.~\cite{bcv2}, 
the GW signal can be neatly written in terms of: (i) parameters 
depending on the observer's location and orientation, 
$\{\Theta, \varphi, \theta, \phi, \psi \}$, which are time 
independent, initial time and initial orbital phase (henceforth denoted 
as extrinsic parameters) and (ii) parameters depending on the details 
of the dynamics, $\{m_1, m_2, S_1, S_2, \theta_{\rm S1}, \theta_{\rm S2}, 
\phi_{\rm S1} - \phi_{\rm S2}\}$ (henceforth denoted as intrinsic 
parameters). The GW signal does not depend on the binary directional 
parameters, $\{ \theta_{\rm L}, \phi_{\rm L}, \phi_{\rm S1} + \phi_{\rm S2}\}$, 
since those parameters can be re-absorbed in the definition of the source 
frame at initial time and in the directional parameters $\{\Theta, 
\varphi, \theta, \phi, \psi \}$ through a rigid rotation of the detector-binary system. 

The distinction between extrinsic and intrinsic parameters is due originally
to Sathyaprakash and Owen~\cite{Sathya,O}. Extrinsic parameters are parameters 
which change the signal shape in such a way that we do not actually need to 
lay down templates in the bank along those parameter directions, saving computational time. 
By contrast we need to lay down templates along the directions 
of the intrinsic parameters. In Refs.~\cite{bcv2, pbcv1},
a semi-analytical method to maximize over the extrinsic 
parameters in precessing binaries has been proposed.

In view of the bad performances of the Taylor-expanded Hamiltonian\footnote{
We have in mind here the absence of LSSO. 
Recall also that when the binary mass ratio is significantly
different from one, one can firmly conclude that the Taylor-expanded
Hamiltonian is a poor representation of the dynamics, while the
EOB-resummed one is  a good one.}, we a priori expect that the
waveforms computed from STHTF models will be significantly
different from the SEHPF ones. It remains, however, interesting
to {\it measure their difference} in the data-analysis sense,
i.e.\ to compute the {\it overlaps} between the two types
of waveforms. If it happened that, after maximization
over all possible parameters, the overlap between the two
types of signals were very close to unity, one could still
consider the Taylor models as {\it effectual} (in the
sense of \cite{DIS98}) representations
of the EOB models. However, for practical reasons, we
did not try to embark on a full maximization of the overlaps.
For simplicity, we {\it only} tackled the maximization over the
 extrinsic parameters, and {\it not} on the intrinsic
ones. The resulting partially maximized overlap is therefore
only a lower bound of the fully maximized overlap. Still, this
result can be considered as a reasonable measure of  the ``closeness'' between the
two sorts of models (especially because we do not wish to
use models which would have significantly different physical
parameters).

\subsection{Lack of ``closeness'' between Taylor and Effective-One-Body models}

In Tables~\ref{overlapmax3pn} and \ref{overlapmax2pn}  
we study the {\it closeness} (in the sense just defined of overlap
maximized only over the extrinsinc parameters\footnote{More precisely,
we do the maximization over the extrinsic parameters of the EOB model. 
Though this introduces an asymmetry in the definition of the {\it closeness},
we do not expect this asymmetry to be physically significant.})
between STHTF(3,3.5) and SEHPF(3,3.5), as well
as  between STHTF(2,2)~\footnote{We use SHT(2,2) instead of STHTF(2,2.5) 
because for equal-mass binaries
the Taylor-approximant for the flux at 2.5PN order
becomes negative for large values of $v$~\cite{DIS98}},
and SEHPF(2,2.5) models.

We consider three typical binary masses $(10 + 10) M_\odot$,
$(15 + 15) M_\odot$ and $(15 + 5) M_\odot$, and several initial 
spin orientations~\footnote{For these data we always refer the initial spin 
directions to the initial direction of the orbital 
Newtonian angular momentum, as specified in Fig. 4 of Ref.~\cite{bcv2},  
and we set the initial direction of the Newtonian orbital angular 
momentum along the $x$-axis of the source frame (i.e., we fix 
$\theta_{\rm L} = \pi/2$ and $\phi_{\rm L}=0$, see Fig. 3 in Ref.~\cite{bcv2}).}.
We always fix the pattern functions $F_+ = 1$, $F_\times = 0$ and GW
propagation parameters $\Theta = \pi/4$ and $\varphi = 
0$ [for notations and definitions see Sec.~IIIC of Ref.~\cite{pbcv1}
and Sec.~II of Ref.~\cite{bcv2}]. 
The initial frequency is always set to $f_{\rm in} = 30\,{\rm Hz}$ and 
the ending frequency $f_{\rm end}$ is determined by one of the 
criteria in Eqs.~\eqref{cond3}--\eqref{cond6}.
In Tables~\ref{overlapmax3pn} and \ref{overlapmax2pn} the two black holes 
are assumed to carry maximal and half-maximal spins, respectively. 
We list the ending frequency, the LSSO frequency  and the BH radial 
separation at $t_{\rm fin}$ for the template and target, 
together with two types of overlaps: the overlaps maximized over the initial
time and initial orbital phase only ($\rho_{\rm max, 2}$),
and the overlaps maximized over those parameters and
$\{\Theta, \varphi, \alpha = f(\theta, \phi, \psi) \}$,
as well, ($\rho_{\rm max, 5}$), using the semi-analytical 
method suggested in Ref.~\cite{pbcv1}.
Table~\ref{overlapmax3pn} and ~\ref{overlapmax2pn} also contains the non-spinning case. 

As these tables show, the two types of models are not
at all ``close to each other''. The overlaps are indeed
quite low, as low as $0.32$. The overlaps evidently increase when
we maximize over five rather than two extrinsic parameters, but
not dramatically, and only for binaries with high and comparable mass, 
with initial spins lying in the half-space 
opposite (with respect to the orbital plane) 
to the initial orbital angular momentum. In this case 
the dynamical evolution is shorter, since the LSSO occurs 
at lower frequency [see also Figs. \ref{fig:eomegaEOB}, \ref{fig:e-a} ], 
and the differences in  STHTF(3,3.5) and SEHPF(3,3.5) can be compensated by an offset in the
extrinsic parameters of the template with respect to the target. 
Moreover, both the conservative dynamics for circular orbits 
and the GW flux predicted by SEHPF-approximants and STHTF-approximants, 
are closer in the anti-aligned case than in the aligned case, 
as can be see in Figs.~\ref{Fig6} and \ref{Fig7}. 

However, when the binary mass ratio is significantly
 different from 1, the number
of modulational cycles increases, and the differences in the 
two models can no longer be compensated by re-adjusting the 
template extrinsic parameters. When the initial spins are lying 
in the same half-space of the orbital angular momentum, the evolution is longer, the LSSO 
happens at high frequency [see also Figs. \ref{fig:eomegaEOB}, \ref{fig:e-a} ], 
and in this case, even for high, comparable masses the differences both in 
the conservative and non-conservative late dymanics in the two models cannot 
be compensated by a bias in the template extrinsic parameters. 

{}From Tables~\ref{overlapmax2pn} we 
notice that all the above considerations apply also at 2PN order,  
where the differences in the conservative 
and non-conservative dynamics of STHTF and SEHPF approximants are
even larger. We checked that these considerations do not change much 
when spins are smaller, say half-maximal. 

Having confirmed that Taylor models cannot be considered
as being effectively close to the EOB ones, we shall
only use in the following the a priori better EOB models.

\subsection{Negligible influence of the of the ``transverse component'' of the
radiation reaction force.}

Having in mind possible simplifications of the models,
we first investigated the relevance of the second term
in the R.H.S. of Eq.~(\ref{n18}), i.e. the component of
the radiation reaction force which is ``transverse'',
in the sense of being directed along $L$, and
therefore orthogonal to the main ``longitudinal term'',
which is parallel to the momentum $\vP$).
In Table~\ref{overlapnoFL} we study the influence of this transverse
component of the RR force on the dynamics and the waveforms.
For the binary masses $(10 + 10) M_\odot$, $(15 + 5) M_\odot$ 
and a few initial spin orientations, we evaluate the same 
quantities of Table~\ref{overlapmax3pn}, when including and not 
including the RR force along $L$ [see second and third term in Eq.~(\ref{n18})]. 
We give here only the overlaps maximized over five extrinsic parameters.
We find that $\rho_{\rm max, 5}$ is larger than $\sim 0.98$
in all cases. We therefore conclude that it would
suffice to use a simplified RR force parallel to the
linear momentum $\vP$ (as in the non-spinning circular case).
One should, however, include, for better accuracy,
 in the coefficient of $P_i$
in Eq.~(\ref{n18}) the spin-dependent terms.

\subsection{Influence of the resummation of the ``longitudinal'' part of
the radiation reaction}

We consider here the ``longitudinal'' part of the radiation reaction,
i.e. the first term on the R.H.S. of Eq.~(\ref{n18}). This term is
given by the flux function, which was written in Eq.~(\ref{n13}) above
as a straightforward PN-expansion.  One can therefore either leave
this longitudinal component in non-resummed, ``Taylor'' form, or
choose to resum it by means of Pad\'e approximants.  In
Table~\ref{overlapflux} we investigate how the choice of the flux
function (Taylor-expanded or Pad\'e-resummed) may affects the dynamics
and the waveforms. Using in all cases the EOB Hamiltonian to describe
the dynamics, we evaluate the overlaps between models using a Taylor
flux and models using a Pad\'e flux. [We maximize over the five
extrinsic parameters of the EOB model.]  We find that, when the initial 
spins are lying in the same half-space of the orbital 
angular momentum, after maximization over the five extrinsic parameters, 
the overlaps are reasonably large (larger than $0.84$), but still lower
than unity. We obtain much higher overlaps when the initial spins are not lying 
in the same half-space of the orbital angular momentum. These results 
are consistent with Figs.~\ref{Fig5} and Fig.~\ref{Fig7}. 

If we assume that the equal-mass flux function is a smooth deformation 
of the test-mass limit one, since previous findings \cite{DIS98,BD2,DIS01,PS} 
in the test-mass limit case pointed out the usefulness of Pad\'e-resumming 
the flux function, we would conclude that Pad\'e-resummed fluxes are better
approximants of the numerically determined flux also in th equal-mass case.

\subsection{Negligible influence of the quadrupole-monopole terms}

Still in the spirit of trying to simplify the models to their crucial elements,
Table~\ref{overlapnoQM} investigates how waveforms are
affected by the quadrupole-monopole terms, and Table~\ref{overlapadiab} 
studies how the evolution obtained by averaging the spin terms over a 
period may differ from the non-adiabatic evolution. 
Considering the high values of $\rho_{\rm max, 5}$ we obtain
in both cases, we can say that the quadrupole-monopole interaction and the adiabaticity 
of the spin terms, have little physical effects over the dynamics and waveforms.
The differences can be compensated by re-adjusting the template extrinsic parameters.

\subsection{Influence of the initial orbital phase}

Finally, we investigated the influence of the initial orbital phase
(all other quantities being fixed) on the waveform.
In an adiabatic evolution in which spin terms are averaged over a 
period the joint evolution of $\hat{\mathbf{L}}_{\rm N}$ and $\mathbf{S}$ is not affected by the
initial orbital phase. As a consequence, two configurations with the
same initial values for $\hat{\mathbf{L}}_{\rm N}$ and $\mathbf{S}$,
but different orbital phases $\phi_{\rm orb}$ will keep the difference
between orbital phases unchanged through the evolution. This may not
be true in our non-adiabatic evolution for two reasons: (i) the spin-spin
interaction Hamiltonian depends explicitly on the separation vector
$\mathbf{N}$ [see Eq.~(\ref{3.3})], and (ii) the evolution 
depends on the canonical orbital angular momentum,
which is not orthogonal to the orbital plane. We illustrate this 
feature by evolving two maximally spinning
$(15+15)M_\odot$ binaries with initial orbital phases (at 40\,Hz)
differing by $\pi/2$, and all other parameters identical:
$(\theta_{S_1},\phi_{S_1};\theta_{S_2},\phi_{S_2})=(60^\circ,90^\circ;60^\circ,0^\circ)$. In
Fig.~\ref{fig:rel-orb-phase}, we plot the difference $\Delta\phi_{\rm
orb}$ between their relative orbital phases measured with respect to
$\hat{\mathbf{L}}_N$. This difference grows in time, and accumulates
around $270^\circ$ by the end of the evolution.  We also show
waveforms detected with $(F_+,F_\times;\Theta,\varphi)= (1,0;\pi/4,0)$
in Fig.~\ref{fig:orbphase}.  Their phases start out to differ by
$180^{\circ}$ as expected, and non-adiabatic effects drive them away
by more than 2 cycles toward the end of the evolution. We notice that 
comparing the waveform is less straightforward than comparing the
relative orbital phase, because the waveform phase can differ from
twice the orbital phase, due to precessions.

\section{Losses of energy and angular-momentum and the waveform including ringdown}
\label{sec4.4}

In the following, we use as model the ``best bet'' we can make, i.e.
the spinning EOB Hamiltonian\footnote{We did not investigate the
``closeness'' between the models derived from the spinning
Hamiltonian used here, and those deduced from
the further resummed, Kerr-like EOB Hamiltonian proposed in \cite{TD}.
In view of the comparison showed in Fig.~\ref{fig:e-a},
we expect that the two models are very close to each other.} 
with a Pad\'e-resummed flux. Both being taken 
to the highest PN-accuracy available, i.e. $n=3, m=3.5$, in the
notation used above.

In Ref.~\cite{BD01}, using the non-spinning EOB Hamiltonian at 2PN order, it was 
found that the energy emitted during the plunge is $ \sim 0.7 \%$ of $M$, 
with a comparable energy loss $ \sim  0.7 \% $ of $ M$ during the ring-down phase.
This gives a total energy released beyond the LSSO in the non-spinning case of 
$\sim 1.4\%$ of $M$ to be contrasted with $4-5\%$ of $M$ 
estimated in Ref.~\cite{BBCLT}, where the authors use a combination of 
numerical and perturbative approximation methods. Note also that
 Flanagan and
Hughes~\cite{FH98} predicted $\sim 10 \% M$ for inspiral and plunge, 
and $\sim 3 \% M$ for ring-down phase. 

Here, to have more confidence in our EOB-based estimates,
we decided to use {\it three different ways  } of evaluating
 the energy radiated in the spinning case.
We used, at once, (i) the change, along the evolution (between
some initial frequency and some final one) in the numerical value
of the Hamiltonian (\ref{Hspineob}) ($\delta E_H$),  (ii) the time-integral
of the square of the third derivative of the
 quadrupole moment $I_{ij}$ with $i,j=1,2,3$, i.e.
\beq
\frac{d E_I}{dt} = \frac{1}{5} \frac{d^3 I_{i j}}{dt^3}\,\frac{d^3 I_{i j}}{dt^3}
\quad \quad I_{ij} = \mu\,\left (X_i\,X_j - \frac{1}{3}\delta_{ij}X^k\,X_k\right)\,.
\eeq
and (iii) the time-integral of the
energy flux carried away by our leading-order quadrupole waveform,
\beq
\frac{d E_h}{dt} = \frac{1}{20}\,\int \sum_{i j} \dot{H}_{i j}^{\rm TF} \dot{H}_{ij}^{\rm TF}\,,
\eeq
with $H_{ij}^{\rm TF}$ being
the trace-free part of the normalized metric perturbation,
$H_{ij}$ [see Eq.~(\ref{hGW0}) above].

In Fig.~\ref{fig:comp:eloss} we compare the accumulated energy release
from these three prescriptions, for a $(15+15)M_{\odot}$ maximally
spinning binary with a generic set of spin orientations when starting
evolution at $f_{\rm GW}=30\,$Hz:
$(\theta_{S_1},\phi_{S_1};\theta_{S_2},\phi_{S_2})=(60^\circ,90^\circ;60^\circ,0^\circ)$. In
the left panel, we keep radiation reaction force at the Newtonian
order, while we use 3.5PN Pad\'e flux in the right panel. As we see
from the figure, these prescriptions differ more from each other when
3.5PN radiation reaction is used instead of Newtonian --- this is
consistent with the fact that both $\delta E_I$ and $\delta E_h$
involve quadrupole radiation only; furthermore, for lower frequencies
the $\delta E_H$ curve lies below those of $\delta E_I$ and $\delta
E_h$, which is consistent with the fact that Post-Newtonian GW
luminosity is in general smaller than the Newtonian prediction. The
difference among $\delta E_I$ and $\delta E_h$ can be attributed to
the difference between PN (in our case EOB at 3PN) and Newtonian
dynamics, which seems to be small till around $f_{\rm
GW}=200$\,Hz in our case (which corresponds to $v\approx 0.45$).

The rather satisfactory agreement between the various ways of
estimating the energy loss gives us some confidence in our
EOB-based estimates. In the following, we shall use the a priori
best estimate (because it is the one which involves the highest
PN accuracy): the one based on the change in the total
EOB  Hamiltonian $H$. [We use  Pad\'e-resummed fluxes,
and the two combined PN-accuracies $(n,m) = (2,2.5)$
and $(n,m) = (3,3.5)$.]

In the upper panels of Fig.~\ref{fig:eloss} we plot
the accumulative energy loss $\delta E_H$
starting from $f_{\rm GW}=40\,$Hz as a
function of the instantaneous GW frequency $f$, for $(15+15)M_{\odot}$
(upper left panel) and $(15+5)M_{\odot}$ (upper right panel) binaries,
each for 4 sets of initial spin orientations with {\it maximal spins}:
aligned (dash-dot-dot curves), antialigned (dotted curves),
$(\theta_{S_1},\phi_{S_1};\theta_{S_2},\phi_{S_2})=(60^\circ,90^\circ;60^\circ,0^\circ)$
(denoted by {\it generic-up}, dash-dot curves), and
$(\theta_{S_1},\phi_{S_1};\theta_{S_2},\phi_{S_2})=(120^\circ,90^\circ;120^\circ,0^\circ)$
(denoted by {\it generic-down}, dashed curves), as well as for the
non-spinning configuration (solid curves).  We use both SEHPF(3,3.5)
(dark curves) and SEHPF(2,2.5) (light curves) models. In each panel, we
also use vertical grid lines to mark LSSO frequencies. For the
SEHPF(2,2.5) model, all our evolutions go beyond their corresponding
LSSO frequencies. The situation is a bit different for the SEHPF(3,3.5).
Indeed, as was shown in \cite{TD} and in Fig.~\ref{fig:e-a} above,
the 3PN-EOB LSSO for mostly aligned fast-spinning BH's is drastically
drawn inwards towards very high orbital frequencies.
So high, indeed, that, for aligned and generic-up configurations,
they fall out of the frequency range plotted in Fig.~\ref{fig:eloss}.
As a consequence, for aligned and generic-up configurations, the dynamical
evolutions become rather non-adiabatic even before the formal LSSOs
is reached.

As we see from the plots, within a given GW frequency interval, binaries tend to
emit more energy in configurations where spins are more aligned with
the orbital angular momentum. This agrees with the results
of \cite{TD} and of Fig.~\ref{fig:eomega} above, showing
that more aligned configurations are drawn towards more deeply bound
states. This, together with the fact that LSSO frequencies are
pushed higher in aligned configurations (as we also see from
Fig.~\ref{fig:eomega}), can make the total energy releases in aligned
configurations several times more than those in anti-aligned
configurations. In Table \ref{tab:EJ}, we list values of $\delta
E_H/M$, accumulated from $40$\,Hz up to LSSO frequency (if reached)
or ending frequency, otherwise, for configurations plotted in
Fig.~\ref{fig:eloss}.  We also list the energy released below 40\,Hz,
and values of energy release when 2PN Hamiltonian and 2.5PN Pad\'e
flux are used.  For $(15+15)M_{\odot}$, maximally spinning binaries,
the energy released from 40\,Hz up to the end of our evolution (determined 
by one of the criteria \eqref{cond3}--\eqref{cond6})
can range from $0.6\%$ of $M$ in the (anti-aligned configuration) to
$5\%$ of $M$ (anti-aligned configuration), with the non-spinning
configuration releasing $1.6\%\sim 1.8\%$ of $M$ (in which $0.8\% \sim
1.1\%$ of $M$ is released before LSSO).  For $(15+5)M_{\odot}$
binaries, the range is similar, from $0.5\%$ to $5\%$ of $M$, with
non-spinning configuration releasing $1.2\% \sim 1.4\%$ of M (in which
$0.7\%\sim 0.8\%$ of M from before the LSSO).  We also note that the
energies of around $ 0.8\%$ of M and $0.5\%$ of M are released below
40\,Hz, for $(15+15)M_\odot$ and $(15+5)M_\odot$ binaries,
respectively. See also Eq. (4.1) of \cite{TD} for an approximate
analytical estimate of the energy released down to the LSSO,
as a function of both $\eta = m_1 m_2/M^2$ and $\chi_L$.

\subsection{Evolution of the dimensionless rotation parameter $J/E^2$}

An important consistency check of the EOB approach concerns
the dimensionless total angular momentum
 ratio $|\mathbf{J}|/E^2$ ultimately reached
by spinning black hole binaries. Indeed, if the EOB method would,
at the end of its validity domain, predict a ratio $|\mathbf{J}|/E^2$
larger than unity, this would preclude to match this end
state with the newly born Kerr black hole expected
from the coalescence of the two initial (spinning) black holes.
This issue was investigated {\it in the adiabatic} approximation
in \cite{TD}. There, it was shown that, when using the
3PN-accurate EOB Hamiltonian, the ratio $|\mathbf{J}|/E^2$
estimated {\it at the LSSO}, was {\it always smaller than unity}.
This result was not at all guaranteed in advance, and resulted
from a delicate  competition between the linear increase of $|\mathbf{J}|$
when increasing the (aligned) spin of the individual BH's,
and its non-linear decrease because of the displacement of the LSSO
towards smaller radii for aligned spinning configurations.
It was found in \cite{TD} that the maximum of
$(|\mathbf{J}|/E^2)_{\rm LSSO}$ was about $0.83$, and was reached for
$\chi_L \simeq + 0.3$.

The fact that this maximum value is significantly below one,
leaves room for not running into any consistency problem even when
taking into account the further changes of both $E$ and $\vJ$
during the plunge that follows the crossing of the LSSO.
Though our present attack on the problem does not properly
consider the final matching between the plunge and the
formation of a final Kerr hole, it goes beyond the previous
treatments in going beyond the adiabatic approximation.
In the lower panels of Fig.~\ref{fig:eloss}, we plot the
continuous time evolution of the ratio $|\mathbf{J}|/E^2$
during the late stages of the inspiral and its subsequent
non-adiabatic ending (which, in many cases, except
in fact for the most dangerous  aligned, is a
post-LSSO plunge). It is convenient to use
the gravitational wave frequency $f_{\rm GW}$ to label
the ``time'' along this evolution. Satisfactorily, we
 observe that, for all the binaries we have considered,
$|\mathbf{J}|/E^2$ decrease to below $1$ before either the
LSSO or the end of our evolution, whichever comes first. This means
there are no a priori obstacles to having a Kerr
black hole form right after the end of the non-adiabatic ``quasi-plunge''.
This means also that, contrary to an early
suggestion \cite{FH98}
based on rather coarse estimates, there is no ground for expecting
 a large emission of gravitational waves between the plunge
 and the merger. In Table~\ref{tab:EJ} we also list the values
of $|\mathbf{J}|/E^2$, at LSSO frequency (if reached) and ending
frequency, for SEP(2,2.5) and SEP(3,3.5) models.

In a pioneering work,
Baker et al.~\cite{BCLT} evaluated by a 3D numerical
simulation the energy radiated from moderately spinning
BH binaries with spins aligned or anti-aligned with the orbital angular-momentum.
They started the (very short) numerical evolution close to the LSSO predicted 
by the effective potential method of Pfeiffer et al.~\cite{PTC}.
As already mentioned above, the numerical initial data chosen in these
works rely only on an Initial Value Problem (IVP)
 formulation, and significantly differ both from the numerical initial data
constructed by the HKV method \cite{GGB, Cook} and from
the predictions made by the EOB method (while the HKV
and EOB results are quite close to each other \cite{DGG,CookPfeiffer}).
For instance, \cite{DGG} estimated that the ratio between the
orbital frequencies $\omega_{\rm IVP}/\omega_{\rm EOB}$ was about 2.
One should probably wait until HKV-type initial data for
spinning BH's are evolved until coalescence to meaningfully
compare their results with the results derived above for
energy releases within the EOB approach. However,
to have an idea of the current distance between analytical
estimates and numerical ones, we
 have determined the energy released between the LSSO and
the final frequency, at 2PN and 3PN order, for two of the spin configurations 
investigated by Baker et al. For spins aligned (anti-aligned) and  
$\chi_1 = \chi_2 = 0.17$ ($\chi_1 = \chi_2 = 0.25$), we find that 
the energy released is $\sim (0.6 - 0.9) \% M$ [$\sim (1 - 3)\% M$]. 
Baker et al. found $\sim (1.7-1.9) \% M$ and $\sim (1.9-2.1) \% M$, 
respectively. It should be noted that the energy released 
evaluated by Baker et al. includes also the ring-down phase. Ours does not.

\subsection{Complete waveforms describing the  non-adiabatic inspiral and coalescence
of precessing binary black holes}

Finally, though we have not yet carefully studied at which stage
we could meaningfully join our ``quasi-plunge'' evolution to the
formation of a ringing BH, we have decided, to show the promise
of a purely analytical EOB-base approach to follow
 Ref.~\cite{BD2} in matching (by
requiring first-order continuity of the emitted waveform)
 the end of our  waveform (here defined by the first violation
 of the ``adiabatic criteria'' \eqref{cond3}--\eqref{cond4}) to
 a ringdown waveform generated from the lowest $l=m=2$
quasi-normal mode of a Kerr black hole. We determine
the  mass and spin parameters of the final hole
 by the energy and angular momentum of the binary at the
end of our evolution:
\begin{equation}
M_{\rm BH} = E_{\rm fin}\,,\quad a_{\rm BH} = \left[{|\mathbf{J}|}/{E^2}\right]_{\rm fin}\,.
\end{equation}
In Fig.~\ref{fig:ringdown}, we plot the complete waveforms,
so obtained,  for non-spinning, and
maximally spinning $(15+15)M_\odot$ binaries in the generic-up and
generic-down configurations (these refer to the configuration at
$f_{\rm GW}=40\,$Hz, the starting point of evolution). We have shifted
these waveforms in time so that the end of inspiral evolutions all
happen at $t=0$.  Notice that at this stage the waveform which includes 
the ring-down phase should be considered as an example.  
Indeed, by restricting ourselves to the quasi-normal mode $l=m=2$, 
we have tacitly assumed that the total angular momentum at the time 
the ring-down phase starts is dominated by the orbital angular momentum. 
However, this is not generally the case when spins are present and 
the quasi-normal modes with $l\neq 2$ might be excited, as well, 
and contribute to the waveform. A more thourough analysis is left for 
the future.

\section{Conclusions}
\label{sec5}

\begin{figure}
\begin{center}
\epsfig{file=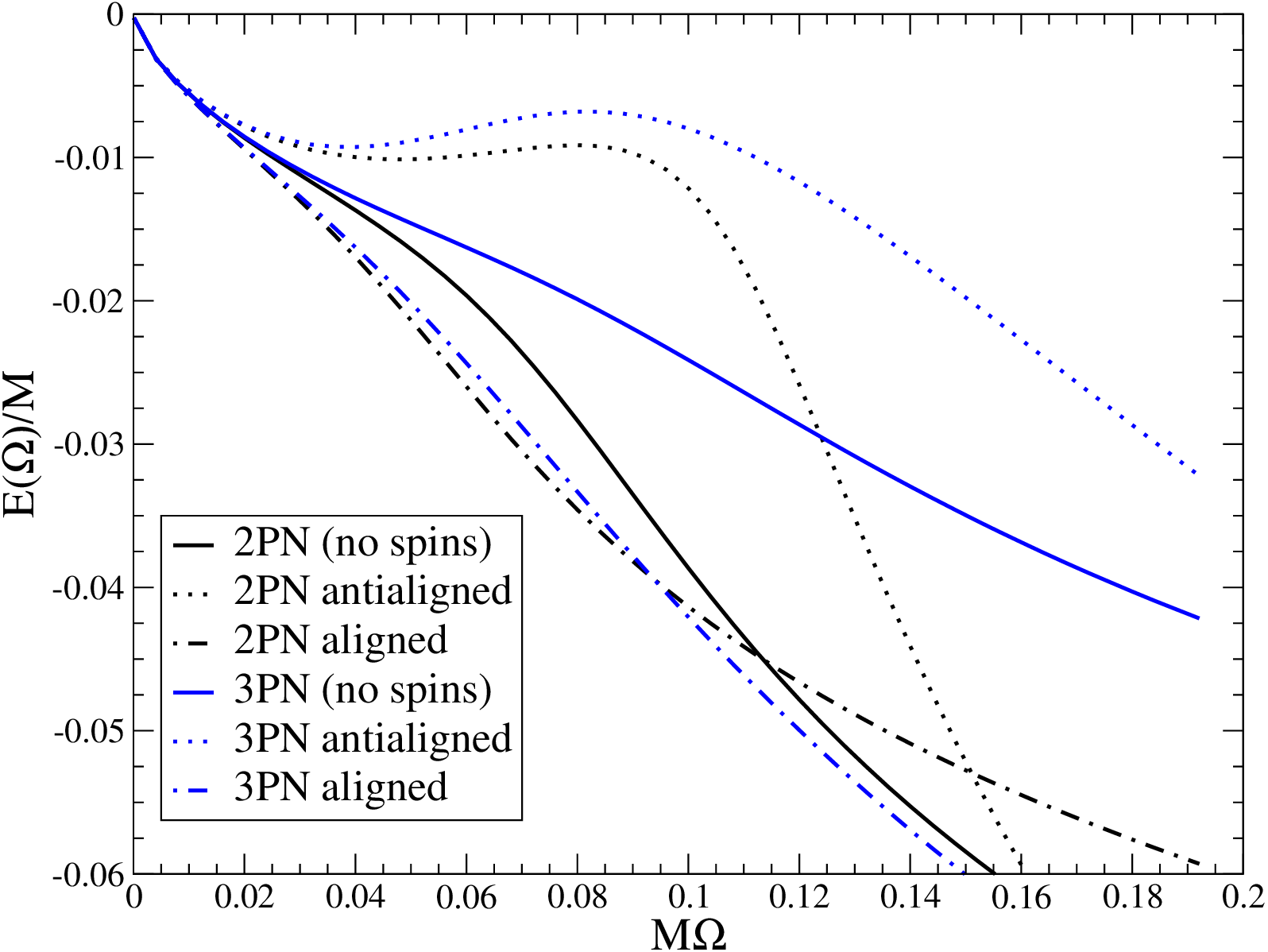,width=0.95\sizetwofig,angle=0} \hspace{0.5cm}
\epsfig{file=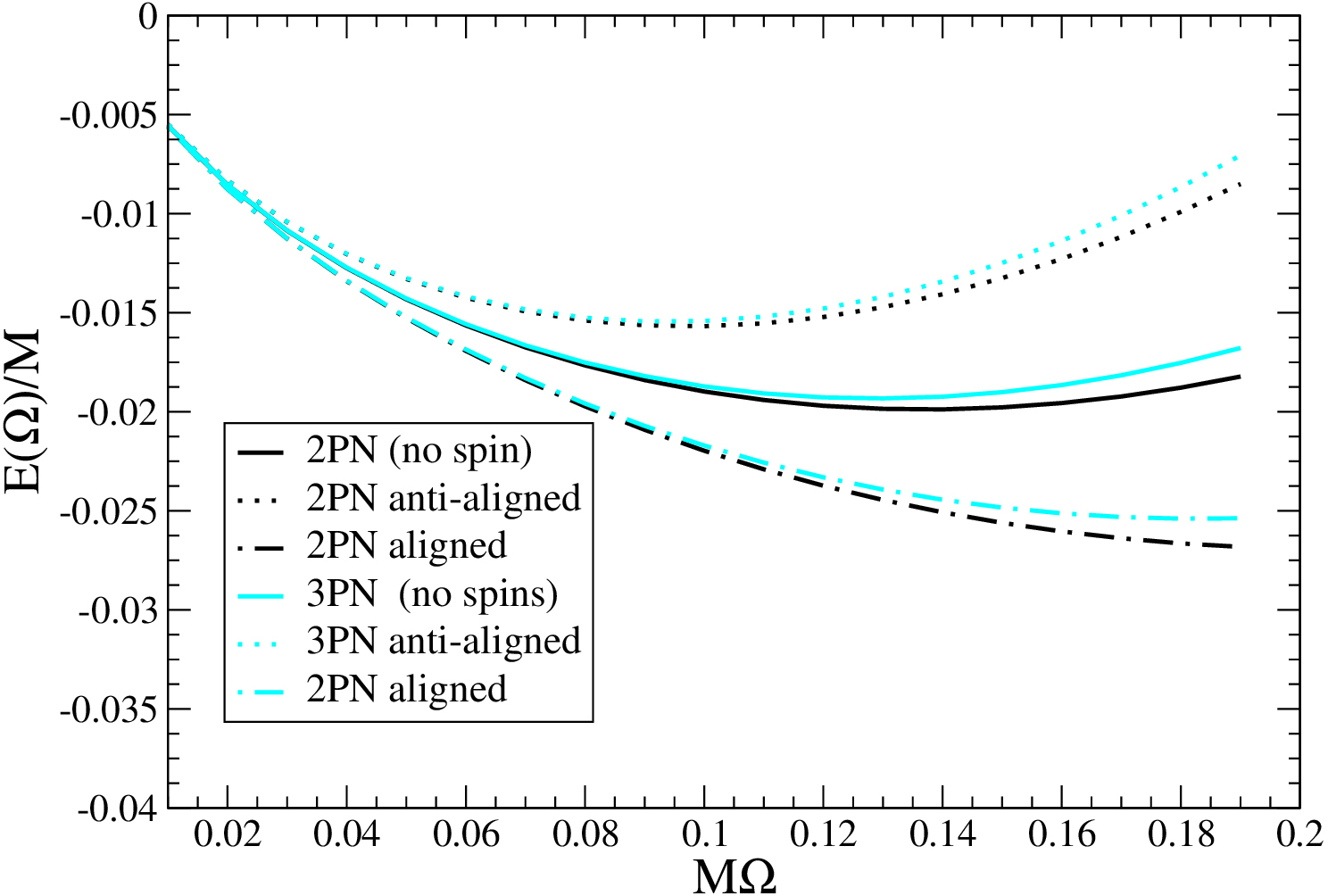,width=0.95\sizetwofig,angle=0} 
\caption{
\label{fig:eomega}
The energy for circular orbits as function of the frequency evaluated using 
the PN-expanded Hamiltonian (left panel) and the 
PN-expansion of the analytically computed function given by Eq.~(\ref{s1}) (right panel)  
at various PN orders for maximal spins and equal mass binaries.}
\end{center}
\end{figure}

\begin{figure}
\begin{center}
\epsfig{file=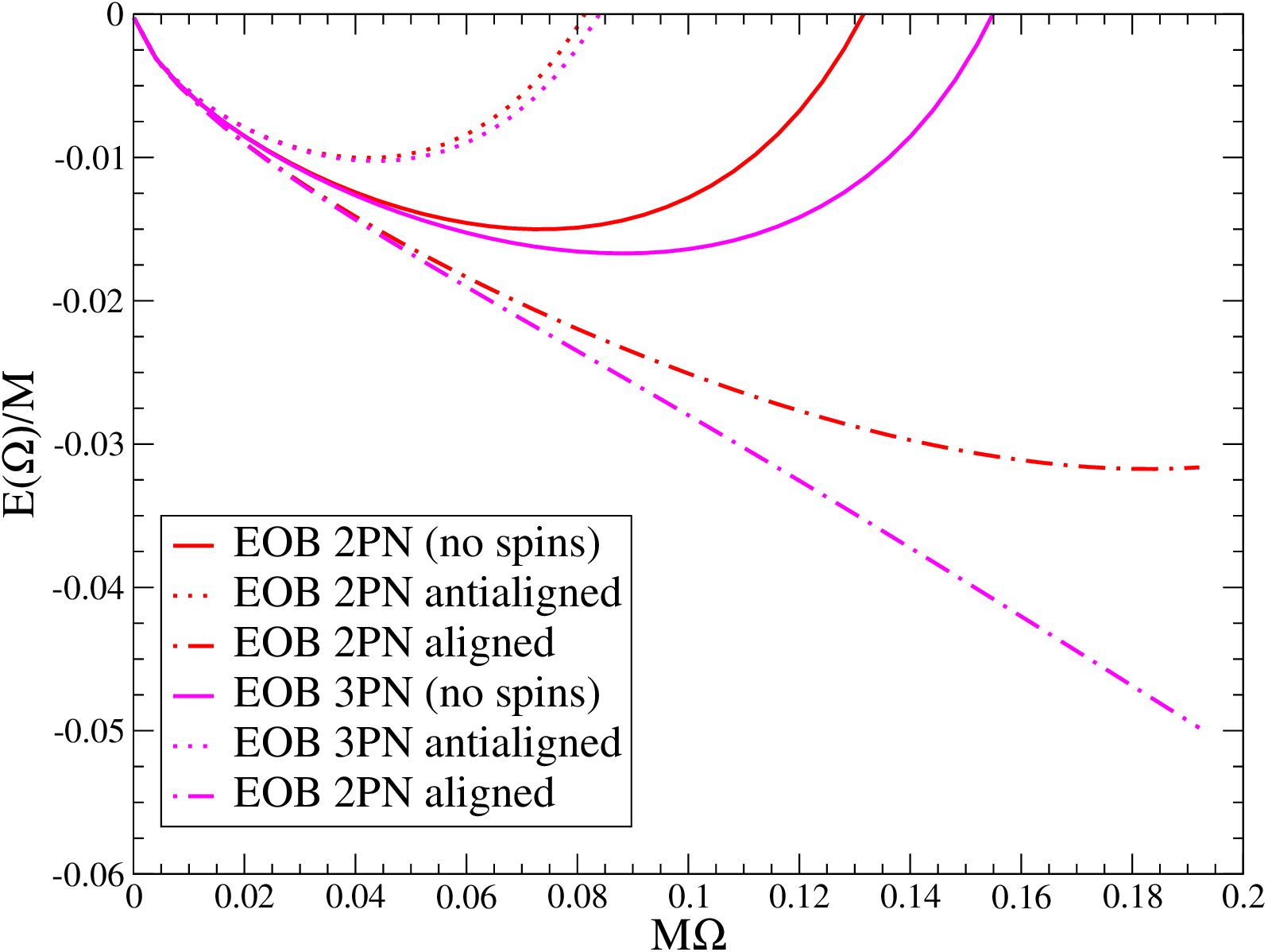,width=0.95\sizetwofig,angle=0}
\caption{
\label{fig:eomegaEOB}
The energy for circular orbits as function of the frequency evaluated from the 
EOB Hamiltonian at various PN orders for maximal spins and equal mass binaries.}
\end{center}
\end{figure}

\begin{figure}
\begin{center}
\epsfig{file=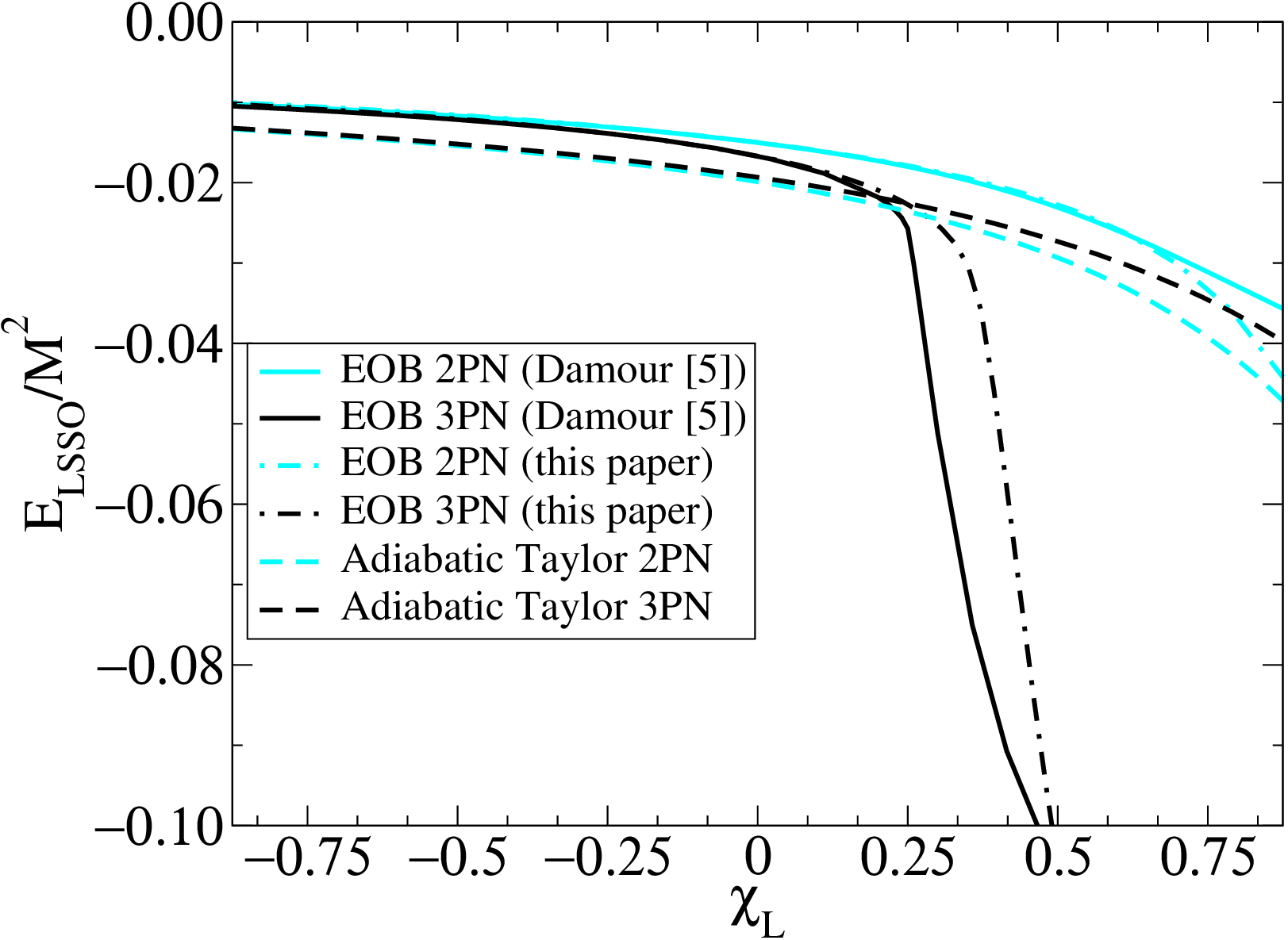,width=0.95\sizetwofig,angle=0} \hspace{0.5cm}
\epsfig{file=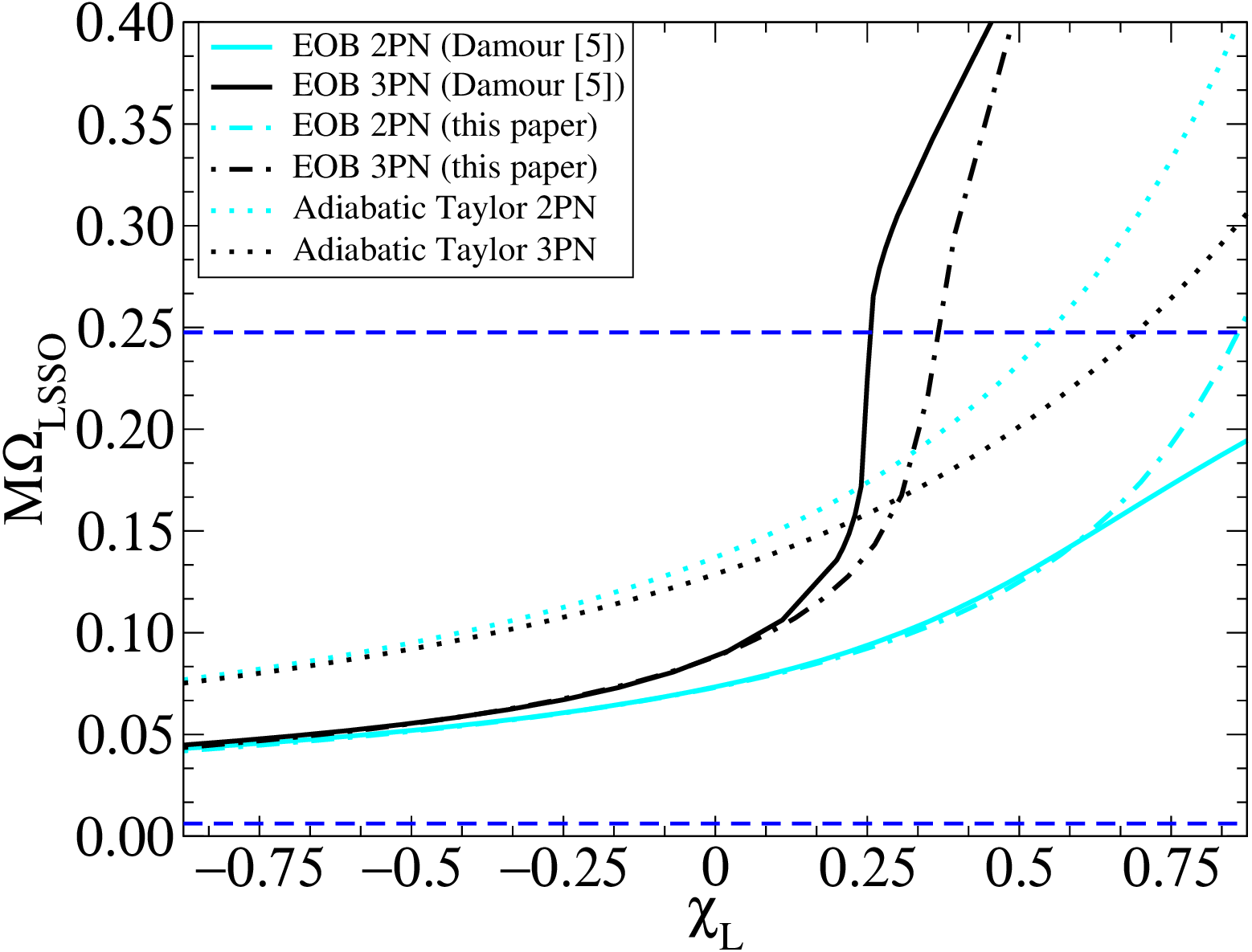,width=0.95\sizetwofig,angle=0}
\caption{
\label{fig:e-a}
The energy (left panel) and the frequency (right panel) 
at the LSSO as function of $\chi_L/M^2 \equiv
S_{\rm eff} \cdot \hL/M^2$ in the
equal mass case for EOB Hamiltonian and PN-expanded 
analytically computed function ${E}(\Omega)$ [see right panel of Fig.~\ref{fig:eomega}]. 
The horizontal dashed line in the right panel marks the highest LSSO angular frequency 
for BBHs with total mass in the range $10\mbox{--}40 M_\odot$, assuming the 
LIGO frequency band $40 \leq f_{\rm GW} \leq 240 Hz$.}
\end{center}
\end{figure}

\begin{figure}
\begin{center}
\epsfig{file=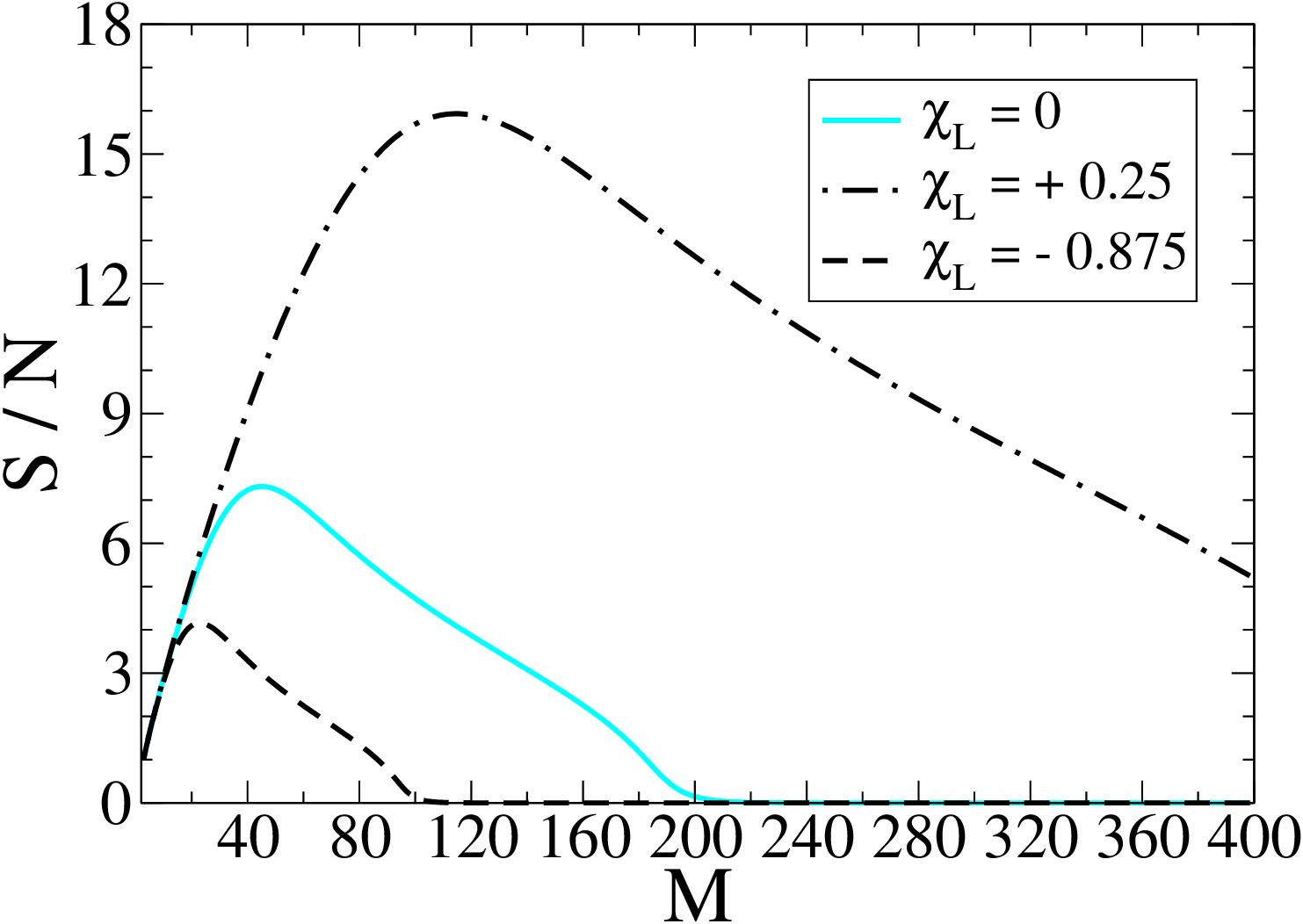,width=0.95\sizetwofig,angle=0} 
\caption{\label{snr}
Signal-to-noise ratio versus binary total mass at 100 Mpc for equal-mass binaries with LSSO 
determined by the 3PN-EOB Hamiltonian.}
\end{center}
\end{figure}

\begin{figure}
\begin{center}
\epsfig{file=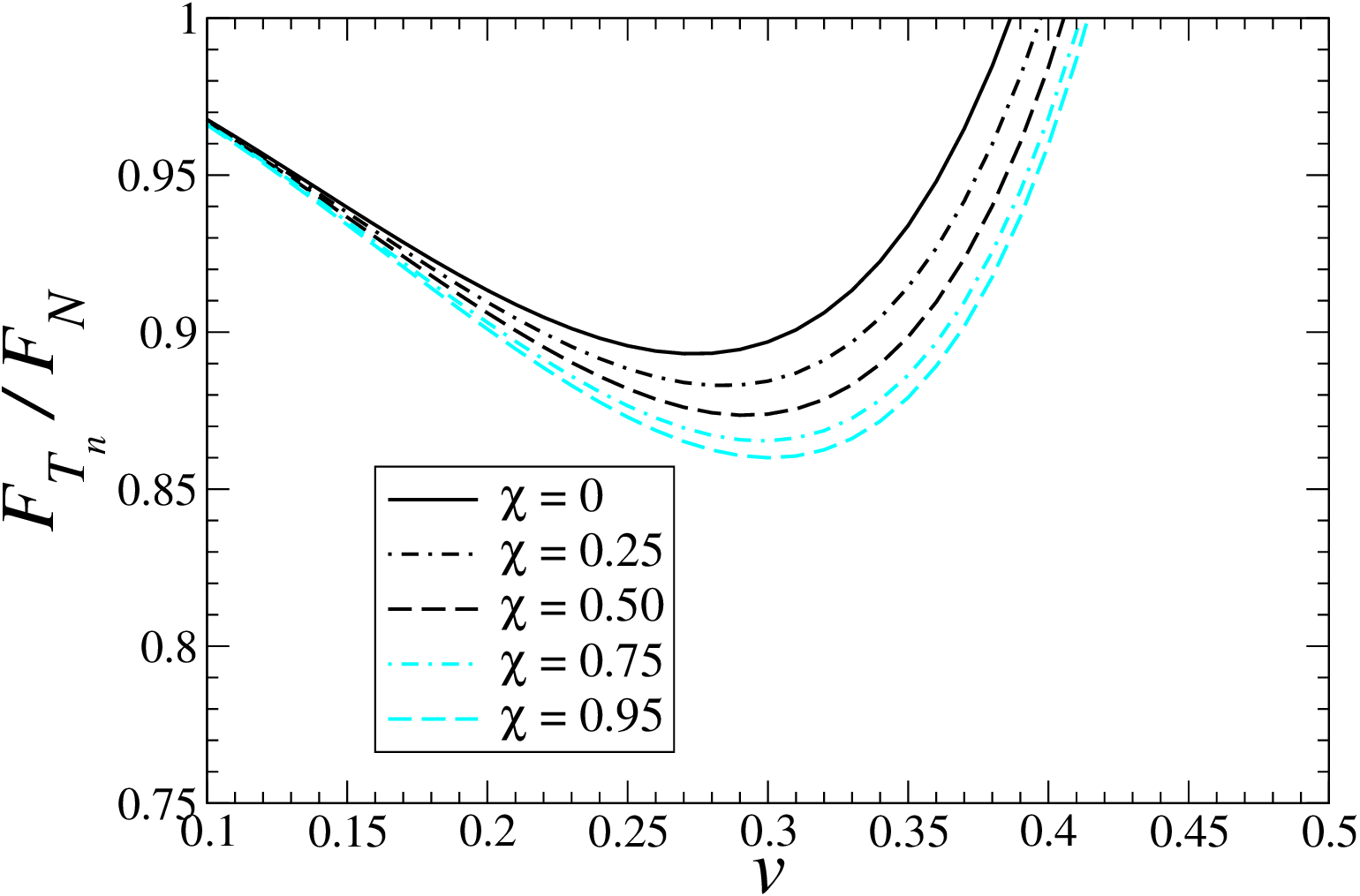,width=0.95\sizetwofig,angle=0} \hspace{0.5cm}
\epsfig{file=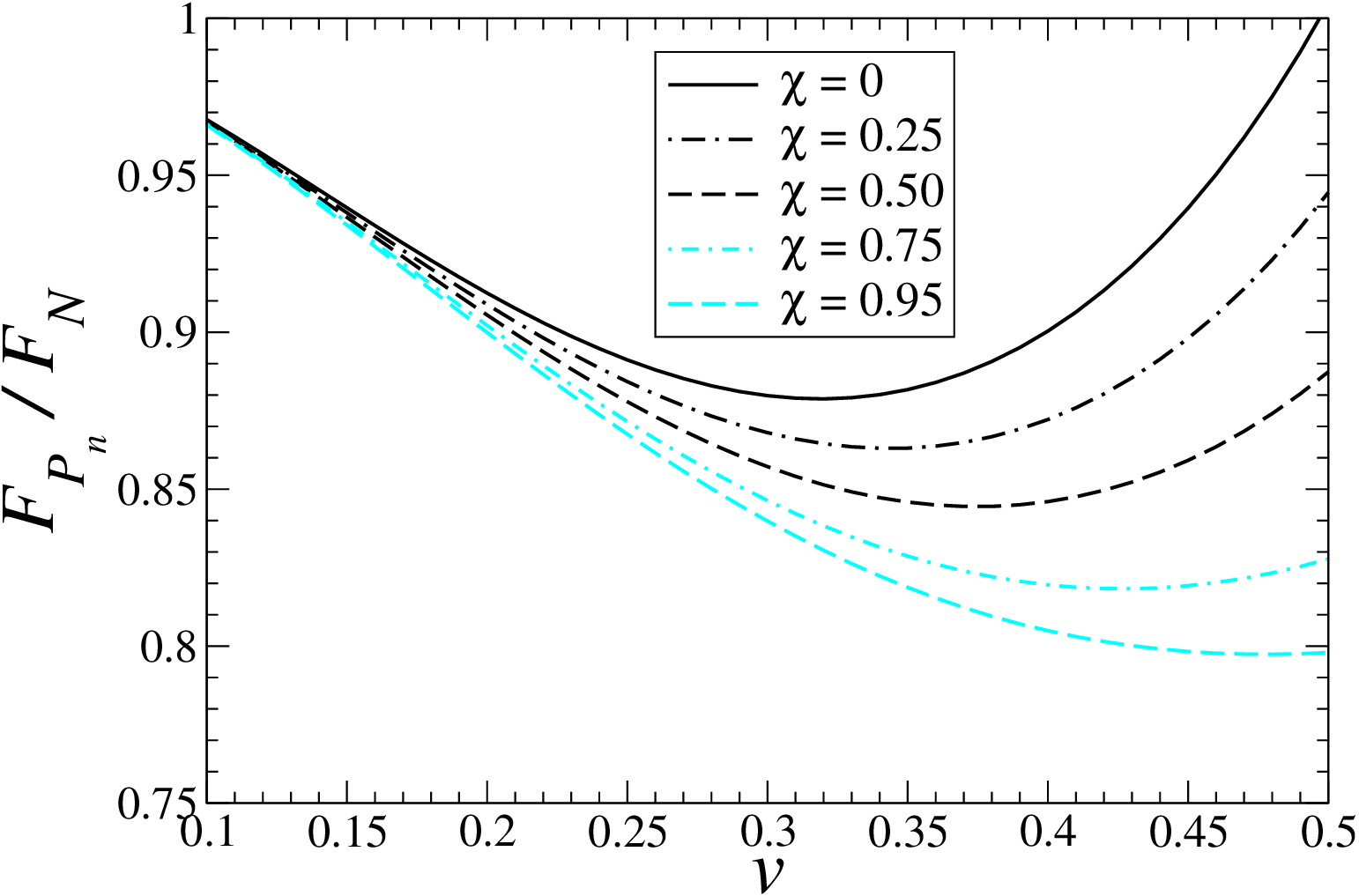,width=0.95\sizetwofig,angle=0}
\caption{Newton-normalized flux in the equal-mass case with both BH spins aligned 
(and maximal $\chi = \chi_1=\chi_2$) 
with orbital angular momentum when T-approximants (left panel) and (upper-diagonal) 
P-approximants (right panel) are used. \label{Fig6}}
\end{center}
\end{figure}

\begin{figure}
\begin{center}
\epsfig{file=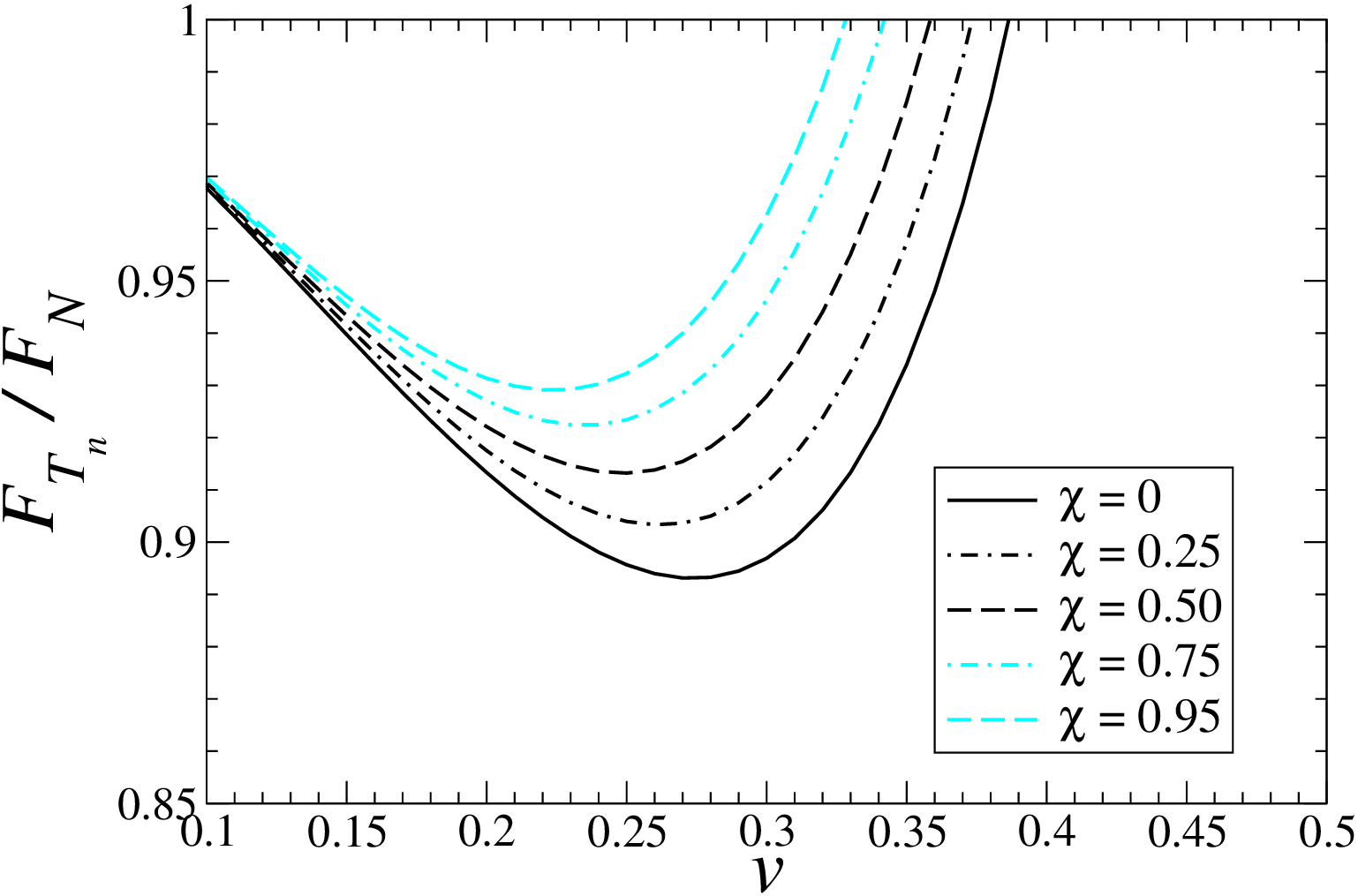,width=\sizetwofig,angle=0} \hspace{0.5cm}
\epsfig{file=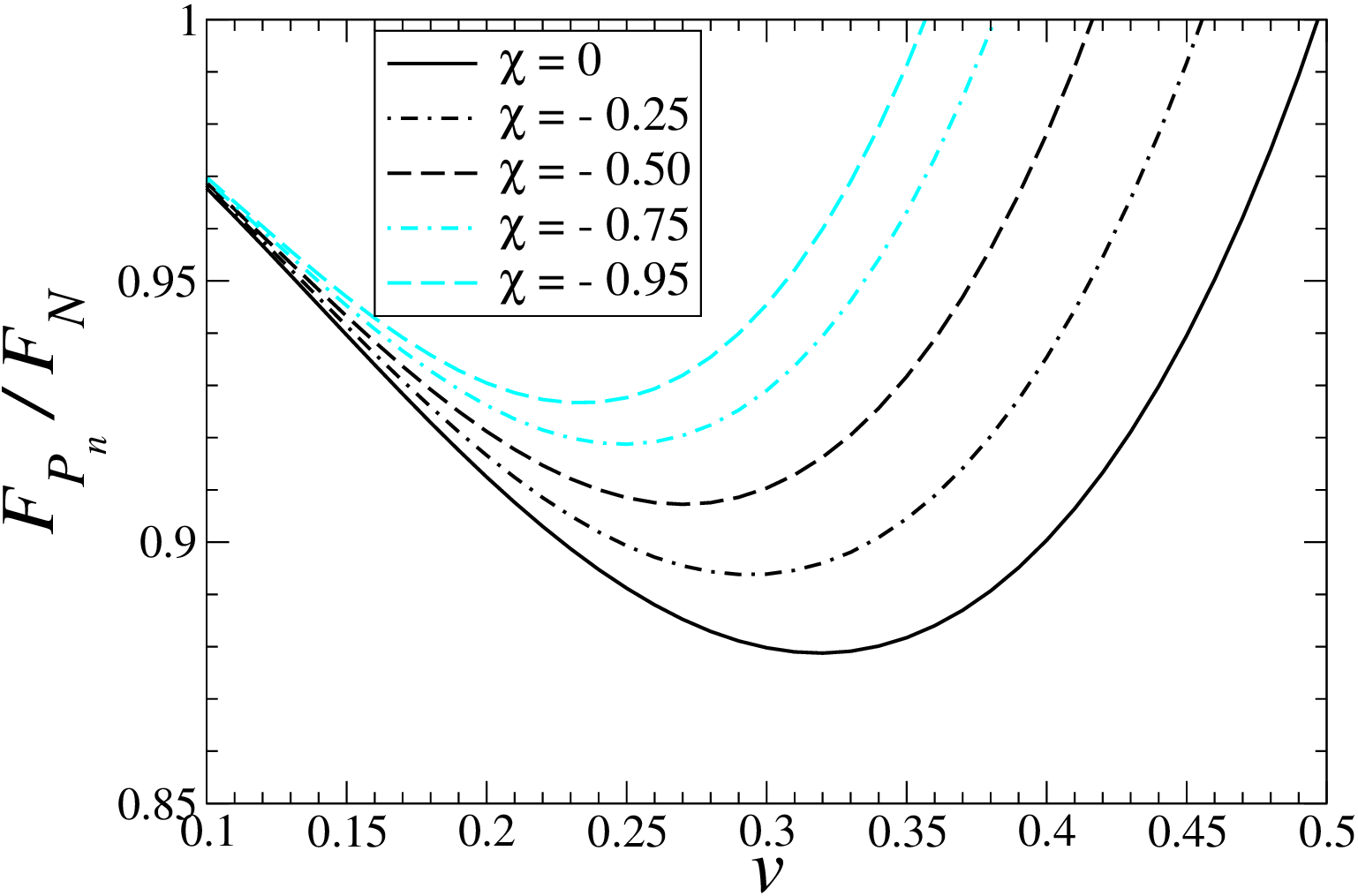,width=\sizetwofig,angle=0}
\caption{Newton-normalized flux in the equal-mass case with both BH spins antialigned
(and maximal $\chi = \chi_1=\chi_2$) with orbital angular momentum 
when T-approximants (left panel) and (upper-diagonal) P-approximants (right panel) are used. 
\label{Fig7}}
\end{center}
\end{figure}

\begin{figure}
\begin{center}
\epsfig{file=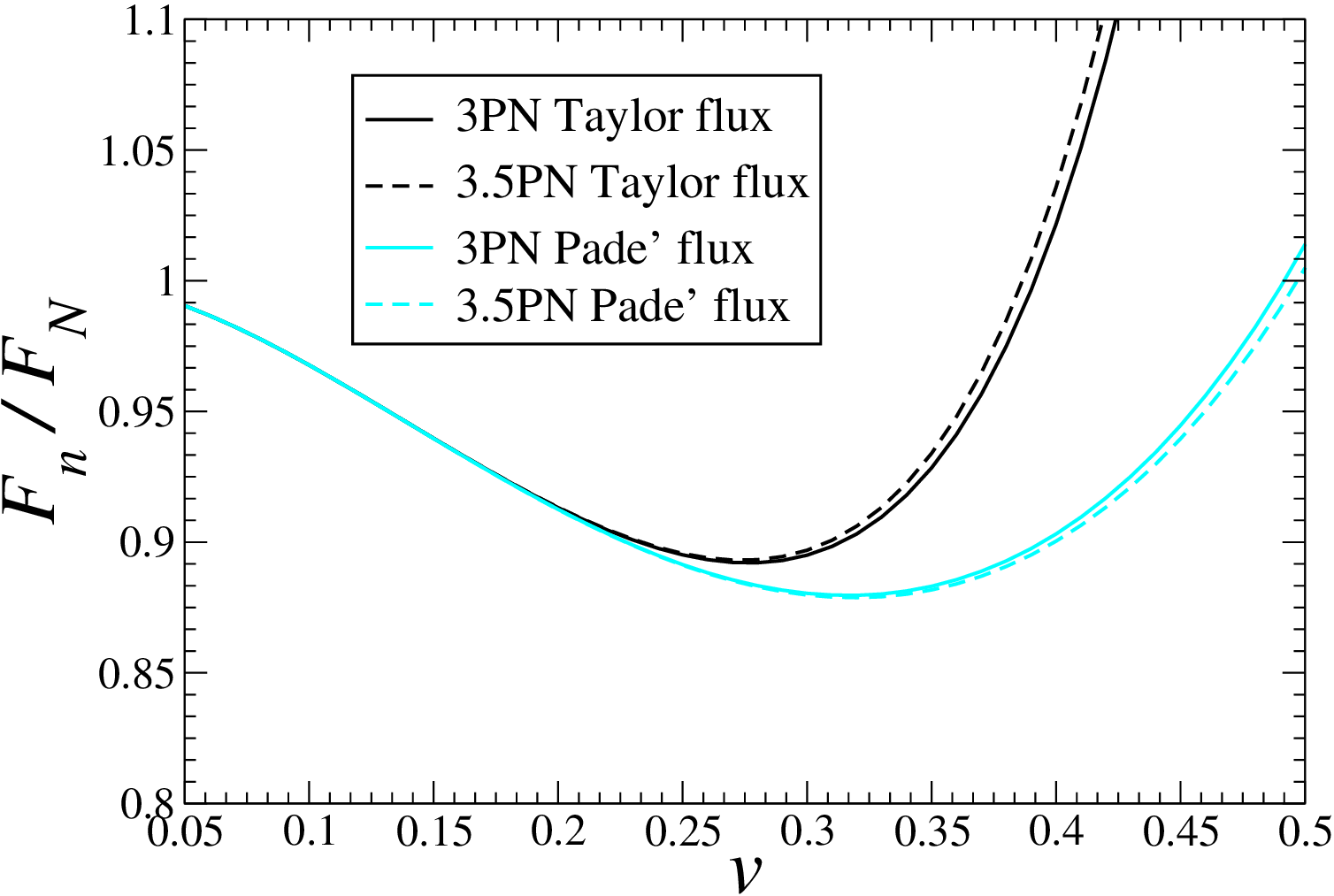,width=\sizetwofig,angle=0} 
\caption{Comparison between the T- and (upper diagonal) P-approximant 
Newton-normalized flux in the equal mass case 
at 3PN and 3.5PN order.\label{Fig5}}
\end{center}
\end{figure}

\begin{figure}
\includegraphics[width=0.9\textwidth]{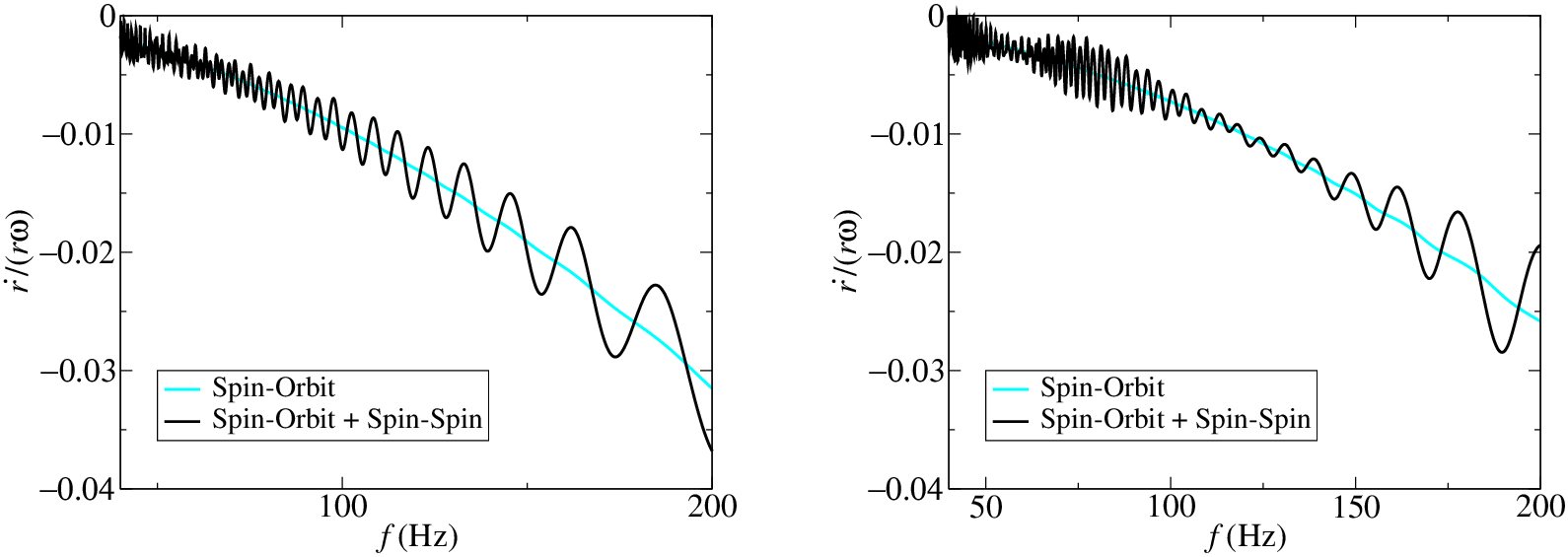}
\caption{Oscillations in $\dot{r}$ when spin-spin interactions are
present, in $(10+10)M_{\odot}$ (left panel) and $(15+5)M_{\odot}$
(right panel) binaries. Dark curves show $\dot{r}/(r\omega)$ as
functions of $f_{\rm GW}$ when both spin-orbit and spin-spin
interactions are take into account, while light curves show the same
quantity when only spin-orbit interactions are included.
\label{fig:rdot}}
\end{figure}

\begin{figure}
\includegraphics[width=0.45\textwidth]{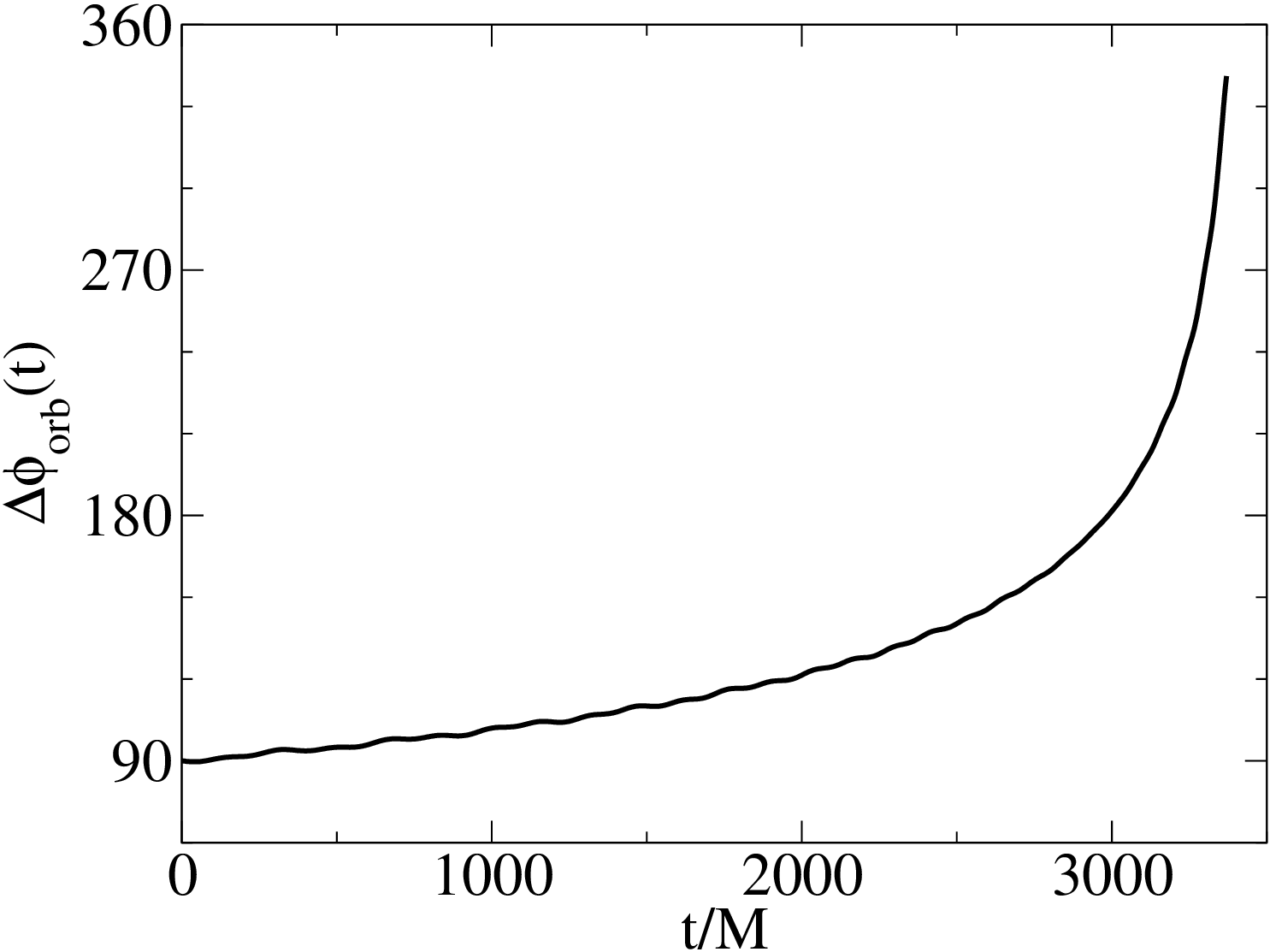}
\caption{\label{fig:rel-orb-phase} Relative orbital phase measured with respect to $\mathbf{S}_{\rm tot}
\times \hat{\mathbf{L}}_N $.  For maximally spinning $(15+15)M_\odot$
binaries, we start evolution at $40\,$Hz, with
$(\theta_{S_1},\phi_{S_1};\theta_{S_2},\phi_{S_2})=(60^\circ,90^\circ;60^\circ,0^\circ)$,
and orbital phases $\phi_{\rm orb}=0$ and $\pi/2$, and plot the
difference $\Delta\phi_{\rm orb}$. }
\end{figure}

\begin{figure}
\includegraphics[width=0.9\textwidth]{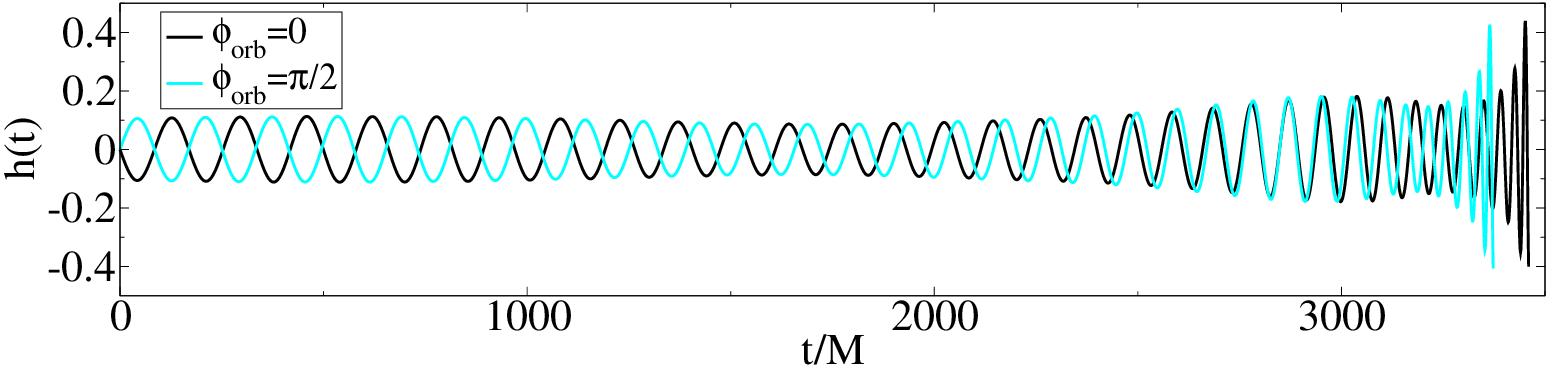}
\caption{Comparison between waveforms from configurations with
different initial orbital phases.  For a $(15+15)M_\odot$ maximally
spinning binary, we start evolution at $40\,$Hz, with
$(\theta_{S_1},\phi_{S_1};\theta_{S_2},\phi_{S_2})=(60^\circ,90^\circ;60^\circ,0^\circ)$,
and orbital phases $\phi_{\rm orb}=0$ and $\pi/2$, and compare the
waveforms detected with $(F_+,F_\times;\Theta,\varphi)=
(1,0;\pi/4,0)$.  \label{fig:orbphase}}
\end{figure}

\begin{figure}
\includegraphics[width=0.9\textwidth]{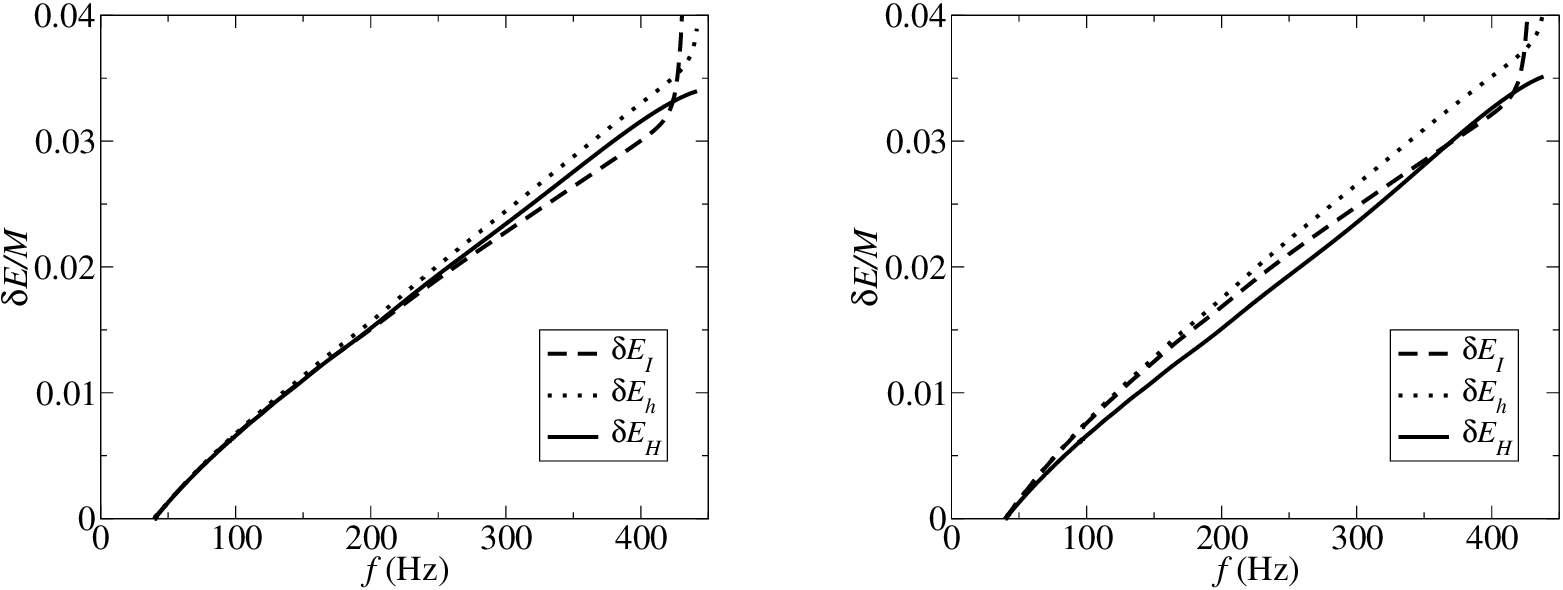}
\caption{
\label{fig:comp:eloss}
Comparison between the three different prescriptions, $\delta E_I$
(dashed curves), $\delta E_h$ (dotted curves) and $\delta E_H$ (solid
curves), for calculating energy losses. We use Newtonian-order
radiation reaction in the left panel, and Pad\'e at 3.5PN order in the
right panel. We use the EOB Hamiltonian at 3PN order}
\end{figure}

\begin{figure}
\begin{tabular}{c}
\includegraphics[width=0.9\textwidth]{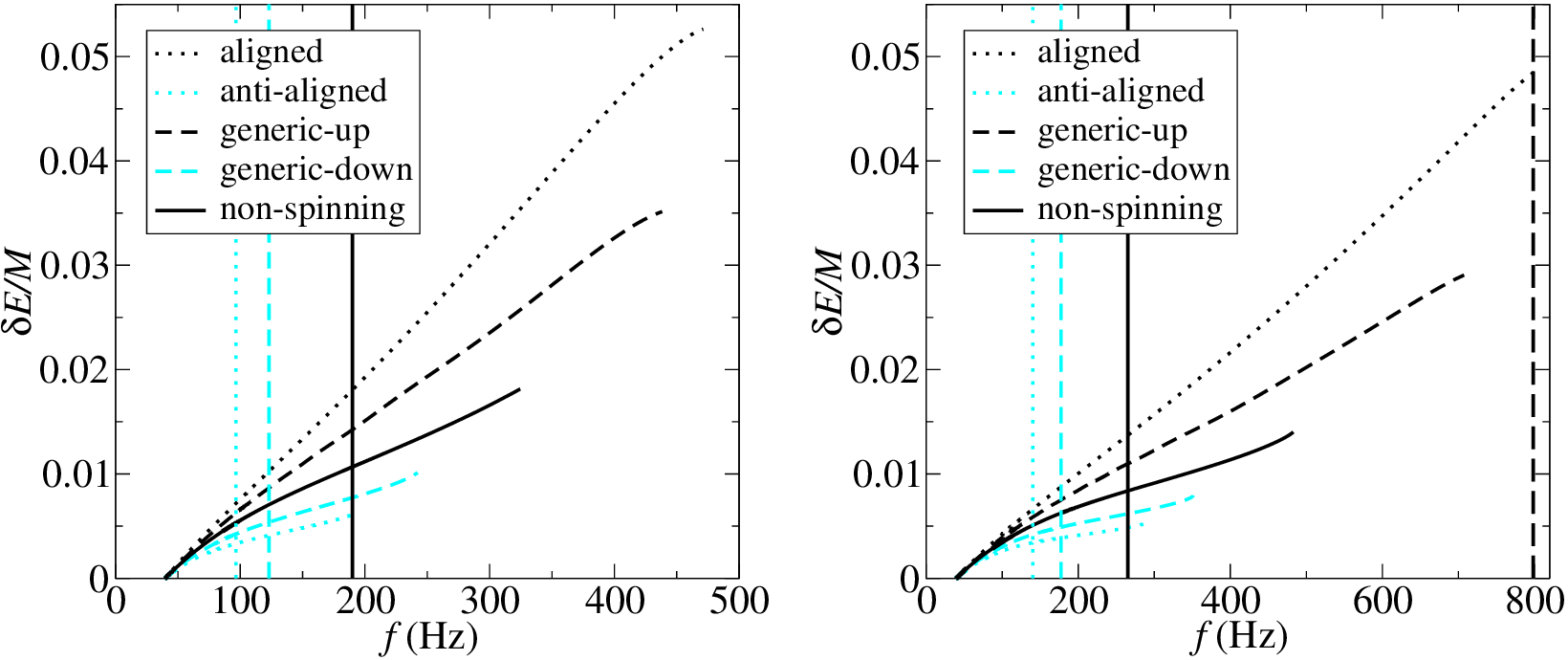} \\
\vspace{0.2cm}\\
\hspace{0.15cm}\includegraphics[width=0.9\textwidth]{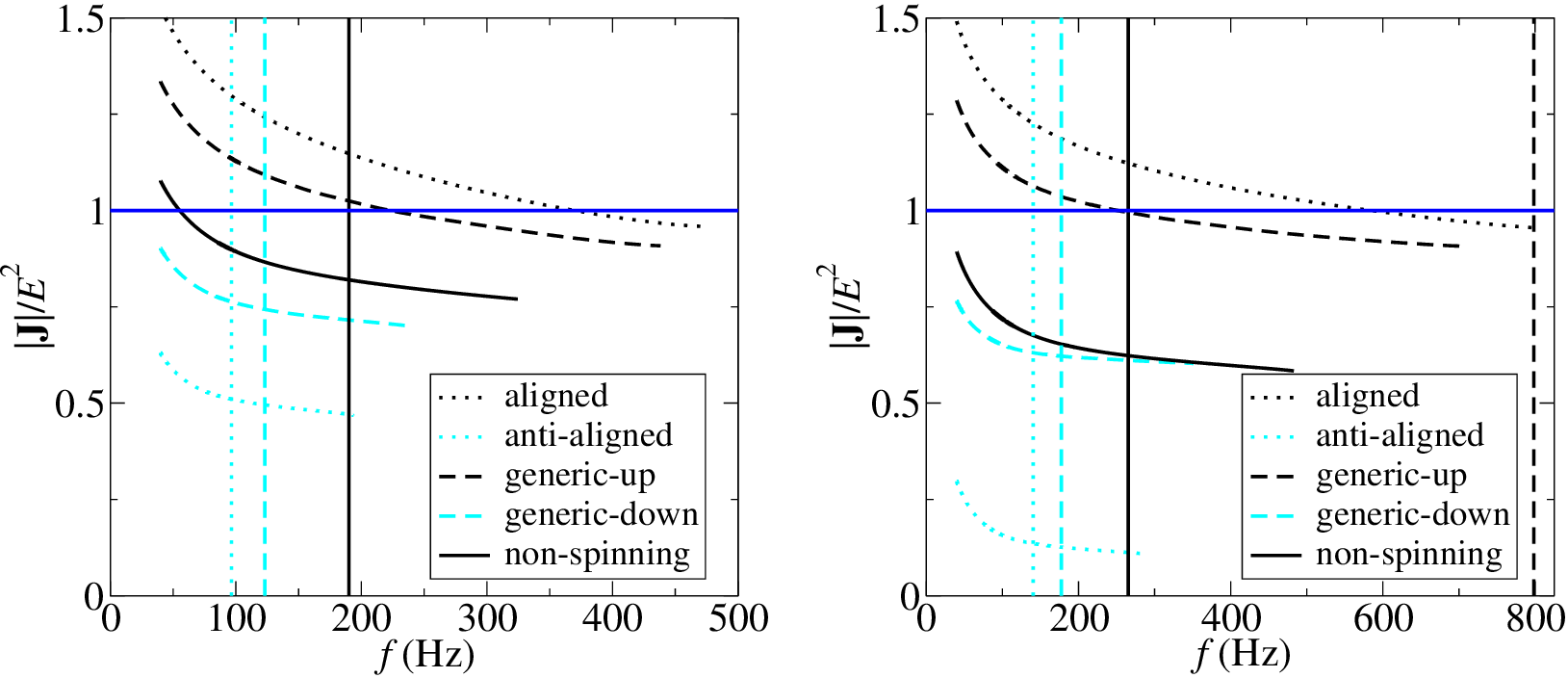}
\end{tabular}
\caption{
Accumulative energy release (upper panels) and instantaneous
values of $|\mathbf{J}|/E^2$ (lower panels) of $(15+15)M_{\odot}$
(left panels) and $(15+5)M_{\odot}$ (right panels) binaries. LSSO
frequencies for the anti-aligned, generic-down, and non-spinning
configurations are shown in vertical grid lines, while LSSOs of
generic-up and aligned configurations are above the ranges of our
plots. [Spin-spin terms are not included in these evolutions.] 
We use the SEHPF(3,3.5) model. 
 \label{fig:eloss}
}
\end{figure}

\begin{figure}
\includegraphics[width=0.9\textwidth]{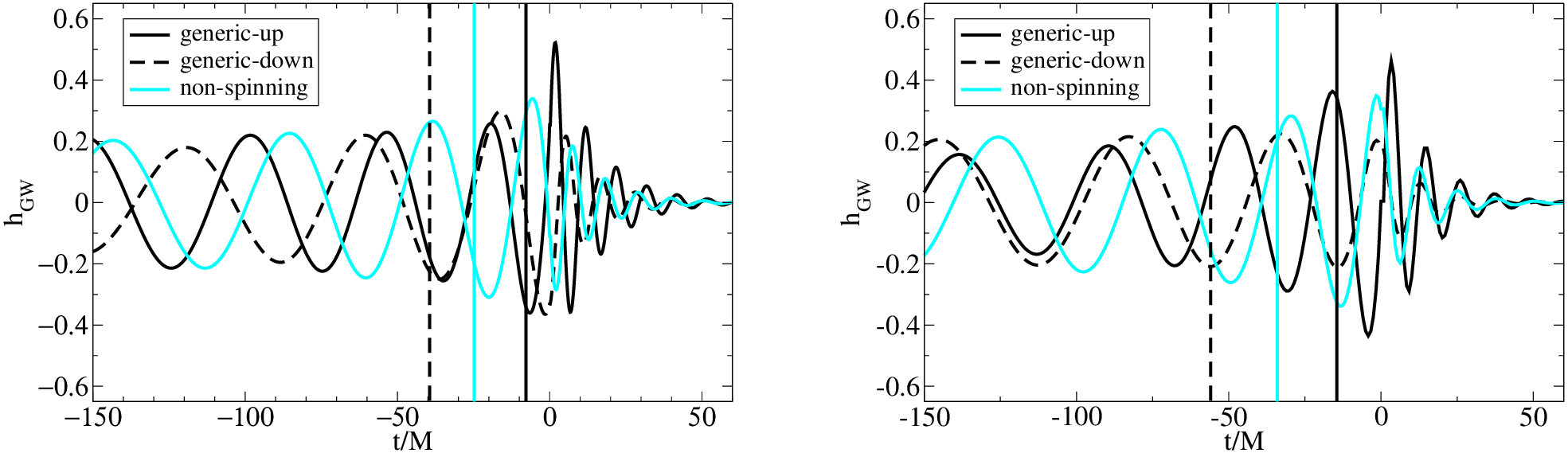}
\caption{
\label{fig:ringdown}
Inspiral waveforms (which end at $t=0$ in our plot) matched to ring-down 
waveforms for non-spinning (light solid curve), and half-maximally
spinning $(15+15)M_\odot$ binaries (left panel) and $(15+5)M_\odot$ binaries (right panel) 
in the generic-up (dark solid curve) and generic-down (dark dashed curve) configurations. We start
our evolutions at $40\,$Hz, and use $(F_+,F_\times;\Theta,\varphi)=(1,0;\pi/4,0)$. 
In the plot we mark the position of the LSSO with solid curves. The waveforms have been 
shifted in time such that the end of the inspiral occurs at $t/M =0$.  
}
\end{figure}

We provided a first attack on the problem of analytically determining
the gravitational waveforms emitted during the last stages of
dynamical evolution of precessing binaries of spinning black holes,
i.e. during the non-adiabatic ending of the inspiral phase, and its
transition to a plunge. We reviewed the various available Hamiltonian
descriptions of the dynamics of spinning black hole (BH) binaries, and
studied (following \cite{TD}) the characteristics of the stable
spherical orbits that exist when spin-spin effects are neglected
compared to spin-orbit ones. We derived the contribution to radiation
reaction (for quasi-circular orbits) which is linear in the spins. Our
results agree with the corresponding recent results of \cite{Will}. We then
used this analytical description of the radiation-reaction-driven
inspiral of spinning binaries to construct non-adiabatic models of
coalescing binary waveforms.  We compared the various models and
concluded, in confirmation of previous results, that our current
``best bet'' for a non-adiabatic model describing the transition from 
adiabatic inspiral to plunge is obtained by combining: (i) an effective one body (EOB)~
\cite{BD1,BD2}, 3PN-accurate \cite{DJS} resummed Hamiltonian, including 
spin-dependent interactions \cite{TD}, with (ii) Pad\'e-resummed radiation reaction force 
(including spin-terms). Conclusion (i) is rather robust, since 
as Fig.~\ref{fig:eomega} shows, the PN-expanded Hamiltonian does not show any 
LSSO and differs significantly from the PN-expanded analytically computed 
function ${E}(\Omega)$; conclusion (ii) is based on the assumption 
that the flux function in the equal-mass case is a smooth deformation of 
the test-mass limit result. Since in the latter case Pad\'e approximants 
were shown \cite{DIS98,PS} to have better agreement with exact 
numerical flux functions, we would conclude that this is also true in the 
equal-mass case. 

Our main results, obtained by means of this ``best bet'' EOB model are:

(1) An estimate of the energy and angular momentum released by the
binary system during its last stages of evolution:  inspiral, transition
from inspiral to plunge, and plunge;

(2) The finding (which confirms the conclusions of \cite{TD}) that
the dimensionless rotation parameter $j/E^2$ is always smaller than unity
at the end of the inspiral;

(3) The construction of complete waveforms, approximately describing
the entire gravitational-wave emission process from
precessing binaries of spinning black holes: adiabatic inspiral,
non-adiabatic transition between inspiral and plunge, plunge,
merger and ringdown. Following \cite{BD2} these waveforms were
constructed by matching a quasi-normal-mode ringdown to the end
of the plunge signal. These tentative complete waveforms are
 preliminary because we did not include here a careful study
of how to join, in a physically motivated manner, the last stages
of the plunge to the merger phase. They extend, however, the
(better justified) complete waveforms constructed in \cite{BD2}
to the more genral case of spinning and precessing binaries.

\acknowledgements

A.B. thanks the Max-Planck Institut f\"ur Gravitationsphysik (Albert-Einstein-Institut) for support during her visit.  Y.C.'s research is supported by Alexander von Humboldt Foundation's Sofja Kovalevskaja Award (funded by the German Federal Ministry of Education and Research), and by the NSF grant PHY-0099568 (during his stay at Caltech); he also thanks the Institut d'Astrophysique de Paris (CNRS) for support during his visit.

\begin{table}
\begin{tabular}{c|cccccccc}
\hline
\hline
$(\theta_{\rm S1}, \phi_{\rm S2}, \theta_{\rm S1}, \phi_{\rm S2}) $
  & $f_{\rm fin}^{\rm SEP}$ (Hz) & $f_{\rm LSSO}^{\rm SEP}$(Hz) & 
${\rm R}_{\rm fin}^{\rm SEP}/M$ & $f_{\rm fin}^{\rm SHT}$(Hz) & 
$f_{\rm LSSO}^{\rm SHT}$(Hz) &
${\rm R}_{\rm fin}^{\rm SHT}/M$ 
& $\rho_{\rm max, 2}$ & $\rho_{\rm max, 5}$ \\
\hline
\hline
\multicolumn{9}{l}{\hspace{6cm}$(10+10)M_\odot$}\\
\hline
\hline
no spin & 289 & 285 & 4.8 & 287 & 285 & 4.2 & 0.9150 & - \\
\hline
$(0^o, 0^o, 0^o, 0^o)$ & 745  & 2145  & 2.5 & 466 & 2145 & 2.8 & 0.3750 &  -   \\	
$(180^o, 0^o, 0^o, 0^o)$  & 290  & 285  & 4.8 & 285 & 285 & 4.2 & 0.9166 &  -   \\	
$(180^o, 0^o, 180^o, 0^o)$  & 145  & 145 & 8.1 & 145 & 145 & 7.4 & 0.5587 &  -   \\	
$(60^o, 90^o, 60^o, 0^o)$ & 633 & 873 & 2.6 & 502 & 502 & 2.6 & 0.4851 & 0.5883 \\ 
$(120^o, 90^o, 60^o, 0^o)$ & 280 & 280 & 4.9 & 269 & 269 & 4.4 & 0.5420 & 0.9472 \\ 
$(120^o, 90^o, 120^o, 0^o)$ & 187 & 187 & 6.7 & 186 & 186 & 6.0 & 0.6096 & 0.9536 \\ 
\hline
\hline
\multicolumn{9}{l}{\hspace{6cm}$(15+15)M_\odot$} \\
\hline
\hline
no spin & 192 & 190 & 4.8 & 192 & 190 & 4.2 & 0.8137 & - \\
\hline
$(0^o, 0^o, 0^o, 0^o)$ & 497  & 1430  & 2.6 & 311 & 1430 & 2.6 & 0.4550 &  -   \\	
$(180^o, 0^o, 0^o, 0^o)$  & 193  & 190  & 4.8 & 191 & 190 & 4.2 & 0.8148 &  -   \\	
$(180^o, 0^o, 180^o, 0^o)$  & 98  & 97 & 8.0 & 97 & 97 & 7.4 & 0.6403 &  -   \\	
$(60^o, 90^o, 60^o, 0^o)$ & 429 & 758 & 2.6 & 347 & 347 & 2.4 & 0.5067 & 0.6290  \\ 
$(120^o, 90^o, 60^o, 0^o)$ & 185 & 185 & 4.9 & 180 & 180 & 4.4 & 0.5112 & 0.9456  \\ 
$(120^o, 90^o, 120^o, 0^o)$ & 124 & 124 & 6.7 & 124 & 124 & 6.7 & 0.6868 & 0.9684 \\ 
\hline
\hline
\multicolumn{9}{l}{\hspace{6cm}$(15+5)M_\odot$}\\
\hline
\hline
no spin & 267 & 265 & 5.2 & 265 & 265  &4.2 & 0.6023 & - \\
\hline
$(0^o, 0^o, 0^o, 0^o)$ & 862  & 1442  & 2.3 & 479 & 1442 & 2.5 & 0.3268 &  -   \\	
$(180^o, 0^o, 0^o, 0^o)$  & 176  & 175  & 7.0 & 177 & 175 & 6.2 & 0.5188 &  -   \\	
$(180^o, 0^0, 180^o, 0^o)$  & 141  & 140 & 8.3 & 140 & 140 & 7.5 & 0.4445 &  -   \\	
$(60^o, 90^o, 60^o, 0^o)$ & 715 & 796 & 2.4 & 425 & 743 & 2.4 & 0.4478 & 0.6111  \\ 
$(120^o, 90^o, 60^o, 0^o)$ & 208 & 207 & 6.2 & 208 & 208 & 5.3 & 0.5471 & 0.7496  \\ 
$(120^o, 90^o, 120^o, 0^o)$ & 224 & 224 & 5.9 & 225 & 225 & 4.9 & 0.5735 & 0.8360  \\ 
\hline
\end{tabular}
\caption{We list the overlaps between STHTF(3,3.5), used as target model, and 
SEHPF(3,3.5), used as template model, for several binary masses and 
initial spin orientations.  
The two black holes are assumed to carry maximal spins $\chi_1=\chi_2=1$, 
but for comparion we also show the results in absence of spins. 
The evolution starts at $f_{\rm in} = 30$ Hz. In the first three columns 
we list the ending frequency, the LSSO frequency  and the 
BH radial separation at $t_{\rm fin}$ for the template model and 
in the second three columns we show the same quantities but for the target 
model. The last two columns contains the overlap maximized only over 2 extrinsic 
parameters $\rho_{\rm max,2}$ and maximed over 5 extrinsic parameters 
$\rho_{\rm max, 5}$, as described in the text. [Notice that for the spin configuration 
$(60^o, 90^o, 60^o, 0^o)$ and masses $(10+10)M_\odot$ and $(15+5)M_\odot$, due to 
a pole in the Pad\'e-approximant flux, we apply the Pad\'e resummation only to 
the non-spinning terms in the flux.]  
\label{overlapmax3pn}}
\end{table}

\begin{table}
\begin{tabular}{c|cccccccc}
\hline
\hline
$(\theta_{\rm S1}, \phi_{\rm S2}, \theta_{\rm S1}, \phi_{\rm S2}) $
  & $f_{\rm fin}^{\rm SEP}$ (Hz) & $f_{\rm LSSO}^{\rm SEP}$(Hz) & 
${\rm R}_{\rm fin}^{\rm SEP}/M$ & $f_{\rm fin}^{\rm SHT}$(Hz) & 
$f_{\rm LSSO}^{\rm SHT}$(Hz) &
${\rm R}_{\rm fin}^{\rm SHT}/M$ 
& $\rho_{\rm max, 2}$ & $\rho_{\rm max, 5}$ \\
\hline
\hline
\multicolumn{9}{l}{\hspace{6cm}$(10+10)M_\odot$}\\
\hline
\hline
no spin & 242 & 237 & 5.6 & 237 & 237 & 4.5 & 0.4691 & - \\
\hline
$(0^o, 0^o, 0^o, 0^o)$ & 628  & 628  & 2.9 & 628 & 628 & 2.6 & 0.3170 &  -   \\	
$(180^o, 0^o, 0^o, 0^o)$  & 237  & 237  & 5.6 & 237 & 237 & 4.5 & 0.4681 &  -   \\	
$(180^o, 0^o, 180^o, 0^o)$  & 139  & 139 & 8.4 & 139 & 139 & 7.6 & 0.6433 &  -   \\	
$(60^o, 90^o, 60^o, 0^o)$ & 367 & 367 & 4.1 & 342 & 342 & 3.5 & 0.4197 & 0.4882 \\ 
$(120^o, 90^o, 60^o, 0^o)$ & 234 & 234 & 5.7 & 229 & 229 & 4.7 & 0.4220 & 0.6015 \\ 
$(120^o, 90^o, 120^o, 0^o)$ & 173 & 173 & 7.1 & 172 & 172 & 6.3 & 0.6681 & 0.9556 \\ 
\hline
\hline
\multicolumn{9}{l}{\hspace{6cm}$(15+15)M_\odot$} \\
\hline
\hline
no spin & 158 & 158 & 5.6 & 158 & 158 & 4.5 & 0.4880 & - \\
\hline
$(0^o, 0^o, 0^o, 0^o)$ & 419  & 419  & 2.9 & 419 & 419 & 2.6 & 0.4044 &  -   \\	
$(180^o, 0^o, 0^o, 0^o)$  & 158  & 158  & 5.6 & 158 & 158 & 4.5 & 0.4885 &  -   \\	
$(180^o, 0^o, 180^o, 0^o)$  & 93  & 93 & 8.4 & 93 & 93 & 7.6 & 0.7140 &  -   \\	
$(60^o, 90^o, 60^o, 0^o)$ & 240 & 238 & 4.2 & 241 & 241 & 4.1 & 0.4549 & 0.5186  \\ 
$(120^o, 90^o, 60^o, 0^o)$ & 156 & 155 & 5.7 & 152 & 152 & 4.7 & 0.4827 & 0.6767  \\ 
$(120^o, 90^o, 120^o, 0^o)$ & 115 & 115 & 7.1 & 116 & 116 & 6.2 & 0.7227 & 0.9442 \\ 
\hline
\hline
\multicolumn{9}{l}{\hspace{6cm}$(15+5)M_\odot$}\\
\hline
\hline
no spin & 232 & 232 & 5.7 & 233 & 232  & 4.7 & 0.6111 & - \\
\hline
$(0^o, 0^o, 0^o, 0^o)$ & 608  & 608  & 2.9  & 608 & 608 & 2.6 & 0.2695 &  -   \\	
$(180^o, 0^o, 0^o, 0^o)$  & 167  & 166  & 7.3 & 166 & 166 & 6.5 & 0.8720 &  -   \\	
$(180^o, 0^0, 180^o, 0^o)$  & 136  & 136 & 8.5 & 136 & 136 & 7.7 & 0.6743 &  -   \\	
$(60^o, 90^o, 60^o, 0^o)$ & 352 & 352 & 4.2 & 367 & 367 & 3.2 & 0.2696 & 0.4978  \\ 
$(120^o, 90^o, 60^o, 0^o)$ & 192 & 192 & 6.6 & 191 & 191 & 5.7 & 0.6566 & 0.8173  \\ 
$(120^o, 90^o, 120^o, 0^o)$ & 169 & 169 & 7.2 & 167 & 167 & 6.5 & 0.6207 & 0.8970  \\ 
\hline
\end{tabular}
\caption{Overlaps between STHTF(2,2), used as target model, and 
SEHPF(2,2.5), used as template model, for several binary masses and 
initial spin orientations.  
The two black holes are assumed to carry maximal spins $\chi_1=\chi_2=1$, 
but for comparion we also show the results in absence of spins. 
The evolution starts at $f_{\rm in} = 30$ Hz. In the first three columns 
we list the ending frequency, the LSSO frequency  and the 
BH radial separation at $t_{\rm fin}$ for the template model and 
in the second three columns we show the same quantities but for the target 
model. The last two columns contains the overlap maximized only over 2 extrinsic 
parameters $\rho_{\rm max,2}$ and maximed over 5 extrinsic parameters 
$\rho_{\rm max, 5}$, as described in the text. 
\label{overlapmax2pn}}
\end{table}

\begin{table}
\begin{tabular}{c|ccccccc}
\hline
\hline
$(\theta_{\rm S1}, \phi_{\rm S2}, \theta_{\rm S1}, \phi_{\rm S2}) $
& $f_{\rm fin}^{\rm SEP, noF_L}$ (Hz) & 
$f_{\rm LSSO}^{\rm SEP, noF_L}$ (Hz) & ${\rm R}_{\rm fin}^{\rm SEP, noF_L}/M$ 
& $f_{\rm fin}^{\rm SEP}$(Hz) & $f_{\rm LSSO}^{\rm SEP}$ (Hz) &
${\rm R}_{\rm fin}^{\rm SEP}/M$ & $\rho_{\rm max, 5}$ \\
\hline
\hline
\multicolumn{8}{l}{\hspace{6cm}$(10+10)M_\odot$}\\
\hline
\hline
$(60^o, 90^o, 60^o, 0^o)$ & 633 & 872 & 2.7 & 660 & 1211 &2.7 & 0.9860 \\ 
$(120^o, 90^o, 120^o, 0^o)$ & 187 & 186 & 6.7 & 186 & 186 &6.7 & 0.9953 \\ 
\hline
\hline
\multicolumn{8}{l}{\hspace{6cm}$(15+5)M_\odot$} \\
\hline
\hline
$(60^o, 90^o, 60^o, 0^o)$ & 743 & 767 & 2.3 & 564 & 564 &2.9 & 0.9839 \\ 
$(120^o, 90^o, 120^o, 0^o)$ & 178 & 177 & 7.0 & 179 & 179 &6.9 & 0.9969 \\ 
\hline
\hline
\end{tabular}
\caption{Effect of radiation-reaction force along $L$ over the binary evolution and 
waveforms by comparing SEHPF(3,3.5) with no $F_L$, used as target model, and 
SEHPF(3,3.5), used as template model, for several binary masses and 
initial spin orientations. 
The two black holes are assumed to carry maximal spins $\chi_1=\chi_2=1$. 
The evolution starts at $f_{\rm in} = 30$ Hz. In the first three columns 
we list the ending frequency, the LSSO frequency  and the 
BH radial separation at $t_{\rm fin}$ for the template model and 
in the second three columns we show the same quantities but for the target 
model. The last two columns contains the overlap maximized only over 2 extrinsic 
parameters $\rho_{\rm max,2}$ and maximed over 5 extrinsic parameters 
$\rho_{\rm max, 5}$, as described in the text. [Notice that for the spin configuration 
$(60^o, 90^o, 60^o, 0^o)$ and masses $(10+10)M_\odot$ and $(15+5)M_\odot$, due to 
a pole in the Pad\'e-approximant flux, we apply the Pad\'e resummation only to 
the non-spinning terms in the flux.]    
\label{overlapnoFL}}
\end{table}

\begin{table}
\begin{tabular}{c|ccccccc}
\hline
\hline
$(\theta_{\rm S1}, \phi_{\rm S2}, \theta_{\rm S1}, \phi_{\rm S2}) $
& $f_{\rm fin}^{\rm SET}$ (Hz) & 
$f_{\rm LSSO}^{\rm SET}$(Hz) & 
${\rm R}_{\rm fin}^{\rm SET}/M$ & $f_{\rm fin}^{\rm SEP}$(Hz) & 
$f_{\rm LSSO}^{\rm SEP}$(Hz) & 
${\rm R}_{\rm fin}^{\rm SEP}/M$ & $\rho_{\rm max, 5}$ \\
\hline
\hline
\multicolumn{8}{l}{\hspace{6cm}$(10+10)M_\odot$}\\
\hline
\hline
$(60^o, 90^o, 60^o, 0^o)$ & 632 & 872 & 2.6 & 616 & 1252 &2.6 & 0.8566 \\ 
$(120^o, 90^o, 120^o, 0^o)$ & 185 & 185 & 6.7 & 186 & 185 &6.7 & 0.9762 \\ 
\hline
\hline
\multicolumn{8}{l}{\hspace{6cm}$(15+5)M_\odot$} \\
\hline
\hline
$(60^o, 90^o, 60^o, 0^o)$ & 743 & 767 & 2.3 & 661 & 772 &2.5 & 0.8232 \\ 
$(120^o, 90^o, 120^o, 0^o)$ & 178 & 177 & 7.0 & 179 & 178 &6.9 & 0.9913 \\ 
\hline
\hline
\end{tabular}
\caption{Effect of Pad\'e and Taylor flux on the binary evolution and 
waveforms by comparing SEHTF(3,3.5), used as target model, and 
SEHPF(3,3.5), used as template model, for several binary masses and initial 
spin orientations. 
The two black holes are assumed to carry maximal spins $\chi_1=\chi_2=1$. 
The evolution starts at $f_{\rm in} = 30$ Hz. In the first three columns 
we list the ending frequency, the LSSO frequency  and the 
BH radial separation at $t_{\rm fin}$ for the template model and 
in the second three columns we show the same quantities but for the target 
model. The last two columns contains the overlap maximized only over 2 extrinsic 
parameters $\rho_{\rm max,2}$ and maximed over 5 extrinsic parameters 
$\rho_{\rm max, 5}$, as described in the text. [Notice that for the spin configuration 
$(60^o, 90^o, 60^o, 0^o)$ and masses $(10+10)M_\odot$ and $(15+5)M_\odot$, due to 
a pole in the Pad\'e-approximant flux, we apply the Pad\'e resummation only to 
the non-spinning terms in the flux.]  
\label{overlapflux}}
\end{table}

\begin{table}
\begin{tabular}{c|ccccccc}
\hline
\hline
$(\theta_{\rm S1}, \phi_{\rm S2}, \theta_{\rm S1}, \phi_{\rm S2}) $
& $f_{\rm fin}^{\rm SEP, noQM}$ (Hz) & $f_{\rm LSSO}^{\rm SEP, noQM}$(Hz) & 
${\rm R}_{\rm fin}^{\rm SEP, noQM}/M$ & $f_{\rm fin}^{\rm SEP}$(Hz) & 
$f_{\rm LSSO}^{\rm SEP}$(Hz) & ${\rm R}_{\rm fin}^{\rm SEP}/M$ 
& $\rho_{\rm max, 5}$ \\
\hline
\hline
\multicolumn{8}{l}{\hspace{6cm}$(10+10)M_\odot$}\\
\hline
\hline
$(60^o, 90^o, 60^o, 0^o)$ & 650 & 1257 & 2.6 & 633 & 872 & 2.6 & 0.9959 \\ 
$(120^o, 90^o, 60^o, 0^o)$ & 185 & 184 & 6.8 & 186 & 186 & 6.7 & 0.9988 \\ 
\hline
\hline
\multicolumn{8}{l}{\hspace{6cm}$(15+5)M_\odot$} \\
\hline
\hline
$(60^o, 90^o, 60^o, 0^o)$ & 702 & 766 & 2.3 & 743 & 767 & 2.3 & 0.9823 \\ 
$(120^o, 90^o, 120^o, 0^o)$ & 178 & 178 & 7.0 & 178 & 177 & 7.0 & 0.9979 \\ 
\hline
\hline
\end{tabular}
\caption{Effect of quadrupole-monopole (QM) interaction on the binary evolution and 
waveforms by comparing SEHPF(3,3.5) with QM interaction, 
used as target model, and SEHPF(3,3.5) without QM terms, 
used as template model, for several binary masses and initial spin orientations. 
The two black holes are assumed to carry maximal spins $\chi_1=\chi_2=1$. 
The evolution starts at $f_{\rm in} = 30$ Hz. In the first three columns 
we list the ending frequency, the LSSO frequency  and the 
BH radial separation at $t_{\rm fin}$ for the template model and 
in the second three columns we show the same quantities but for the target 
model. The last two columns contains the overlap maximized only over 2 extrinsic 
parameters $\rho_{\rm max,2}$ and maximed over 5 extrinsic parameters 
$\rho_{\rm max, 5}$, as described in the text. [Notice that for the spin configuration 
$(60^o, 90^o, 60^o, 0^o)$ and masses $(10+10)M_\odot$ and $(15+5)M_\odot$, due to 
a pole in the Pad\'e-approximant flux, we apply the Pad\'e resummation only to 
the non-spinning terms in the flux.]   
\label{overlapnoQM}}
\end{table}

\begin{table}
\begin{tabular}{c|ccccccc}
\hline
\hline
$(\theta_{\rm S1}, \phi_{\rm S2}, \theta_{\rm S1}, \phi_{\rm S2}) $
& $f_{\rm fin}^{\rm SEP, adiab}$ (Hz) & 
$f_{\rm LSSO}^{\rm SEP, adiab}$(Hz) & 
${\rm R}_{\rm fin}^{\rm SEP, adiab}/M$ & $f_{\rm fin}^{\rm SEP}$(Hz) & 
$f_{\rm LSSO}^{\rm SEP}$(Hz) & 
${\rm R}_{\rm fin}^{\rm SEP}/M$ & $\rho_{\rm max, 5}$ \\
\hline
\hline
\multicolumn{8}{l}{\hspace{6cm}$(10+10)M_\odot$}\\
\hline
$(60^o, 90^o, 60^o, 0^o)$ & 636 & 1185 & 2.6 & 633 & 872 & 2.7 & 0.9666 \\ 
$(120^o, 90^o, 120^o, 0^o)$ & 186 & 185 & 6.7 & 186 & 186 & 6.7 & 0.9932  \\ 
\hline
\hline
\multicolumn{8}{l}{\hspace{6cm}$(15+5)M_\odot$} \\
\hline
\hline
$(60^o, 90^o, 60^o, 0^o)$ & 699 & 827 & 2.3 & 743 & 767 & 2.3 & 0.9665  \\ 
$(120^o, 90^o, 120^o, 0^o)$ & 177 & 177 & 7.0 & 178 & 177 & 7.0 &  0.9914  \\ 
\hline
\hline
\end{tabular}
\caption{Effect of assuming that spins evolve adiabatically. 
We compare SEHPF(3,3.5), used as target model, and SEHPF(3,3.5) obtained 
by averaging the spin couplings over an orbit, 
as template model, for several binary masses and initial spin orientations. 
The two black holes are assumed to carry maximal spins $\chi_1=\chi_2=1$. 
The evolution starts at $f_{\rm in} = 30$ Hz. In the first three columns 
we list the ending frequency, the LSSO frequency  and the 
BH radial separation at $t_{\rm fin}$ for the template model and 
in the second three columns we show the same quantities but for the target 
model. The last two columns contains the overlap maximized only over 2 extrinsic 
parameters $\rho_{\rm max,2}$ and maximed over 5 extrinsic parameters 
$\rho_{\rm max, 5}$, as described in the text. [Notice that for the spin configuration 
$(60^o, 90^o, 60^o, 0^o)$ and masses $(10+10)M_\odot$ and $(15+5)M_\odot$, due to 
a pole in the Pad\'e-approximant flux, we apply the Pad\'e resummation only to 
the non-spinning terms in the flux.]   
\label{overlapadiab}}
\end{table}

\begin{table*}
\begin{tabular}{c|c|ccc|ccc}
\hline
\hline
$(\theta_{\rm S1}, \phi_{\rm S1}, \theta_{\rm S2}, \phi_{\rm S2}) $
  & $[\delta E_H]_{f<40\,{\rm Hz}}/M$  & $f_{\rm LSSO}$ (Hz) & $[\delta E_H]_{\rm LSSO}^{40\,{\rm Hz}}/M$ 
  & $\left[|\mathbf{J}|/E^2\right]_{\rm LSSO}$ 
  & $f_{\rm fin}$ &
   $\left[\delta E_H\right]_{\rm fin}^{40,{\rm Hz}}/M$
&$\left[|\mathbf{J}|/E^2\right]_{\rm fin}$\\
\hline
\hline
\multicolumn{8}{l}{\hspace{6cm}$(15+15)M_\odot$, 3PN}\\
\hline
\hline
nospin &   0.0082 & 190 & 0.0107 & 0.82 & 325 & 0.0182 & 0.77  \\
\hline
($0^\circ$,$0^\circ$,$0^\circ$,$0^\circ$) &
0.0086 & (1430) &   $-$ & $-$ & 474 & 0.0527 &  0.96 \\
($180^\circ$,$0^\circ$,$180^\circ$,$0^\circ$) &
0.0077 & 97 &   0.0033 & 0.51 & 194 & 0.0064 &  0.47 \\
($60^\circ$,$90^\circ$,$60^\circ$,$0^\circ$) &
0.0084 & (760) &  $-$ & $-$ & 440 & 0.0352 &  0.91 \\
($120^\circ$,$90^\circ$,$120^\circ$,$0^\circ$) &
0.0079 & 123 &   0.0054 & 0.74 & 242 & 0.0101 &  0.70 \\
\hline\hline
\multicolumn{8}{l}{\hspace{6cm}$(15+5)M_\odot$, 3PN}\\
\hline\hline
nospin & 0.0048 &  265  & 0.0084 & 0.62 & 484 & 0.0141 & 0.58 \\
\hline
($0^\circ$,$0^\circ$,$0^\circ$,$0^\circ$) &
0.0049 & (1442) &   $-$ & $-$ & 819 & 0.0493 &  0.95 \\
($180^\circ$,$0^\circ$,$180^\circ$,$0^\circ$) &
0.0046 & 140 &   0.0034 & 0.14 & 289 & 0.0054 &  0.11 \\
($60^\circ$,$90^\circ$,$60^\circ$,$0^\circ$) &
0.0049& (793) &  $-$ & $-$ & 719  & 0.0294 &  0.91 \\
($120^\circ$,$90^\circ$,$120^\circ$,$0^\circ$) &
0.0047 & 
177 &   0.0049 & 0.62 & 351 & 0.0080 &  0.60 \\
\hline
 \end{tabular}
\caption{Energy released and the magnitude of angular momentum
through the evolution (with spin-spin terms ignored). For non-spinning
binaries, and four configurations of maximally spinning binaries, we
give the energy released below 40\,Hz, from 40\,Hz up to the LSSO, and
from 40\,Hz up to to the end of the evolution. [In some cases the
evolution stops before LSSO can be reached.] We also show the
corresponding values of $|\mathbf{J}|/E^2$. Note that these results 
do not include the ring-down phase. 
\label{tab:EJ}}
\end{table*}

\clearpage

\vspace{0.3cm}

\vspace{0.1cm}

\vspace{0.1cm}

\vspace{0.1cm}

\vspace{0.1cm}

\vspace{0.1cm}

\vspace{0.1cm}

\vspace{0.1cm}

\vspace{0.1cm}

\vspace{0.1cm}

\vspace{0.1cm}

\vspace{0.1cm}

\end{document}